\documentclass[%
 reprint,
 amsmath,amssymb,
 aps,
]{revtex4-2}
\usepackage{multirow}
\usepackage{graphicx}
\usepackage{dcolumn}
\usepackage{bm}
\usepackage{amsmath,amssymb,amsfonts}
\usepackage{mathrsfs}
\usepackage{bbm}
\usepackage{slashed}
\usepackage{verbatim}
\usepackage[T1]{fontenc}
\usepackage[utf8]{inputenc}
\usepackage{hyperref}
\usepackage{ifthen}
\usepackage{forloop}
\usepackage[english]{babel}
\usepackage{mathbbol}
\usepackage[usenames,dvipsnames]{xcolor}
\usepackage[normalem]{ulem}
\usepackage{tabularx, booktabs}
\usepackage{subcaption}

\newcommand{\bea}{\begin{eqnarray}}
\newcommand{\eea}{\end{eqnarray}}
\newcommand{\be}{\begin{equation}}
\newcommand{\ee}{\end{equation}}

\definecolor{darkgreen}{rgb}{0,0.4,0}

\graphicspath{{figures/}}

\begin{document}


\title{Dynamical Dark Energy from Lattice Quantum Gravity}

\author{Mingwei Dai}
\email{mdai07@syr.edu}
\affiliation{Department of Physics, Syracuse University, Syracuse, NY 13244}
\author{Walter Freeman}
\email{wafreema@syr.edu}
\affiliation{Department of Physics, Syracuse University, Syracuse, NY 13244}
\author{Jack Laiho}
\email{jwlaiho@syr.edu}
\affiliation{Department of Physics, Syracuse University, Syracuse, NY 13244}
\author{Marc Schiffer}
\email[]{mschiffer@perimeterinstitute.ca}
\affiliation{Perimeter Institute for Theoretical Physics, 31 Caroline St. N., Waterloo, ON N2L 2Y5, Canada}
\author{Judah Unmuth-Yockey}
\email{jfunmuthyockey@gmail.com}
\affiliation{Fermi National Accelerator Laboratory, Batavia, IL 60510}

\begin{abstract}
We study the behavior of the vacuum in Euclidean dynamical triangulations (EDT).  Algorithmic improvements and better lattice spacing determinations allow us to test the properties of the emergent de Sitter geometries of our simulations to higher precision than previously possible.  Although the agreement with de Sitter is good, the improved precision reveals deviations that can be interpreted as non-trivial vacuum dynamics, well-described by a cosmological constant that runs with scale.  The simulations show that the dominant running is quadratic and that the scale can be identified with the Hubble rate.  Several key cross-checks support this picture, including consistent results across multiple lattice spacings and the fact that the null energy condition is not violated.  The parameters of the running are fully determined by simulations, enabling predictions when  extrapolated to the scales relevant for our universe.  This leads to a model for dark energy that is compatible with current observations, but which predicts deviations from $\Lambda$CDM at the ${\cal O}(10^{-3})$ level in cosmological observables that could be tested with future improvements in  precision measurements.
\end{abstract}
\maketitle

\section{\label{sec:intro}Introduction}

The formulation of quantum gravity is one of the great outstanding problems in theoretical physics.  The most straightforward quantization of general relativity as a field theory using a small coupling expansion is not renormalizable by power counting, and explicit calculation shows that there are counterterms at two-loop order in pure gravity \cite{Goroff:1985th} and at one-loop order when including matter fields \cite{tHooft:1974toh}.  Although the theory can be viewed as a low energy effective theory \cite{Donoghue:1994dn}, there is a loss of predictive power due to the need to introduce new parameters at each order in the perturbative expansion.     

Another problem with gravity is the difficulty explaining the smallness of the vacuum energy.  The vacuum energy should see contributions from quantum fluctuations of all fields in a theory, including graviton degrees of freedom, and these should contribute at scales up to the Planck scale, absent any special symmetries.  Although this leads to an unobservable overall shift in the energy of a flat-space quantum field theory, gravity should be sensitive to all energy sources, including the vacuum energy.  A bare cosmological constant would have to be extraordinarily fine-tuned to cancel these vacuum fluctuations if it is to produce the very small renormalized cosmological constant observed in our universe \cite{Weinberg:1988cp}.  This is the cosmological constant fine-tuning problem.

Weinberg introduced the asymptotic safety scenario \cite{Weinberg:1980gg} as a possible solution to the first problem, where a quantum field theory of general relativity, possibly supplemented by higher curvature operators, would be renormalizable nonperturbatively.  
This scenario would be realized if there existed a non-trivial fixed point, with a finite-dimensional ultraviolet critical surface of trajectories attracted to the fixed point at short distances.  Investigations of this scenario require nonperturbative methods, since the couplings at the fixed point may not be small.  Efforts in this direction have mainly included functional renormalization group \cite{Wetterich:1992yh, Reuter:1996cp, Percacci:2007sz, Reuter:2012id, Eichhorn:2018yfc, Bonanno:2020bil} and lattice \cite{Ambjorn:2005db, Ambjorn:2005qt, Ambjorn:2007jv, Laiho:2016nlp} methods.  This work focuses on Euclidean dynamical triangulations (EDT), one of the first attempts at a lattice formulation of gravity \cite{Agishtein:1991cv, Ambjorn:1991pq}.  In order to successfully realize the asymptotic safety scenario, the phase diagram of EDT must have a continuous phase transition, the approach to which would define a continuum limit.  The physics of the universality class associated with the phase transition must also lead to general relativity in the classical limit if it is to reproduce our world.
A final requirement for any successful theory of gravity is that it must lead to predictions beyond the classical theory that can be tested against observations.

Recent simulations of EDT suggest that it has a semiclassical limit consistent with general relativity \cite{Laiho:2016nlp, Asaduzzaman:2022kxz, Dai:2023tud}.  It was shown that the fine-tuning of a parameter associated with a local measure term \cite{Bruegmann:1992jk} is necessary to recover physical geometries, and a prescription for performing this tuning and for taking a continuum limit was suggested \cite{Laiho:2016nlp}.
Evidence for the emergence of de~Sitter space in EDT was presented in Refs.~\cite{Laiho:2016nlp, Asaduzzaman:2022kxz, Dai:2023tud}, where it was shown that the global Hausdorff dimension is compatible with four and that the agreement with the classical de~Sitter solution improves as the lattice spacing gets smaller.  Ref.~\cite{Bassler:2021pzt} showed that the Euclidean partition function is dominated by the de~Sitter instanton, and using a saddle-point approximation, used this to determine Newton's constant. This sets the Planck length in lattice units, thus determining the absolute lattice spacing of the simulated ensembles.  Calculations with quenched matter, including scalars \cite{Dai:2021fqb} and K\"ahler-Dirac fermions \cite{Catterall:2018dns}, are consistent with the expectations for gravitational interactions of matter fields.

Although there is no obvious barrier to taking the continuum limit in EDT with the appropriate tuning (other than the practical issues confronting a numerical simulation), more work is needed.  To this end, Ref.~\cite{Dai:2023tud} introduced a new rejection free algorithm that leads to a significant speedup of the simulations in the region of the phase diagram where finer lattices are expected.  Ref.~\cite{Dai:2023tud} also provided a first look at the new, finer ensembles that became feasible to generate using the new algorithm.  In this work we turn to a detailed study of these new ensembles.  We find that some of our previous methods for determining the lattice spacing within EDT lead to ambiguous results on our new ensembles, which makes these methods insufficient for meeting our target precision.  In previous work \cite{Laiho:2016nlp, Bassler:2021pzt}, a diffusion process on the geometries was used to set various relative lattice spacings.  
We revisit these determinations here, especially the ratio between the bare direct lattice spacing and the renormalized dual lattice spacing.  We show that this ratio can be computed in multiple ways and that the good agreement between these determinations gives confidence both in the results and in the assumptions that go into the determination.  

The machinery introduced here and in Ref.~\cite{Dai:2023tud} thus enables a strong test of our simulation results against expectations from the semiclassical de Sitter solution.  This new precision allows us to show that despite the good overall agreement with this picture, there are deviations from semiclassical de Sitter that can be associated with nontrivial vacuum dynamics.  This dynamics is well-described by a power-law running of the cosmological constant, where the renormalization scale is associated with the Hubble scale.  This coincides with one of the simplest of a family of running vacuum models studied by Sol\`a and collaborators \cite{Shapiro:2009dh, Sola:2013gha, Gomez-Valent:2015pia, Sola:2015rra, Sola:2016jky, Moreno-Pulido:2022phq,  Moreno-Pulido:2023ryo}. These models have been compared to observations, thus constraining the free parameters of the models and testing whether the new physics implied by these models can resolve the emerging tensions in observational cosmology \cite{Sola:2016jky, SolaPeracaula:2021gxi}.  We are able to turn this around by fitting lattice data directly to a specific running vacuum model, thus determining the free parameters of the model from first principles via our calculations.  We find that only one parameter, a dimensionless coupling, is compatible with a nonzero value. 
The evidence for this parameter being different from zero is compelling, with high statistical significance and good agreement across four lattice spacings, making it unlikely that this effect is merely a lattice artifact.  

The sign of the effect seen here 
ensures that it does not violate the null-energy condition.  There are strong constraints on theories that are inconsistent with this condition \cite{Carroll:2003st, Dubovsky:2005xd, Anber:2009qp}, so the fact that our lattice result does not run afoul of 
it is a significant point in its favor.  Although our results are found on lattices that are no larger than several Planck volumes, it is interesting to ask what would happen if the model singled out by this work was assumed to hold throughout cosmic history.  One consequence is that the running of the vacuum energy would lead to a large vacuum energy density in the early universe compared to the current era.  It turns out that the computed size of the coupling of the model ensures that the effect of the new vacuum dynamics on the evolution of the universe remains subdominant throughout cosmic history, thus making it compatible with present observational bounds.  Nonetheless, the corrections to standard cosmology would not be negligible, with the effects of the running vacuum giving rise to deviations from the $\Lambda$CDM model at the part-per-mil level across many cosmological observables, including the dark energy equation of state.

This paper is organized as follows.  In Section~\ref{sec:edt} we review the EDT formulation and discuss the details of the simulations.  In Section~\ref{sec:desitter} we introduce two methods for determining the ratio of the direct to (renormalized) dual lattice spacing using the classical de Sitter solution.  Details and results from our numerical analysis of this ratio are presented.  We also use the semiclassical approximation, where the de Sitter instanton dominates the Euclidean partition function, to extract Newton's constant at each of our four lattice spacings, thus fixing the lattice scale in terms of the Planck scale.  In Section~\ref{sec:RVM} we show that the emergent behavior of the EDT geometries is consistent with a cosmological constant that exhibits a power law running with scale.  We use our lattice calculation to constrain this behavior, to provide a number of non-trivial cross-checks, and to determine the parameters of the running.  Subsection~\ref{sec:theorymotive} provides an introduction to the running vacuum model that we consider, while subsections~\ref{sec:modelParams}, \ref{sec:nuPrime_determination}, \ref{sec:nutilde}, and \ref{sec:alt_nuPrime} give an in depth treatment of the numerical analysis used to constrain the parameters of this model.  Subsection~\ref{sec:Implications_RVM} summarizes the model and provides the final parameter values that we obtain from our calculation.  In Section~\ref{sec:cosmology} we discuss the implications of the running vacuum model picked out by EDT for observational cosmology.  We conclude in Section~\ref{sec:conclude}.

\section{\label{sec:edt}Lattice Formulation}

\subsection{Euclidean Dynamical Triangulations}

The Euclidean quantum gravity partition function is formally given by a path integral over all geometries
\bea\label{eq:part}  Z_E = \int {\cal D}[g] e^{-S_{EH}[g]},
\eea
where the Euclidean Einstein-Hilbert action is
\bea  \label{eq:ERcont} S_{EH} =  -\frac{1}{16 \pi G}\int d^4x \sqrt{g} (R - 2\Lambda),
\eea
with $R$ the Ricci curvature scalar, $g$ the determinant of the metric tensor, $G$ Newton's constant, and $\Lambda$ the cosmological constant.
    
The path integral of dynamical triangulations is formulated directly as a sum over geometries, without the need for gauge fixing or the introduction of a metric.  The dynamical triangulations approach is based on the conjecture that the path integral for Euclidean gravity is given by the partition function \cite{Ambjorn:1991pq, Bilke:1998vj}
\bea\label{eq:ZE} Z_E = \sum_T \frac{1}{C_T}\left[\prod_{j=1}^{N_2}{\cal O}(t_j)^\beta\right]e^{-S_{ER}},
\eea
where $N_i$ is the number of simplices of dimension $i$, and $C_T$ is a symmetry factor that divides out the number of equivalent ways of labeling the vertices in the triangulation $T$.  The term in brackets in Eq.~(\ref{eq:ZE}) is a nonuniform measure term \cite{Bruegmann:1992jk, Bilke:1998vj}, where the product is over all triangles, and ${\cal O}(t_j)$ is the order of triangle $j$, i.e. the number of four-simplices to which the triangle belongs.  This corresponds in the continuum to a nonuniform weighting of the measure in Eq.~(\ref{eq:part}) by $\prod_x \sqrt{g}^{\beta}$, where we leave $\beta$ an adjustable parameter in the simulations.  

  In four dimensions the discretized version of the Einstein-Hilbert action is the Einstein-Regge action \cite{Regge:1961px}
  \begin{equation} \label{eq:GeneralEinstein-ReggeAction}
S_{E}=-\kappa\sum_{j=1}^{N_2} V_{2}\delta_j+\lambda\sum_{j=1}^{N_4} V_{4},
\end{equation}
  \noindent where $\delta_j=2\pi-{\cal O}(t_j)\arccos(1/4)$ is the deficit angle around a triangular hinge $t_j$, with ${\cal O}(t_j)$ the number of four-simplices meeting at the hinge, $\kappa=\left(8\pi G \right)^{-1}$, $\lambda=\kappa\Lambda$, and the volume of a $d$-simplex is 
\begin{equation} \label{eq:SimplexVolume}
V_{d}=\frac{\sqrt{d+1}}{d!\sqrt{2^{d}}}a_{\rm lat}^d,
\end{equation}
\noindent where the equilateral $d$-simplex has a side of length $a_{\rm lat}$.  After performing the sums in Eq.~(\ref{eq:GeneralEinstein-ReggeAction}) one finds
\begin{equation}\label{eq:DiscAction}
S_{E}=-\frac{\sqrt{3}}{2}\pi\kappa N_{2}+N_{4}\left(\kappa\frac{5\sqrt{3}}{2}\mbox{arccos}\frac{1}{4}+\frac{\sqrt{5}}{96}\lambda\right).
\end{equation}
We rewrite the Einstein-Regge action in the simple form 
\bea\label{eq:ER}  S_{ER}=-\kappa_2 N_2+\kappa_4N_4,
\eea
introducing the parameters $\kappa_4$ and $\kappa_2$ in place of $\kappa$ and $\lambda$ for convenience in the numerical simulations.

\subsection{Details of the simulations}


Geometries are constructed by gluing together four-simplices along their ($4-1$)-dimensional faces.  The four-simplices are equilateral, with fixed edge length $a_{\rm lat}$.  The set of all four-geometries is approximated by gluing together four-simplices, and the dynamics is encoded in the connectivity of the simplices.  We work with a set of degenerate triangulations in which the combinatorial manifold constraints are relaxed.  The set of degenerate triangulations is larger than that of combinatorial triangulations, and the degeneracy allows distinct four-simplices to share the same 5 distinct vertex labels \cite{Bilke:1998bn}.  It was shown that the finite size effects of degenerate triangulations are a factor of $\sim$10 smaller than those of combinatorial triangulations \cite{Bilke:1998bn}, giving it some advantages over combinatorial triangulations.  A comparison between the two approaches shows no essential difference in the phase diagram between degenerate and combinatorial triangulations in four dimensions \cite{Ambjorn:2013eha, Coumbe:2014nea}.

We use Markov Chain Monte Carlo methods to evaluate the partition function in Eq.~(\ref{eq:ZE}).  In particular, we use the new rejection free algorithm introduced in Ref.~\cite{Dai:2023tud}, which is tailored for EDT simulations.  This algorithm mimics the Metropolis-Hastings algorithm in that the partition function is sampled using a Markov chain constructed from a series of local moves, but the moves are always accepted, with a weight that is used to reconstruct the proper distribution.  This has allowed us to push to finer lattice spacings than were previously accessible, since the region of the phase diagram that corresponds to finer lattice spacings is characterized by a very low Metropolis acceptance rate.  This has led to a speedup of one to two orders of magnitude compared to standard Metropolis in the relevant part of the phase diagram.  Thus, almost all of the ensembles used in this work have been generated with the new rejection free algorithm.

The simulations are done at approximately fixed lattice
four-volumes. The local moves that evolve the lattices in Monte Carlo time are the same Pachner moves \cite{Gross:1991je} that are used in the standard Metropolis-Hastings algorithm.  The Pachner moves are topology preserving, making it straightforward to restrict the sum over geometries to those with fixed topology.  In this and previous works \cite{Laiho:2016nlp, Dai:2023tud} the topology is fixed to $S_4$.  The Pachner moves require the volume to
vary in order for them to be ergodic, though in practice
the volume fluctuations about the target four-volume are
constrained to be small. This is done by introducing a
volume preserving term in the action $\delta \lambda|N_4^f - N_4|$, which
keeps the four-volume close to a fiducial value $N_4^f$.  
In principle one
should take the limit in which $\delta\lambda$ goes to zero, but setting it too small leads to large volume fluctuations and long autocorrelation times.  We choose $\delta\lambda$ sufficiently large so that it does not lead to very long simulation times and sufficiently small that it does not lead to significant systematic errors; in this work we take
$\delta\lambda=0.04$.  Once the volume is specified,
the parameter $\kappa_4$ is constrained to a particular value, as discussed
in the following section. The other parameters, $\kappa_2$ and
$\beta$, form a two-dimensional parameter space for the phase
diagram in which to search for a fixed point.  

\begin{figure}
    \centering
    \includegraphics[width=\linewidth]{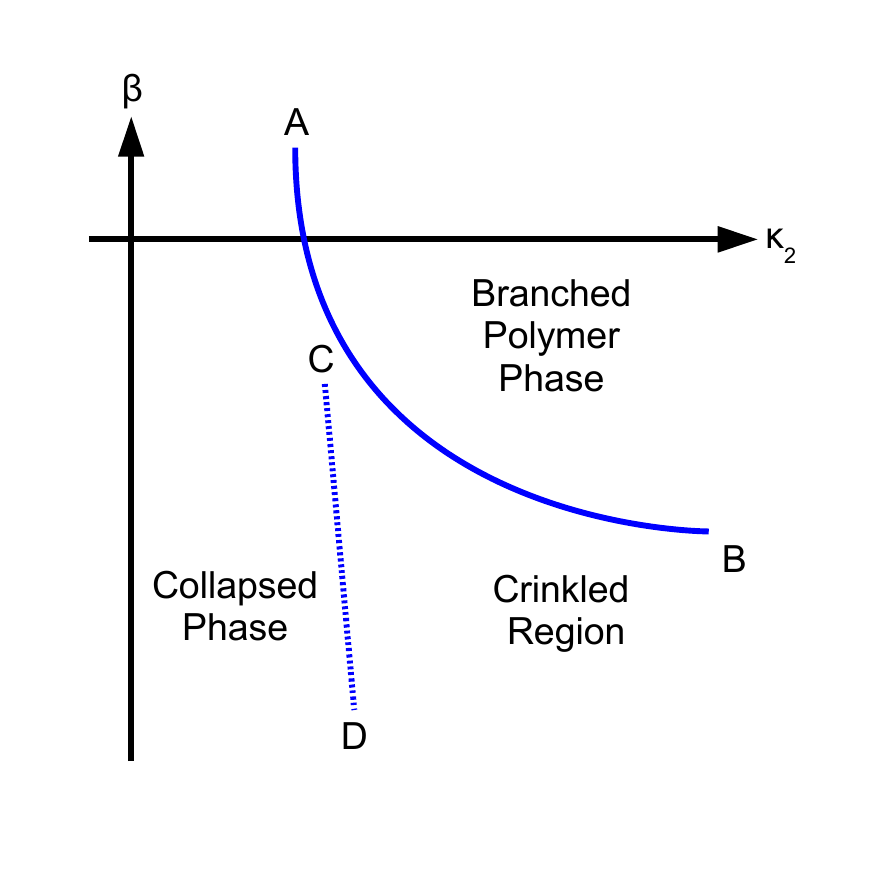}
    \caption{Schematic of the phase diagram as a function of $\kappa_2$ and $\beta$.}
    \label{fig:phaseDiagram}
\end{figure}

\subsection{\label{sec:phase}The phase diagram and recovering physical results}

The phase diagram for this model has been studied
in previous work \cite{Coumbe:2014nea, Laiho:2016nlp, Dai:2023tud}, and a schematic is shown in Fig. \ref{fig:phaseDiagram}.  There are two unphysical phases, the collapsed phase and the branched polymer phase, which were identified already in early EDT simulations \cite{Agishtein:1991cv, Ambjorn:1991pq, Catterall:1994pg, deBakker:1994zf, Ambjorn:1995dj, Egawa:1996fu}.  The
solid line AB is a first order line that separates
the two phases \cite{Bialas:1996wu, deBakker:1996zx, Ambjorn:2013eha, Coumbe:2014nea, Rindlisbacher:2015ewa}.  The phases are not smooth, with emergent fractal dimensions that differ from four.  One useful definition of fractal dimension is the Hausdorff dimension, which is a measure of geometry and is defined by the scaling of the volume of a sphere with its radius.
The branched polymer phase has Hausdorff dimension 2,
while the collapsed phase has a large dimension that may diverge in the infinite volume limit. The crinkled region and the collapsed phase
do not appear to be distinct phases; rather the crinkled
region appears to be connected to the collapsed phase
by an analytic crossover.  The crinkled region requires
large volumes to see the characteristic behavior of
the collapsed phase, suggesting that it is a part of the
collapsed phase with large finite-size effects \cite{Ambjorn:2013eha, Coumbe:2014nea}.

In the EDT formulation of gravity, one must tune to close to the first-order line in order to recover semiclassical physics \cite{Laiho:2016nlp}.  This tuning was originally motivated by an analogy to lattice QCD with Wilson fermions \cite{Wilson:1975id}, where one must tune the quark mass parameter to a critical value in order to take the zero-mass limit.  At the critical mass value of Wilson fermions there is a first-order phase transition (when using Symanzik improved lattice QCD actions \cite{Sharpe:1998xm}), and this softens to a continuous phase transition as one follows the first order line to zero gauge coupling, allowing the continuum limit to be taken.  Ref.~\cite{Laiho:2016nlp} showed that similar behavior is seen in EDT when one applies an analogous prescription.  Thus, tuning to the first-order transition line of EDT is needed to recover physical, semiclassical behavior, and following that line out to large, possibly infinite $\kappa_2$, is needed to approach the continuum limit.  The reason this tuning is necessary in EDT is still under debate.  In lattice QCD the need for the fine-tuning is well-understood to be due to the lattice regulator breaking the chiral symmetry of the continuum fermion action, thus requiring a fine-tuning of the bare fermion mass term in the lattice action.  There are still remnants of that explicit symmetry breaking at finite lattice spacing, and chiral symmetry is only fully restored in the continuum limit.  

Reference \cite{Laiho:2016nlp} argued that the analogy to QCD can be taken further and that the symmetry that is broken by the regulator in the case of EDT is continuum diffeomorphism invariance, where a fine-tuning of the local measure term is needed to recover the correct continuum theory.  Further evidence that this is the case is provided by the need to perform a subtraction in the bare couplings in order to recover the physical running, and by the presence of long-distance lattice artifacts that appear to vanish as the continuum limit is approached.  These are all signs that a symmetry is broken by the regulator.  If this is the case, and the associated symmetry is continuum diffeomorphism invariance, then there would be more relevant parameters in the lattice theory than in the symmetry preserving theory, since diffeomorphism invariance is expected to be exact even in the quantum theory, and additional parameters would need to be included and fine-tuned in order to restore the symmetry on the lattice.  This would mean that in the symmetry preserving theory there would be less than three relevant couplings.  In fact, Ref.~\cite{Laiho:2016nlp} argued that the ultraviolet (UV) critical surface of the symmetry preserving theory is one-dimensional, which would be the maximally predictive case.  This is in tension with results coming from truncations of the functional renormalization group, where calculations using many different truncations find a three-dimensional UV critical surface \cite{Codello:2007bd, Codello:2008vh, Benedetti:2009rx, Falls:2014tra} (but see the recent work \cite{Baldazzi:2023pep}, where a one-dimensional UV critical surface of essential operators was found). 

A different motivation for why the parameter $\beta$ needs to be tuned to recover a continuum limit while maintaining correct classical behavior comes from a comparison to FRG studies.  There, indications are found that a higher-curvature operator might become relevant at the asymptotically safe fixed point, see, e.g., \cite{Benedetti:2009rx, Denz:2016qks, Knorr:2021slg}\footnote{More recently, the \textit{essential scheme} has been proposed, which distinguishes essential couplings that enter scattering amplitudes from inessential couplings, which can be removed by field redefinition \cite{Baldazzi:2021ydj, Baldazzi:2021orb, Knorr:2022ilz, Baldazzi:2023pep}. For quantum gravity, all curvature squared operators are inessential, and hence the notion of relevant or irrelevant does not exist for those operators.}. To approach that fixed point on the lattice, the lattice action would need to have overlap with the relevant operators, and the corresponding lattice couplings would therefore need to be tuned. Exponentiating the local measure term in \eqref{eq:Z}, it can be interpreted as a part of the action with coupling $\beta$. Since the triangle-order ${\cal O}(t_j)$ is an ingredient of the Regge curvature (see \eqref{eq:GeneralEinstein-ReggeAction}) the resulting term in the action might correspond to a resummation of some higher-order curvature operators. Hence, tuning this term might lead to an approach to the fixed point discovered in FRG studies, see~\cite{Percacci:2017fkn, Reuter:2019byg, Saueressig:2023irs} for overviews.  If this  is the case, then the fixed point of the theory would have several physically realizable relevant directions, so that it would not be maximally predictive.  It is thus of high importance to decide between these possible scenarios.

The tuning of the bare lattice parameters to the phase transition is performed as follows:  The emergent shape of the lattice geometries differs markedly between the two phases, allowing us to use a measurement of the shape to identify the transition.  To do this, we introduce the shelling function $N_4^{\rm shell}(\tau)$, which is defined as the number of four-simplices lying exactly a geodesic distance $\tau$ away from a source four-simplex, comprising a spherical shell with the source at its center.  This function is averaged over multiple random sources on a given configuration and over all configurations of an ensemble.  We account for autocorrelation errors present in the data for the shelling function by blocking the data before averaging.  We study the variation of the error with block size, increasing the block size until the standard error obtained from a jackknife average over configurations no longer increases. The height of the peak (as a function of $\tau$) of the shelling function $N_{4,\mathrm{peak}}^{\rm shell}$ varies significantly across the phase transition and thus serves as a good order parameter for studying the phase diagram.

\begin{figure*}[htb]
\begin{minipage}{0.475\linewidth}
    \centering
    \includegraphics[width=\linewidth]{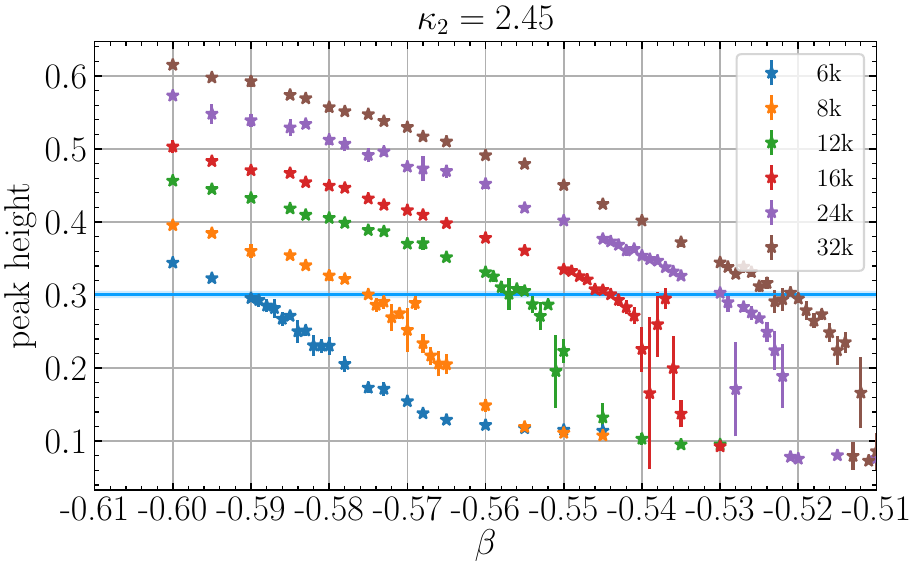}
\end{minipage}
\begin{minipage}{0.475\linewidth}
    \centering
    \includegraphics[width=\linewidth]{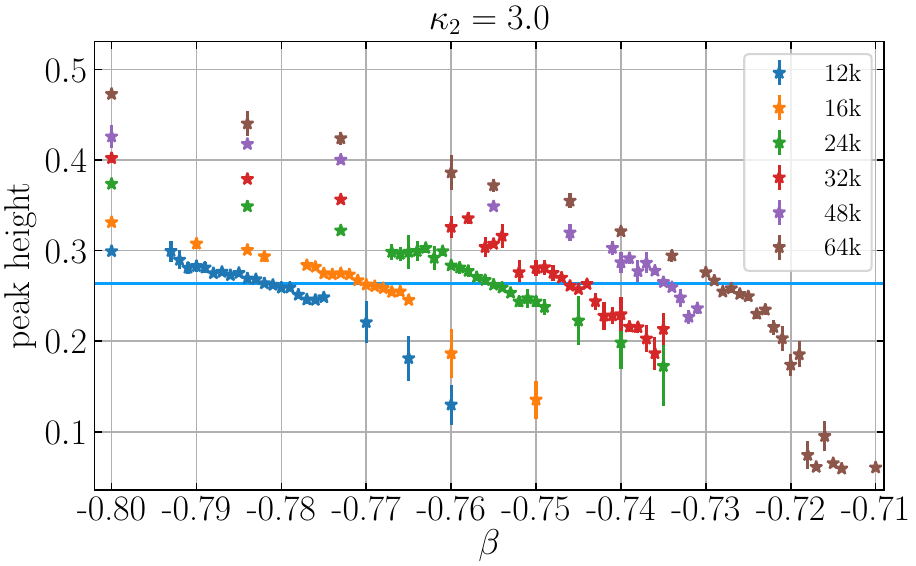}
\end{minipage}
\begin{minipage}{0.475\linewidth}
    \centering
    \includegraphics[width=\linewidth]{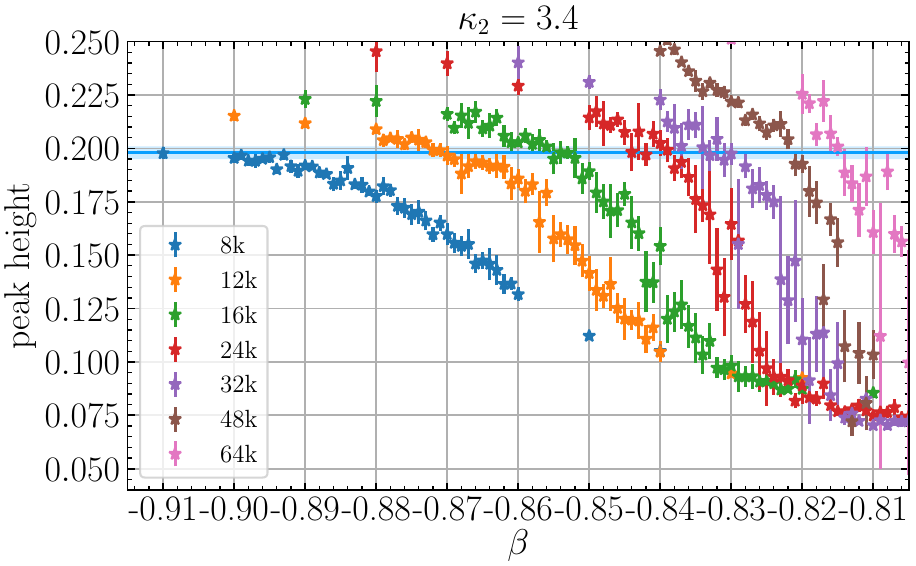}
\end{minipage}
\begin{minipage}{0.475\linewidth}
    \centering
    \includegraphics[width=\linewidth]{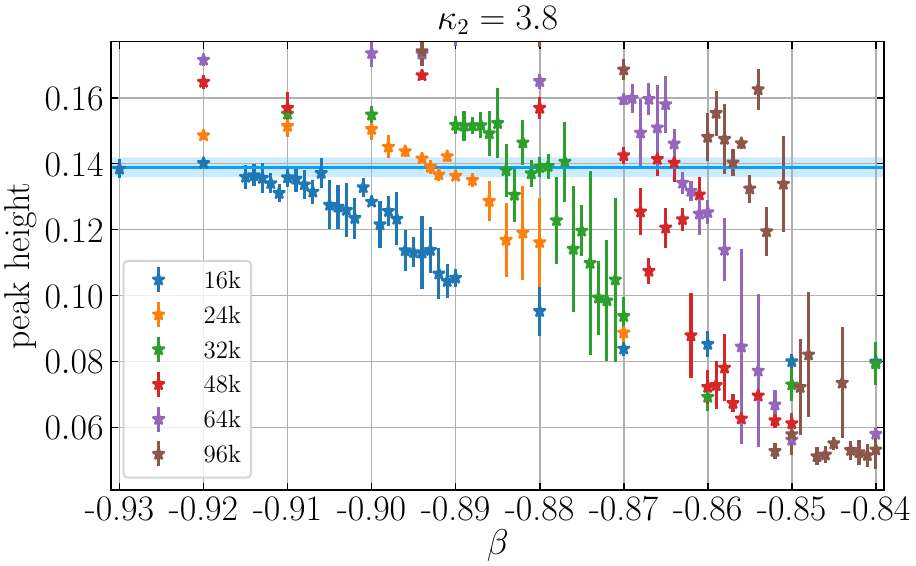}
\end{minipage}
\caption{Rescaled peak heights for multiple volumes at four lattice spacings. The horizontal cyan line represents the peak height to which the tuning is performed for all volumes at that lattice spacing. The volume picked for each lattice spacing is the 8k volume at $\kappa_2=2.45$, the 12k volume at $\kappa_2=3.0$, the 16k volume at $\kappa_2=3.4$, and the 32k volume at $\kappa_2=3.8$.
} 
\label{fig:peaktuning}
\end{figure*}

We consider the rescaled shelling function 
\bea n_4(\rho) = \frac{1}{N_4^{1-1/D_H}}N_4^{\rm shell}(N_4^{1/D_H}\rho),
\eea
where $\rho=\tau/N_4^{1/D_H}$ is the rescaled Euclidean distance, and $D_H$ is the Hausdorff dimension.  Fig. \ref{fig:peaktuning} shows a series of plots of the peak height of the rescaled shelling function $n_4(\rho)$ for ensembles with many different $\kappa_2$ and $\beta$ values in the vicinity of the phase transition.  In the rescaling of all of the peak heights, $D_H$ is set to four.  In Fig. \ref{fig:peaktuning}, the phase to the left (more negative beta) is the collapsed phase, and the phase to the right is the branched polymer phase.

As described in Ref.~\cite{Dai:2023tud}, there is a ``knee'' in the plots at each volume, just before the slope becomes large and negative, that marks the onset of the phase transition.  We take a fixed value of $\kappa_2$ to define our nominal lattice spacing, and we then tune $\beta$ by matching the rescaled peak heights in the region just to the left of the phase transition.   The $\beta$ value is chosen so that the tuned point is sufficiently far from the phase transition that there is no evidence of the tunneling between metastable states that is expected in the vicinity of a first-order phase transition.  Adjusting the rescaling of the peak height so that the location of the ``knee'' is at the same height across different volumes leads to a $D_H$ close to the physical value of four.  We thus assume that the peak height rescaling is four-dimensional, and we choose $\beta$ as a function of volume so that the rescaled peak heights match the rescaled peak height of one of our smaller volumes near the transition.  The tuned value is chosen so that a broad range of volumes can be adjusted so as to match the peak height after the rescaling.  If the volume is chosen too small at a given lattice spacing, it is not possible to reach the tuned peak height, and indeed, for too small of a physical volume, the identification with semiclassical geometry with $D_H=4$ is not possible.  The choice of the tuned peak height varies smoothly across the different lattice spacings, with its value given by the cyan band in each plot of Fig.~\ref{fig:peaktuning}.  It is the 8k volume at $\kappa_2=2.45$, the 12k volume at $\kappa_2=3.0$, the 16k volume at $\kappa_2=3.4$, and the 32k volume at $\kappa_2=3.8$.  As we show in Section~\ref{sec:desitter}, this collection of ensembles passes a broad set of consistency checks, including the recovery of the semiclassical limit, suggesting that this tuning procedure is a valid one. 

\subsection{Lattice ensembles}
\label{sec:ensembles}

We have generated a number of tuned lattice ensembles across a range of volumes and many different lattice spacings.  Table \ref{tab:ensembles} gives the parameters for the ensembles used in this work.  Almost all of these ensembles are new, having been generated by the rejection free algorithm introduced in Ref. \cite{Dai:2023tud}.  In order to have controlled calculations of the various quantities that we consider here, we require several volumes at each lattice spacing so that the large volume limit can be taken.  

The tuning of the bare lattice parameters described in the previous subsection requires that we simulate close to the phase transition, which means that the cost of generating ensembles grows rapidly with increasing volume.  One simple approach to speeding up the lattice generation, which we adopt, is to use multiple streams.  We branch multiple streams off of a single thermalized run and then allow the separate streams to run for a full autocorrelation time before we begin to take measurements on them.  As noted in the previous subsection we estimate the error using a jackknife resampling procedure, and to account for autocorrelation errors we block our data sets, increasing the block size until the jackknife error no longer increases.  The integrated autocorrelation time $\tau_{\rm int}$ can then be estimated using 
\bea \sigma^2=2 \tau_{\text {int }} \sigma_0^2,
\eea
where $\sigma_0$ is the naive jackknife error (assuming independent measurements) and $\sigma$ is the maximal binned jackknife error. Note that the integrated autocorrelation time is not the same quantity as the block size needed to account for the autocorrelation. Usually, the block size is much larger than the integrated autocorrelation time.

\begin{figure*}[ht]
\begin{minipage}{0.495\linewidth}
\centering
    \includegraphics[width=1\linewidth]{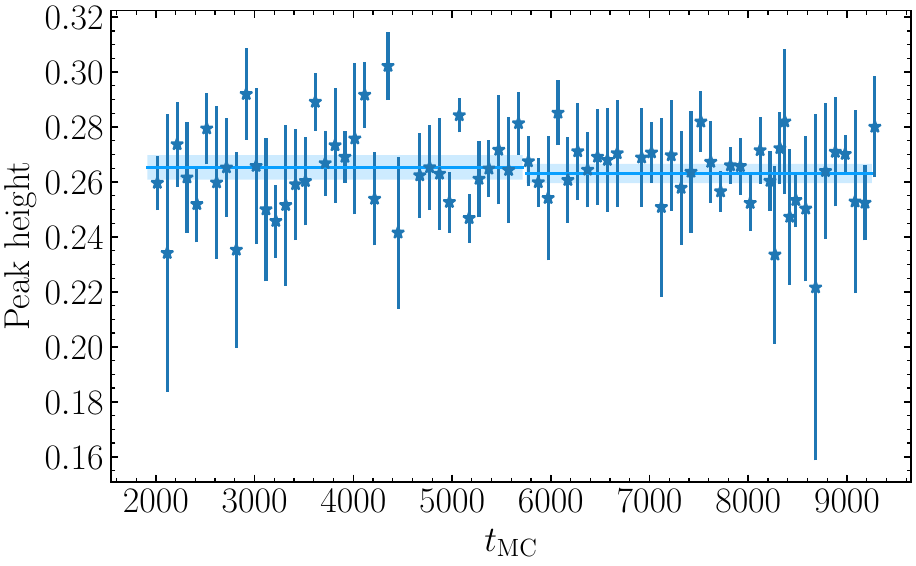}
\end{minipage}
\begin{minipage}{0.495\linewidth}
\centering
    \includegraphics[width=1\linewidth]{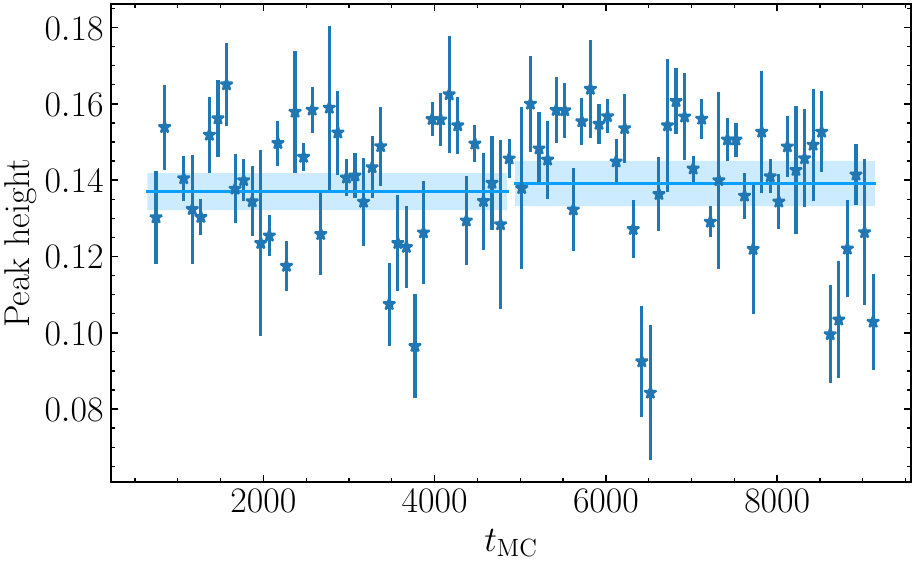}
\end{minipage}
    \caption{The rescaled peak height evolution for $N_4=16000$, $\kappa_2=3.0$ and $\beta=-0.771$ (left) and $N_4=32000$, $\kappa_2=3.8$ and $\beta=-0.88$ (right) ensembles in Monte Carlo time $t_{\rm MC}$ where $t_{\rm MC}$ is measured in number of configurations saved.}
    \label{fig:phmcevolution}
\end{figure*}

We choose an observable that is representative of the long-distance behavior of our geometries when we estimate the autocorrelation time and perform our thermalization tests, because those are expected to be the slowest to evolve through Monte Carlo time.  We use the peak height of the shelling function, since this is also closely related to the analysis that is the subject of this work. Fig. \ref{fig:phmcevolution} shows the evolution of the peak height with Monte Carlo time, measured in units of the number of saved configurations.  Each point in the plot corresponds to the mean of 100 consecutive saved configurations.  The correct block size determined from the larger data set is found to exceed 100 configurations, such that these points do not represent independent measurements. Thus, the error bars of the individual data points in the plots of Fig.~\ref{fig:phmcevolution} are expected to be underestimated. However, the main purpose of this plot is to view the evolution of the peak height with respect to MC time in order to ensure that the runs have thermalized. The horizontal bars overlaid on the plots are the mean and absolute error of the first half and the second half of the data sets. The agreement between the two halves of the runs indicate that the runs are thermalized and that the statistical errors are properly estimated.

\begin{center}
\begin{table}
\caption{Ensembles used in this work.  The first column is the $\kappa_2$ value of a given lattice spacing.  The second is the tuned $\beta$ value of the ensemble.  The third is the target volume in four-simplices of the ensemble. The fourth column is the number of configurations generated, and the fifth column is the autocorrelation time on that ensemble as determined by the peak height of the shelling function.  The sixth column is the ratio of the direct to renormalized dual lattice spacing at the given $\kappa_2$ value, and the seventh is the (renormalized dual) relative lattice spacing at the given $\kappa_2$ value. }
\label{tab:ensembles}
\begin{tabular}{c c c c c c c} 
\hline \hline
$\kappa_2$& $\beta$ & $N_4$ & $N_{\rm config}$ & $\tau_{\rm int}$ & $a_{\rm lat}/\ell$ & $\ell_{\rm rel}$\\
\hline
\multirow{6}{4em} {\centering 2.45}& -0.590 & 6000 & 2275 & 5 & \multirow{6}{4em} {\centering 9.02(19)} &  \multirow{6}{4em} {\centering 1.45(12)}\\
& -0.575 & 8000 & 4748 & 9 \\
& -0.555 & 12000 & 5145 & 15 \\
& -0.544 & 16000 & 8233 & 14 \\
& -0.530 & 24000 & 9041 & 27 \\
& -0.520 & 32000 & 10074 & 50 \\
\hline
\multirow{6}{4em} {\centering 3.0}& -0.800 & 8000 & 1486 & 67 & \multirow{6}{4em} {\centering 11.66(16)} &  \multirow{6}{4em} {\centering 1.00}\\
& -0.782 & 12000 & 17350 & 3 &\\
& -0.771 & 16000 & 21114 & 6 &\\
& -0.756 & 24000 & 30555 & 19 &\\
& -0.746 & 32000 & 36142 & 45 &\\
& -0.735 & 48000 & 24295 & 99 &\\
\hline
\multirow{9}{4em} {\centering 3.4} & -0.910 & 8000 & 2533 & 8 & \multirow{9}{4em} {\centering 16.34(19)} & \multirow{9}{4em} {\centering 0.618(49)} \\
& -0.870 & 12000 & 2781 & 21 &\\
& -0.853 & 16000 & 2282 & 8 &\\
& -0.839 & 24000 & 3242 & 49 &\\
& -0.830 & 32000 & 3431 & 36 &\\
& -0.822 & 48000 & 2832 & 16 \\
& -0.821 & 48000 & 2721 & 33 \\
& -0.815 & 64000 & 1430 & 15 \\
& -0.814 & 64000 & 1052 & 40 \\
\hline
\multirow{9}{4em} {\centering 3.8} & -0.920 & 16000 & 6096 & 10 & \multirow{9}{4em} {\centering 23.67(52)} & \multirow{9}{4em} {\centering 0.386(51)} \\
& -0.906 & 16000 & 2229 & 26 \\
& -0.893 & 24000 & 4866 & 24 \\
& -0.880 & 32000 & 13137 & 49 \\
& -0.864 & 48000 & 5805 & 154 \\
& -0.863 & 64000 & 10959 & 62 \\
& -0.864 & 64000 & 1902 & 49 \\
& -0.857 & 96000 & 37684 & 187 \\
& -0.855 & 96000 & 42803 & 1020 \\
\hline\hline
\end{tabular}
\end{table}
\end{center}

Table~\ref{tab:ensembles} shows the bare parameters $\kappa_2$ and $\beta$ used in the simulations, as well as the target number of four-simplices, $N_4$.  Once $N_4$ is chosen, $\kappa_4$ is tuned so that the four-volume fluctuates about the target volume.  Table~\ref{tab:ensembles} also gives the number of saved configurations in each ensemble, as well as the autocorrelation time $\tau$ of an ensemble, as measured by the Monte Carlo time series for the peak height of the shelling function. The ratio of the direct to renormalized dual lattice spacings $a_{\rm lat}/\ell$ and the relative (renormalized dual) lattice spacings $\ell_{\rm rel}$ are also given, including the errors in their determinations, for future reference. At our two finest lattice spacings for larger volumes there is some ambiguity in determining the tuned $\beta$ value, so more than one ensemble is used to estimate a tuning error. We discuss in Sections~\ref{sec:desitter} and \ref{sec:RVM} how we propagate this tuning error to our analyses.

\section{\label{sec:desitter}de Sitter space and setting the lattice spacing}

This section presents the method and numerical analysis used to determine the lattice spacing in our calculation.  The large number of new ensembles at finer lattices used in this work, as well as our current target precision, makes it necessary to revisit the methods of determining relative and absolute lattice spacings used in previous work.  Given that our lattices are not hypercubic, and the geometry at long distance scales is emergent, it is necessary to determine the ratio, $a_{\rm lat}/\ell$, where $\ell$ is a distance scale that we identify as a renormalized dual lattice spacing, from the simulations.  Some quantities are more naturally measured in terms of $a_{\rm lat}$ or in terms of $\ell$, requiring that we know their relative sizes at each nominal lattice spacing (i.e., at fixed $\kappa_2$).  We also require the relative lattice spacing $\ell_{\rm rel}$ across different $\kappa_2$ values.  In previous work \cite{Laiho:2016nlp, Bassler:2021pzt} we used the return probability of a diffusion process to determine sets of $a_{\rm lat}/\ell$ and $\ell_{\rm rel}$ values.  However, we find that there are large finite-volume and discretization effects in the return probability, making it difficult to reach a useful precision with this quantity.  In order to improve on this situation, we introduce two new methods for determining $a_{\rm lat}/\ell$ based on the identification of our lattice geometries with de Sitter space.  The good agreement between these determinations provides a strong foundation for using our values of $a_{\rm lat}/\ell$ as inputs to determinations of other quantities. 

We also use the semiclassical expansion of the Euclidean partition function about the de Sitter instanton to compute the absolute and relative lattice spacings, following the method introduced in Ref.~\cite{Bassler:2021pzt} for computing Newton's constant.  We revisit this determination here with new data at several lattice spacings and many different volumes.  This establishes the Planck length of our simulations and allows us to normalize our calculations in physical units.

\subsection{The classical de Sitter solution}

In the absence of matter, the homogeneous, isotropic Einstein equations reduce to a single equation for the scale factor $a$,
\bea \frac{\dot{a}^2}{a^2} + \frac{1}{a^2} = \frac{\Lambda}{3},
\eea
in the presence of a positive cosmological constant.  This equation leads to the de Sitter solution, where the scale factor undergoes exponential expansion,
\bea \label{eq:deSitter} a(t) = \sqrt{\frac{3}{\Lambda}} \cosh\left(\sqrt{\frac{\Lambda}{3}}t\right).
\eea
Since the lattice calculation is done in Euclidean signature, for comparison we should analytically continue the classical solution, which becomes
\bea \label{eq:EucliddeSitter}  a(\tau) = \sqrt{\frac{3}{\Lambda}} \cos\left(\sqrt{\frac{\Lambda}{3}}\tau\right),
\eea
under $t \rightarrow -i\tau$.

Given that the shape of the lattice is an emergent property of the numerical simulations, we would like to relate the numerical computation of the shelling function characterizing that shape (described in Section~\ref{sec:phase}), to Eq.~(\ref{eq:EucliddeSitter}).  To do this, we consider the convenient parameterization \cite{Ambjorn:2008wc},

\bea\label{eq:deSitterLat}  N_4^{\rm shell} = \frac{3}{4}N_4 \frac{1}{s_0 N_4^{\frac{1}{4}}}\sin^3\left(\frac{\tau/\ell}{s_0 N_4^{\frac{1}{4}}}\right),
\eea
where $s_0$ is a free parameter and $\tau/\ell$ is the Euclidean time in renormalized dual lattice ($\ell$) units.  $N_4^{\rm shell}$ is normalized so that a sum over all time slices gives the total number of four-simplices of the geometry.  This expression can be interpreted as Euclidean de Sitter space.  It differs from Eq.~(\ref{eq:EucliddeSitter}) by a shift in origin of the Euclidean time $\tau$, and by the fact that the shelling function is a three-volume and therefore corresponds to the scale factor cubed.  This expression is a fair description of the lattice result for the shelling function, but the presence of discretization effects leads to difficulties in fitting the data, given its precision.  In order to obtain good fits, we need to add additional parameters to the expression,  
\bea\label{eq:deSitterMod}  N_4^{\rm shell} = \frac{3}{4}\eta N_4 \frac{1}{s_0 N_4^{\frac{1}{4}}}\sin^3\left(\frac{\tau/\ell}{s_0 N_4^{\frac{1}{4}}} + b\right),
\eea
where the parameter $b$ is an offset in the Euclidean time, and $\eta$ is the fraction of the volume of the universe that is actually well-described by the classical solution.

Fits to lattice data for the shelling function thus use the functional form
\bea N_4^{\rm shell}= A_{s} \sin^3(B_{s}\tau/\ell + C_{s}),
\label{eq:desitterfit}
\eea
where $A_{s}$, $B_{s}$, and $C_{s}$ are free parameters.  Because the Euclidean time in the shelling function is measured in dual lattice units, and the lattice scale factor is a measure of the three volume obtained by counting four-simplices in concentric shells, they are not determined in the same units.  We can use the fact that we must make contact with a solution to the classical Einstein equations in the appropriate limit in order to find a conversion factor between these two sets of lattice units.  If the correct solution is Euclidean de Sitter space, then Eq.~\ref{eq:EucliddeSitter} provides a relationship between the constant multiplying the trigonometric function and the coefficient in the argument of that function. Then the distance units $\ell$ and $\ell_V$ must satisfy
\bea \label{eq:AB} \frac{1}{A_{s}^{1/3}B_{s}} = \frac{\ell_V}{\ell}
\eea
where $\ell$ is the renormalized dual lattice spacing and $\ell_V$ is the unit length derived from taking the cube root of the normalization of the shelling function.  These units of measurement can be related via a comparison of the total four-volume of a thin slice of de Sitter space measured either by summing the volumes of four-simplices making up that slice, or by taking the integral over a four-dimensional shell.  This leads to the relation
\bea\label{eq:slice_relation}  N_4^{\rm shell}(\tau) C_4 a_{\rm lat}^4 = 2\pi^2 N_4^{\rm shell}(\tau) \ell_V^3 \ell,
\eea
where $C_4 = \sqrt{5}/96$, and $N_4^{\rm shell}(\tau)$ is the number of four-simplices in a given shell.  Eqs.~(\ref{eq:AB}) and (\ref{eq:slice_relation}) may be taken to implicitly define $\ell$ and $\ell_V$.  Eq.~(\ref{eq:slice_relation}) can then be solved for the ratio of direct and renormalized dual lattice spacings
\bea \label{eq:aoverell}  \frac{a_{\rm lat}}{\ell} = \left(\frac{2\pi^2}{C_4}\left(\frac{\ell_V}{\ell}\right)^3\right)^{\frac{1}{4}}.
\eea
We refer to the calculation of $a_{\rm lat}/\ell$ via Eq.~(\ref{eq:aoverell}) as Method 1.  

Note that $\ell$, and thus $a_{\rm lat}/\ell$, is an emergent quantity not necessarily given by the expectation from short-distance simplicial geometry.  If we denote the short-distance dual lattice spacing by $\tilde{\ell}$, then it can be shown that $\tilde{\ell}$ is related to $a_{\rm lat}$ as follows.  Given two equilateral $d$-simplexes with edge length $a_{\mathrm{lat}}$ glued together, the distance between the centers is $\sqrt{\frac{2}{d(d+1)}} a_{\mathrm{lat}}$, so that the ratio $a_{\rm lat}/\tilde{\ell}$, is a fixed number $\sqrt{\frac{d(d+1)}{2}}$. The effective long-distance $\ell$ differs from $\tilde{\ell}$, and its value must be determined from the simulations at each lattice spacing.  With $\ell$ given implicitly by Eqs.~(\ref{eq:AB}) and (\ref{eq:slice_relation}), it is defined in an operational way that is not altered by the use of degenerate triangulations.

In summary, $a_{\rm lat}/\ell$ is determined in Method 1 using the fact that the de Sitter radius is measured in two different ways when we construct the shelling function, each in different units.  Setting these measures of the radius equal gives a value for the conversion factor $a_{\rm lat}/\ell$.  
Thus, we obtain $a_{\rm lat}/\ell$ using Eqs.~(\ref{eq:AB}) and (\ref{eq:aoverell}) and fits of lattice data to the de Sitter solution.

\subsection{Semiclassical fluctuations about the de Sitter solution}
\label{sec:semiclflucdS}

The shape of the lattice can be matched to Euclidean de Sitter space; this allows for a determination of the renormalized cosmological constant, which is fixed once the four-volume is specified.  However, the Newton coupling does not appear in the de Sitter solution, Eq.~(\ref{eq:EucliddeSitter}), so matching lattice data to this classical solution cannot by itself determine the absolute lattice spacing in Planck units.  However, by examining the semiclassical fluctuations about de Sitter space, we are able to make contact with an expression involving the Planck scale and thus determine the absolute lattice spacing.  A consideration of the semiclassical fluctuations also provides us with an alternative expression for $a_{\rm{lat}}/\ell$, leading to a different strategy for obtaining this quantity.  

\subsubsection{Determining the Newton constant}
\label{sec:GNewtonanalysis}

We revisit the theory of the semiclassical fluctuations about de Sitter space, reviewing the approach to computing the Planck scale on our lattices that was presented in Ref.~\cite{Bassler:2021pzt} and introducing an additional method for obtaining $a_{\rm lat}/\ell$.  The Euclidean partition function takes a simplified form after integrating out all degrees of freedom except for the four-volume of the geometry \cite{Ambjorn:2012jv},
\bea \label{eq:Z}  Z(\kappa_4, \kappa_2)= \sum_{N_4} e^{-(\kappa_4 - \kappa_4^c)N_4}f(N_4, \kappa_2),
\eea
where $f(N_4, \kappa_2)$ is sub-exponential in $N_4$, and $\kappa_4^c$ is the pseudo-critical value of the coupling $\kappa_4$.  The limit $\kappa_4 \to\kappa_4^c$ allows one to take the infinite lattice-volume limit $N_4\to\infty$.  However, this does not necessarily correspond to the infinite physical-volume limit, since this procedure is equally valid in the unphysical crumpled phase, where the numerical simulations show that the emergent geometries are on the order of the size of the cutoff.  The critical value $\kappa_4^c$ is not known a priori, but emerges from the nonperturbative sum over triangulations.  In practice it is determined by adjusting the constant $\kappa_4$ at a particular target volume until the moves are equally likely to cause an upward fluctuation in volume as a downward one.  

The term in the exponential corresponds in the continuum to the renormalized cosmological constant term in the classical action, such that we identify
\bea \label{eq:cc} (\kappa_4-\kappa_4^c)N_4 = \frac{\Lambda}{8\pi G}V,
\eea
with $V=C_4 N_4 a_{\rm lat}^4$.  Once the bare parameters $\kappa_2$ and $\beta$ are chosen so that the simulations are in the physical region of the phase diagram, the size of the semiclassical universe is specified when we input the target volume $N_4$.  The size of the de~Sitter universe at a given $\kappa_2$ and $\beta$ uniquely fixes $\kappa_4$, and thus the renormalized cosmological constant $\Lambda$.

 In order for the partition function in Eq.~(\ref{eq:Z}) to reproduce semiclassical gravity, the subleading exponential behavior should be given by the Einstein-Hilbert term.  By power counting, the 4-volume dependence of this term should scale like 
 \bea \frac{1}{16\pi G}\int d^4x\sqrt{g}R \propto \frac{\sqrt{V}}{G}.
 \eea
Thus, the partition function with all other degrees of freedom integrated out except for the four-volume should take the form \cite{Ambjorn:2012jv}
\bea \label{eq:part_func}   Z(\kappa_4, \kappa_2)=\sum_{N_4} e^{-(\kappa_4 - \kappa_4^c)N_4 + k(\kappa_2)\sqrt{N_4}},
\eea
where the expected scaling of $k$ is
\bea \label{eq:k} k(\kappa_2) \propto \frac{a_{\rm lat}^2}{G}.
\eea
In order for a continuum limit to exist, $k$ must go to zero as $N_4$ goes to infinity, so that the volume in physical units remains finite.  

The value of $k$ at a given lattice spacing must be determined from the simulations, as it is an emergent, long-distance quantity.  It can be obtained using the expectation value of the number of four-simplices $\langle N_4\rangle$, which can be approximated using a saddle-point expansion of the partition function in Eq.~(\ref{eq:part_func}),     
\bea \label{eq:saddlepoint} \langle N_4 \rangle = \frac{\sum_{N_4}N_4 e^{-(\kappa_4 - \kappa_4^c)N_4 + k(\kappa_2)\sqrt{N_4}}}{\sum_{N_4}e^{-(\kappa_4 - \kappa_4^c)N_4 + k(\kappa_2)\sqrt{N_4}}} \nonumber \\  \approx \frac{k^2(\kappa_2)}{4(\kappa_4-\kappa_4^c)^2}.
\eea
In our simulations we fix $N_4$, so the expectation value $\langle N_4 \rangle=N_4$ is an input to our simulations.  Solving Eq.~(\ref{eq:saddlepoint}) for $k$ we find
\bea \label{eq:slope}  k=2|\kappa_4-\kappa_4^c|\sqrt{N_4},
\eea
and from this we see that a plot of $\kappa_4$ as a function of $1/\sqrt{N_4}$ should be linear if the semiclassical limit is realized in the simulations.  This was found to be the case in Ref.~\cite{Bassler:2021pzt}; we extend that study to finer-lattice-spacing ensembles generated for this work.  

Once we have determined $k$ we can obtain $G/a_{\rm lat}^2$, since by Eq.~(\ref{eq:k}), they are inversely proportional.  We review here the derivation of that proportionality constant.  The same saddle-point expansion used in Eq.~(\ref{eq:saddlepoint}) gives for the partition function
\bea \label{eq:HawkingMoss} Z(\kappa_4, \kappa_2) \approx \exp\left( \frac{k^2(\kappa_2)}{4(\kappa_4-\kappa_4^c)}\right) =\exp\left(\frac{3\pi}{G\Lambda}\right),
\eea
where the equality follows from the semiclassical approximation in the continuum.  This continuum result for the partition function is the well-known Hawking-Moss instanton production amplitude \cite{Hawking:1981fz}, and in making this identification we are assuming that de Sitter space is the semiclassical solution that dominates the saddle-point approximation.  The emergence of semiclassical geometry can be tested by showing that $\kappa_4$ plotted versus $1/\sqrt{N_4}$ is linear; this behavior was seen in Ref.~\cite{Bassler:2021pzt}.    

The renormalized Newton's constant $G$ can be obtained from the partition function by combining Eqs.~(\ref{eq:cc}), (\ref{eq:slope}), and (\ref{eq:HawkingMoss}) to get
\bea G= \frac{5^{\frac{1}{4}}a_{\rm lat}^2}{16\sqrt{N_4}|\kappa_4-\kappa_4^c|},
\label{eq:Galat}
\eea
which implies
\bea  \frac{G}{a_{\rm lat}^2}=\frac{5^{\frac{1}{4}}}{16|s_G|},
\label{eq:NCa}
\eea
with $s_G$ the slope determined by a fit to $\kappa_4$ as a function of $1/\sqrt{N_4}$.  
A comparison of $G/a_{\rm lat}^2$ at different lattice spacings allows us to determine the relative (direct) lattice spacing $a_{\rm rel}$ with reference to a fiducial lattice spacing.  The ratios $a_{\rm lat}/\ell$ allow us to compute as well the relative renormalized dual lattice spacing $\ell_{\rm rel}$. 

\subsubsection{\label{sec:aellmethod2}Determination of $a_{\rm lat}/\ell$ from Method 2}

An alternative determination of $a_{\rm lat}/\ell$ can be obtained as follows. Combining Eqs.~(\ref{eq:cc}), (\ref{eq:slope}), and (\ref{eq:HawkingMoss}), and solving for the renormalized cosmological constant in terms of the four-volume, we find
\bea \Lambda = \sqrt{\frac{24 \pi^2}{C_4 N_4}}\frac{1}{a_{\rm lat}^2}.
\eea
We can turn this into a relation between the renormalized cosmological constant in $\ell$ units and the ratio $a_{\rm lat}/\ell$ by multiplying through by $\ell^2$,
\bea \label{eq:Lambdaell2} \Lambda \ell^2 = \sqrt{\frac{24 \pi^2}{C_4 N_4}}\frac{\ell^2}{a_{\rm lat}^2}.
\eea
This relation can be corrected to account for the fact that the de Sitter solution is not a good description of the lattice data over the entire range of the shelling function for coarse lattices, especially at long distances.  The fraction of the four-volume that appears under the classical curve is given by the parameter $\eta$ introduced in Eq.~(\ref{eq:deSitterMod}) and can be obtained from fits of lattice data to the classical de Sitter curve.  In terms of the fit parameters $A_s$ and $B_s$ of Eq.~(\ref{eq:desitterfit}), $\eta$ is given by
\bea  \eta = \frac{4A_s}{3B_s N_4}.
\label{eq:eta}
\eea
With the substitution $\eta N_4$ in place of $N_4$, Eq.~(\ref{eq:Lambdaell2}) becomes
\bea  \label{eq:Lambdaell2eta} \Lambda \ell^2 = \sqrt{\frac{24 \pi^2}{C_4 \eta N_4}}\frac{\ell^2}{a_{\rm lat}^2}.
\eea

We can extract the slope $s_{\Lambda}$ from a linear fit to $\Lambda \ell^2$ versus $1/\sqrt{N_4}$
\bea  \Lambda \ell^2 = \frac{s_{\Lambda}}{\sqrt{N_4}},
\label{eq:LambdaSlope}
\eea 
and from this definition and Eq.~(\ref{eq:Lambdaell2eta}), we find an expression for $a_{\rm lat}/\ell$,
\bea  \frac{a_{\rm lat}}{\ell} = \left(\frac{24}{C_4 \eta}\right)^{\frac{1}{4}}\left(\frac{\pi}{s_\Lambda} \right)^{\frac{1}{2}}.
\label{eq:aell2}
\eea
We refer to the calculation of $a_{\rm lat}/\ell$ from Eq.~(\ref{eq:aell2}) as Method 2.  Note that this formula could also be obtained by comparing the total four-volume of the lattice (corrected by $\eta$) to the time integral of the classical expression for the cube of the scale factor.  Thus, both methods for determining $a_{\rm lat}/\ell$ can be derived from the classical solution.  Even so, they require different numerical procedures and have different systematic errors, allowing for a useful cross-check.

\subsection{Numerical results for $a_{\rm lat}/\ell$}
\label{sec:aellboth}

\subsubsection{$a_{\rm lat}/\ell$ from Method 1}
\label{sec:aell1}

\begin{figure}[ht]
    \centering
    \includegraphics[width=1\linewidth]{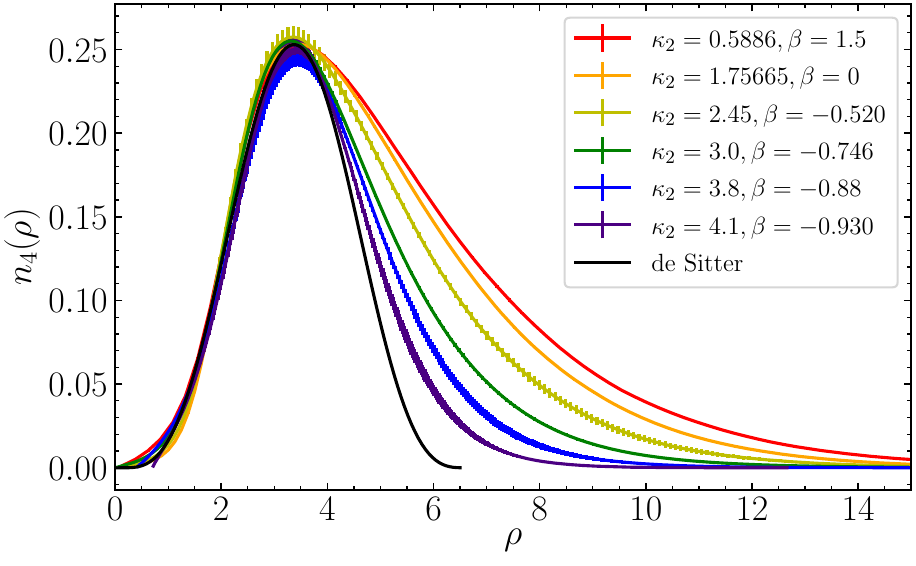}
    \caption{The rescaled shelling function $n_4(\rho)$ at different lattice spacings.  The black curve is the classical de Sitter solution.  All curves have been rescaled to overlap in the region that matches the classical solution.  The asymmetry at large Euclidean time decreases as the lattice spacing gets finer, so that the lattice results approach the classical curve.  The $\beta = 1.5$ ensemble shown here has a volume of 4k, and all the others have a volume of 32k.}
    \label{fig:deSitteraAymTail}
\end{figure}

For our first method of determining the ratio $a_{\rm lat}/\ell$, we fit the lattice data to the functional form of the classical de Sitter solution Eq.~(\ref{eq:desitterfit}).  Figure~\ref{fig:deSitteraAymTail} shows the theoretical expectation Eq.~(\ref{eq:desitterfit}) along with lattice data for the shelling function at various lattice spacings. The data has been rescaled along the vertical and horizontal axes so that it coincides with the classical curve to the left of the peak for all lattice spacings.  There is a large discrepancy between the classical curve and the lattice data at large Euclidean time, as Fig.~\ref{fig:deSitteraAymTail} shows, but this discrepancy gets smaller as the continuum limit is approached.

The computation of the shelling function requires that we choose a 4-simplex source randomly on a given lattice.  For each configuration, we choose between 1 and 60 sources to compute the shelling function, depending on the size and lattice spacing of the ensemble. The average of the shelling function over different sources on a given configuration is first taken, and then the ensemble average over configurations is performed.  Here, as in the rest of this work, the error is computed using single-elimination jackknife resampling, and the autocorrelation error is accounted for by blocking configurations until the jackknife error stops increasing.  Based on the number of blocks that are required before the errors saturate, we can estimate the autocorrelation time for the shelling function in the vicinity of the peak.  From this, we estimate that we have between 15 and 5000 independent samples, depending on ensemble, with the smallest number corresponding to the largest volume at our finest lattice spacing.

\begin{figure}[htb]
    \centering
    \includegraphics[width = \linewidth]{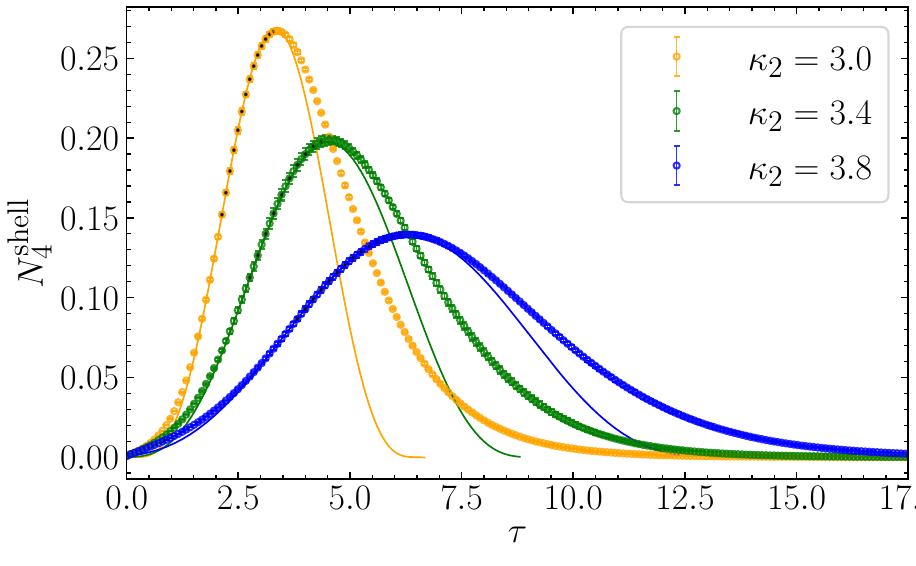}
    \caption{The fit of the shelling function to Eq. \eqref{eq:desitterfit} on $16$k ensembles at three lattice spacings. The horizontal and vertical axes have been rescaled by $N_4^{1/4}$ and $N_4^{3/4}$, respectively.}
    \label{fig:dsfit}
\end{figure}
Examples of fits to the de Sitter form are shown in Fig. \ref{fig:dsfit} for ensembles across three different lattice spacings.  The fit is to Eq.~(\ref{eq:desitterfit}), and the range of the fit is chosen to be in the vicinity of the peak, with a starting point around the inflection point to the left of the peak and the final point at, or slightly to the right of, the peak.  On the finer lattice spacings, we can include a few points to the right of the peak and still get decent fits, as measured by the fits' $p$-values.  The fits are performed under a jackknife, where the estimate of the $\chi^2/{\rm d.o.f.}$ uses the full correlation matrix, which is remade for every jackknife entry.  The correlation matrix is estimated from the data sample, and the fits are done on the unblocked data in order to provide more points from which to sample the correlation matrix, but the unblocked errors are inflated to account for the autocorrelation
determined from our blocking studies.  The estimation of $p$-values includes a correction due to the finite-sample size of the ensemble \cite{Toussaint:2008ke}.  In order to estimate a fitting systematic error, we vary the fit range and, keeping only the fit ranges with an average $p$-value>0.01, we use the spread of the different fit results to estimate the error.  We choose as our central value the results of a fit that sits in the middle of the fit variations, and a fit systematic error is determined by taking the standard error of the distribution of all acceptable fits on a given ensemble.  At our finest lattice spacing ($\kappa_2 = 3.8$), we find that it is necessary to thin the number of data points included in the fits, given the large number of points and the difficulty of measuring the correspondingly large correlation matrix.  At this lattice spacing our fitting procedure only includes every other $\tau$ value to the left of the peak and every point to the right of the peak.  

\begin{figure}[htb]
    \centering
    \includegraphics[width = \linewidth]{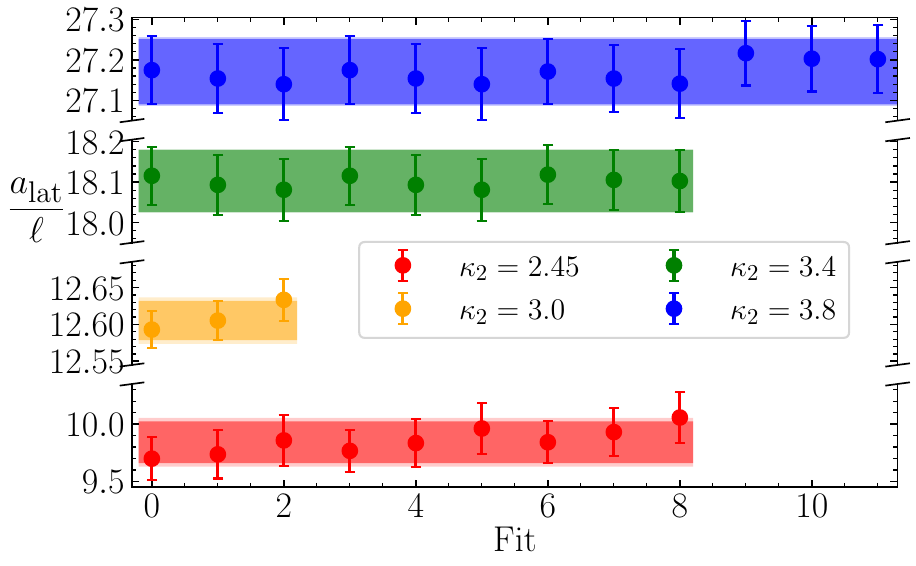}
    \caption{Stability plot showing results for different fit ranges on the $N_4=16000$ ensembles at four lattice spacings. The horizontal axis enumerates the fits. Only acceptable fits, which we take to be those fits with $p$-value $\geq 0.01$, are shown here. All twelve of our fits at $\kappa_2=3.8$ are above our $p$-value cutoff, and all nine of our $\kappa_2=3.4$ and $\kappa_2=2.45$ fits are above the threshold, but only three out of nine of our fits at $\kappa_2=3.0$ are. The darker bands indicate the central fits with 1$\sigma$ statistical error. The lighter bands also include the fit error added in quadrature.}
    \label{fig:dsfitrange}
\end{figure}

Given the fit parameters $A_s$, $B_s$, and $C_s$ determined from fits to the Euclidean de Sitter form, Eq.~(\ref{eq:desitterfit}), we can reconstruct $a_{\rm lat}/\ell$ using Eqs.~(\ref{eq:AB}) and (\ref{eq:aoverell}).
Fig. \ref{fig:dsfitrange} shows the resulting $a_{\rm lat}/\ell$ values for the different acceptable fit windows on representative ensembles at four lattice spacings, some of which correspond to the fits shown in Fig. \ref{fig:dsfit}. This figure gives a sense of the variation of $a_{\rm lat}/\ell$ with the chosen fit range.  For some of our tuned $\beta$ values, the uncertainty is large enough that we choose two ensembles to estimate a tuning systematic error. For example, at $N_4=96$k, $\kappa_2=3.8$, we do the usual analysis on ensembles with two different $\beta$ values close to the tuned point, and then we take the weighted average of the $a_{\rm lat}/\ell$ values obtained from the two separate ensembles.  The error of the combined point is taken to be the maximum of the weighted standard error and the difference between the weighted average and the original two points. This is to make sure that both of the original $a_{\rm lat}/\ell$ values are encompassed by the error of the combined result.  The values of $a_{\rm lat}/\ell$ determined from this procedure are given in Table~\ref{tab:etaaell}.

The construction of $a_{\rm lat}/\ell$ using Eqs.~(\ref{eq:AB}) and (\ref{eq:aoverell}) requires that the lattice geometries be well-described by the classical de Sitter solution.  As presented in Sect.~\ref{sec:RVM}, small but significant deviations from classical de Sitter space can be resolved in our data even in the region to the left of the classical peak, and these effects are more pronounced for smaller volume ensembles.  Thus, we expect our procedure for constructing $a_{\rm lat}/\ell$ to be valid only in the infinite volume limit, where corrections to the classical de Sitter solution should be small.  We therefore extrapolate $a_{\rm lat}/\ell$ to infinite volume, determining a value for this ratio independently at each lattice spacing.  We estimate a systematic error for this extrapolation by dropping either the largest or smallest volume from the extrapolation.  For ensembles with $\kappa_2 = 3.4$ and $3.8$, the extrapolation error estimate involves dropping the two largest volumes or the smallest single volume from the extrapolation.  This is because our finer lattice spacings suffer more than the others from the difficulty of estimating autocorrelation errors, especially for the largest volumes. 
The largest deviation of these fit choices from the central fit is taken as an additional systematic error for the infinite-volume extrapolation.  

\begin{figure*}[htb]
\begin{minipage}{0.475\linewidth}
    \centering
    \includegraphics[width=\linewidth]{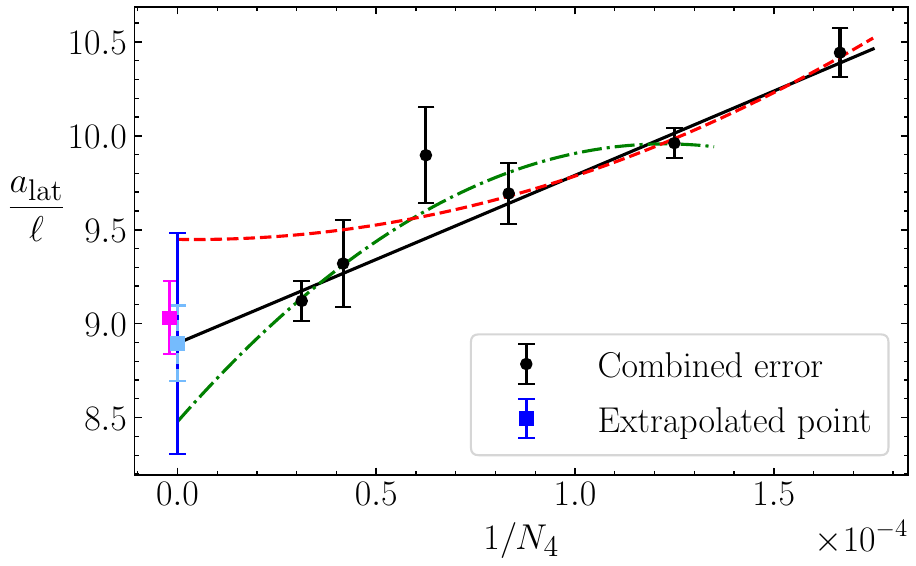}
\end{minipage}
\begin{minipage}{0.475\linewidth}
    \centering
    \includegraphics[width=\linewidth]{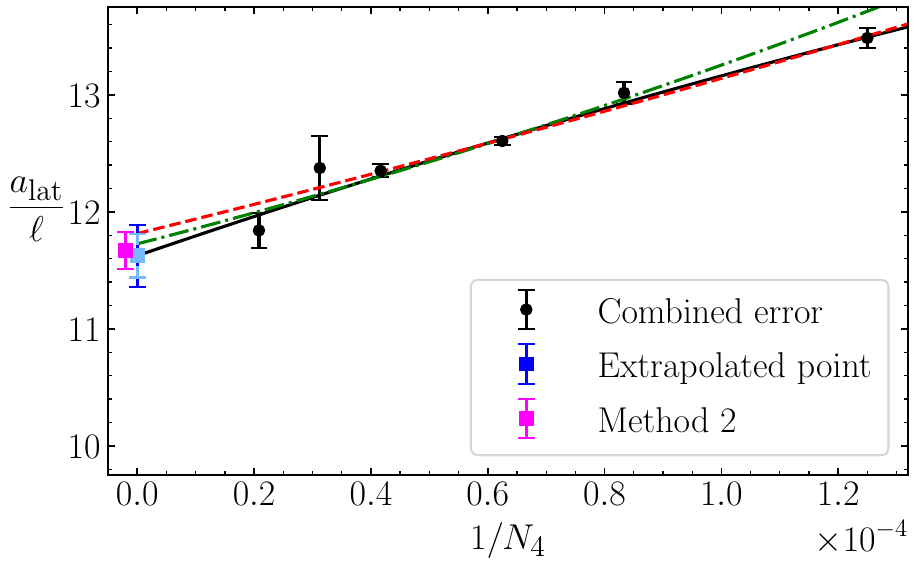}
\end{minipage}
\begin{minipage}{0.475\linewidth}
    \centering
    \includegraphics[width=\linewidth]{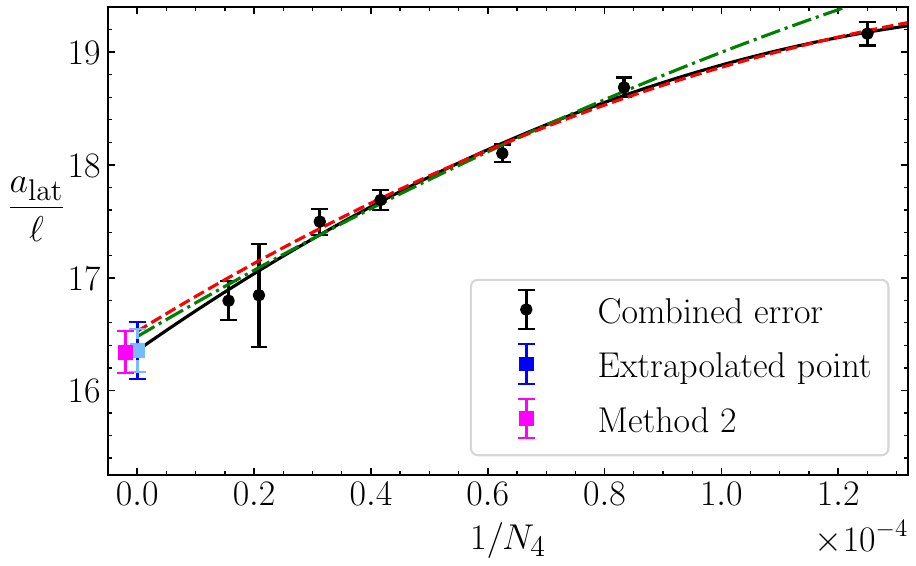}
\end{minipage}
\begin{minipage}{0.475\linewidth}
    \centering
    \includegraphics[width=\linewidth]{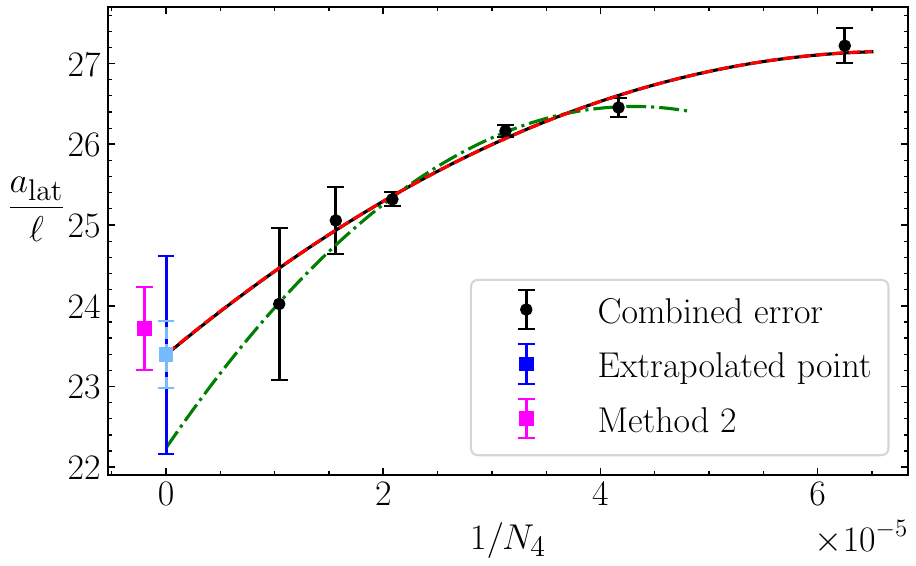}
\end{minipage}
\caption{Plots of $a_{\rm lat}/\ell$ versus $1/N_4$ for four different lattice spacings, which from top left to bottom right are $\kappa_2=2.45$, $\kappa_2=3$, $\kappa_2=3.4$, and $\kappa_2=3.8$. The errors on the individual data points are the statistical and systematic error (associated with varying the fit window of the shelling function) combined in quadrature.
The central fit curve for the extrapolation to infinite volume is shown in black, with the statistical error on the extrapolated point of the central fit shown in light blue.  Alternative extrapolations dropping the largest or smallest volume ensembles from the fit are shown in dashed (red) and dot-dashed (green), respectively.  The infinite-volume extrapolated point (dark blue) includes a systematic error coming from these alternate fit variations.
The magenta points are the results of the alternate (Method 2) determination of $a_{\rm lat}/\ell$, which is discussed in Section \ref{sec:aell2}, shown for comparison.  They are offset horizontally to stand out from the Method 1 value. \newline
} 
\label{fig:aellex}
\end{figure*}

The infinite volume extrapolations for $a_{\rm lat}/\ell$ are shown in Fig.~\ref{fig:aellex} and the extrapolated values are given in Table~\ref{tab:aellex}, along with the $\chi^2/{\rm d.o.f.}$ and the $p$-value of each fit.  Extrapolations assuming a three parameter fit quadratic in $1/N_4$ give a good description of the extrapolation, as can be seen by the $p$-values of the central fits in Fig.~\ref{fig:aellex}.  Figure~\ref{fig:aellex} also shows the values of $a_{\rm lat}/\ell$ obtained from Method 2 outlined in subsection~\ref{sec:aellmethod2}.  The details of the numerical analysis for this alternative $a_{\rm lat}/\ell$ determination are given in the following subsection.

\begin{table}[ht]
\caption{Fit results for the $a_{\rm lat}/\ell$ infinite-volume extrapolation at four lattice spacings.}
\label{tab:aellex}
\centering
\begin{tabular}{m{3em} m{5em} m{5em} m{5em} }
\hline \hline
$\kappa_2$ & $a_{\rm lat}/\ell$ & $\chi^2/\mathrm{d.o.f.}$& $p$-value \\ \hline
2.45&  8.90(59) & 1.35 & 0.26\\
3.0& 11.62(26) & 1.13 & 0.34\\
3.4& 16.35(25) & 0.90 & 0.46\\
3.8& 23.39(122) & 1.26 & 0.29\\
\hline\hline
\end{tabular}
\end{table}

\subsubsection{$a_{\rm lat}/\ell$ from Method 2}
\label{sec:aell2}

Our second method of determining $a_{\rm lat}/\ell$ from lattice data makes use of Eq.~(\ref{eq:aell2}), which itself makes use of the relation between the number of four-simplices and the effective cosmological constant of the lattice ``universe''.  This requires as input $s_\Lambda$, which can be obtained from Eq.~(\ref{eq:LambdaSlope}), as well as $\eta$, which can be determined via Eq.~(\ref{eq:eta}) from the results of our fits to the classical de Sitter solution.  The determination of $s_\Lambda$ requires as input the renormalized cosmological constant in $\ell$ units, $\Lambda \ell^2$, which can be obtained as follows.  We relate the lattice shelling function to the scale factor $a$ and its time derivatives.  Given this, we can infer the value of the emergent cosmological constant at long distances by constructing the Einstein equation for the scale factor from our lattice data.  In Euclidean space, the homogeneous, isotropic Einstein equations in the presence of a positive cosmological constant reduce to
\bea \label{eq:CC} -\frac{\dot{a}^2}{a^2} + \frac{1}{a^2} = \frac{\Lambda}{3},
\eea
where $\dot{a}$ is now understood to be the derivative $\frac{da}{d\tau}$, i.e. the derivative with respect to Euclidean time.  The shelling function Eq.~(\ref{eq:deSitterLat}) is identified as the cube of the scale factor, so the shelling function obtained from our lattice simulations allows us to reconstruct the scale factor $a$ as a function of Euclidean time $\tau$, as well as its time derivatives.

The scale factor measured in $\ell$ units is obtained directly from the lattice scale factor 
\bea \label{eq:lattice_a}  \frac{a(\tau)}{\ell} = \frac{\ell_V}{\ell}\frac{A(\tau)}{\ell_V},
\eea
where we introduce the notation 
\bea A(\tau)/\ell_V\equiv (N_4^{\rm shell}(\tau))^{\frac{1}{3}},
\eea
to relate the lattice scale factor to the shelling function, emphasizing the dependence on the lattice spacing conversion factors.  Then Eq.~(\ref{eq:CC}) can be written in terms of $A$,
\bea \label{eq:latdeSitter}  -\frac{\dot{A}^2}{A^2} + \frac{\ell^2}{\ell_V^2}\frac{\ell_V^2}{A^2} = \frac{\Lambda \ell^2}{3},
\eea
where the ratio $\ell_V/\ell$ converts the scale factor $A/\ell_V$ into renormalized dual lattice ($\ell$) units, with $\ell_V/\ell$ related to the ratio $a_{\rm lat}/\ell$ by 
\bea\label{eq:ellVoverell}  \frac{\ell_V}{\ell}=\left(\frac{C_4}{2 \pi^2}\right)^{\frac{1}{3}} \left(\frac{a_{\rm lat}}{\ell} \right)^{\frac{4}{3}}.
\eea 
The Euclidean time in the time derivative of the first term of Eq.~\ref{eq:latdeSitter} is already in $\ell$ units, so the linear combination of terms on the left side of Eq.~\ref{eq:latdeSitter} gives the cosmological constant also in $\ell$ units.

\begin{figure}[htb]
    \centering
    \includegraphics[width = \linewidth]{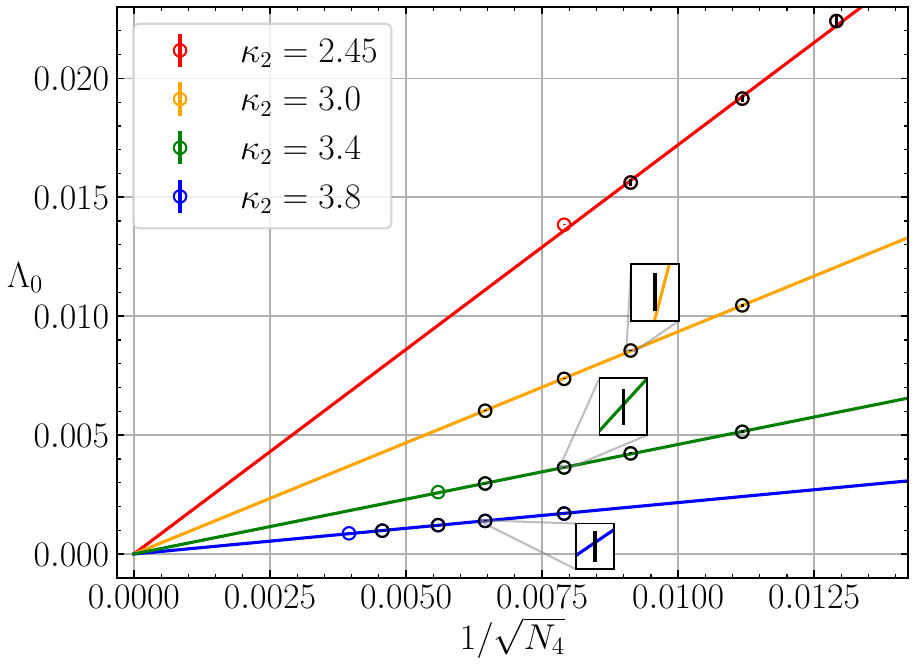}
    \caption{Data for $\Lambda_0$ versus $1/\sqrt{N_4}$ at four lattice spacings.  Also shown are linear fits to the functional form $\Lambda_0=\frac{s_\Lambda}{\sqrt{N_4}}$.  The inset figures show sample points with enough resolution to see the error bars compared to the fit line.  Data points highlighted in black are included in the fits.  The results of the linear fits to this data are given in Table \ref{tab:l0-n4-intercept}.}
    \label{fig:lambda-n4}
\end{figure}

In practice, we find that the cosmological constant inferred from this method is not strictly a constant as a function of $\tau$, but acquires some dynamics that are well-described by the model \cite{Sola:2013gha}
%
\bea\label{eq:RVM_gen_1} \Lambda(H) = \Lambda_0 + 3\nu H^2,
\eea
where $\Lambda_0$ is a fixed constant, $\nu$ is a dimensionless running coupling constant of the model, and $H=\dot{A}/A$ is obtained from the (numerical) time derivative of the lattice scale factor.  A discussion of the motivation for this model and the analysis of our lattice data in terms of it is given in the following section.  For the purposes of this subsection, it is enough to observe that the parameter $\Lambda_0$ plays the role of the classical cosmological constant, while the second term contains non-trivial dynamics, similar to quintessence models of cosmology \cite{Caldwell:1997ii}.  It is $\Lambda_0$ that enters the derivation of Eq.~(\ref{eq:LambdaSlope}), as this derivation assumes a constant term for the vacuum energy.  Using the values of $\Lambda_0$ extracted from our analysis at different four-volumes (discussed in more detail in the following section), we find that $\Lambda_0$ versus $1/\sqrt{N_4}$ is nicely described by a linear function that passes through zero, as can be seen in Fig.~\ref{fig:lambda-n4} for all four of our lattice spacings.  Thus, our data for $\Lambda_0$ is well-described by Eq.~(\ref{eq:LambdaSlope}) and can be used to obtain $s_\Lambda$, which in turn allows us to compute $a_{\rm lat}/\ell$.

Note, however, that this determination of $a_{\rm lat}/\ell$ from the slope of $\Lambda_0$ versus $1/\sqrt{N_4}$ is complicated by the fact that in order to construct $\Lambda_0$ using Eqs.~(\ref{eq:latdeSitter}) and (\ref{eq:RVM_gen_1}), we need as input $\ell_V/\ell$, which itself depends on $a_{\rm lat}/\ell$.  Thus, we must determine $a_{\rm lat}/\ell$ in a self-consistent way, such that the input to Eq.~(\ref{eq:latdeSitter}) matches the output of Eq.~(\ref{eq:aell2}).  This is done by scanning over input values of $a_{\rm lat}/\ell$ until a near match with the output is found, and then refining the window used for the scan until the input and output values match to the third decimal place.  The fit results with or without the origin fixed to zero are presented in Table \ref{tab:l0-n4-intercept}.  When the $y$-intercepts are allowed to vary, they are consistent with zero.  Since the one-parameter linear fits with the y-intercept fixed to zero describe the data well and are compatible with the semiclassical theory expectation, we choose the results of those fits as input to our $a_{\rm lat}/\ell$ determinations in order to minimize the errors in the rest of the analysis.
\begin{table*}[]
\begin{tabular}{c c c c||c c c c}
\multicolumn{4}{c}{$\Lambda_0 = s_\Lambda/\sqrt{N_4}$}                                                                                 & \multicolumn{4}{c}{$\Lambda_0 = s_\Lambda/\sqrt{N_4}+b$}       \\ \hline \hline
 $\kappa_2$  & $s_\Lambda$ & $\chi^2/\mathrm{d.o.f.}$ & $p$-value & $s_\Lambda$ & b & $\chi^2/\mathrm{d.o.f.}$ & $p$-value \\ \hline
$2.45$& 1.72(1) & 0.813 & 0.444 & 1.79(7)  & -0.007(7) & 0.571 & 0.565 \\ 
$3.0$  & 0.935(1) & 1.510 & 0.210 & 0.949(10) & -0.00012(8) & 1.296 & 0.274 \\
$3.4$  & 0.461(2) & 0.155 & 0.927  & 0.464(13)  & -0.00003(11) & 0.194 & 0.900\\
$3.8$  & 0.216(1) & 0.190 & 0.903  & 0.208(7)   & 0.00004(4) & 1.275 & 0.281\\ \hline \hline

\end{tabular}

\caption{Fit results for $\Lambda$ versus $1/\sqrt{N_4}$ for two different choices of fit function, shown above (see Eq.~\eqref{eq:LambdaSlope}). 
}
\label{tab:l0-n4-intercept}
\end{table*}

\begin{figure*}[htb]
\begin{minipage}{0.475\linewidth}
    \centering
    \includegraphics[width=\linewidth]{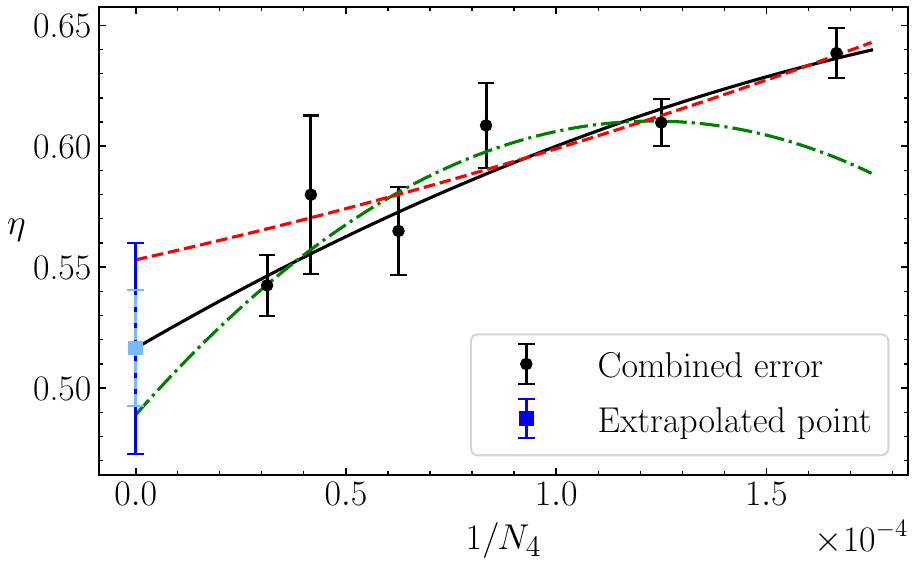}
\end{minipage}
\begin{minipage}{0.475\linewidth}
    \centering
    \includegraphics[width=\linewidth]{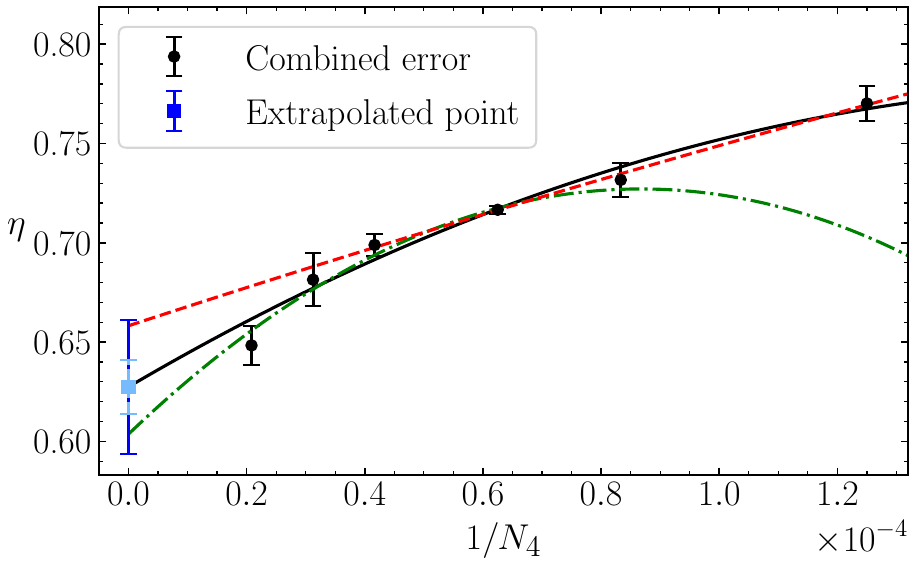}
\end{minipage}
\begin{minipage}{0.475\linewidth}
    \centering
    \includegraphics[width=\linewidth]{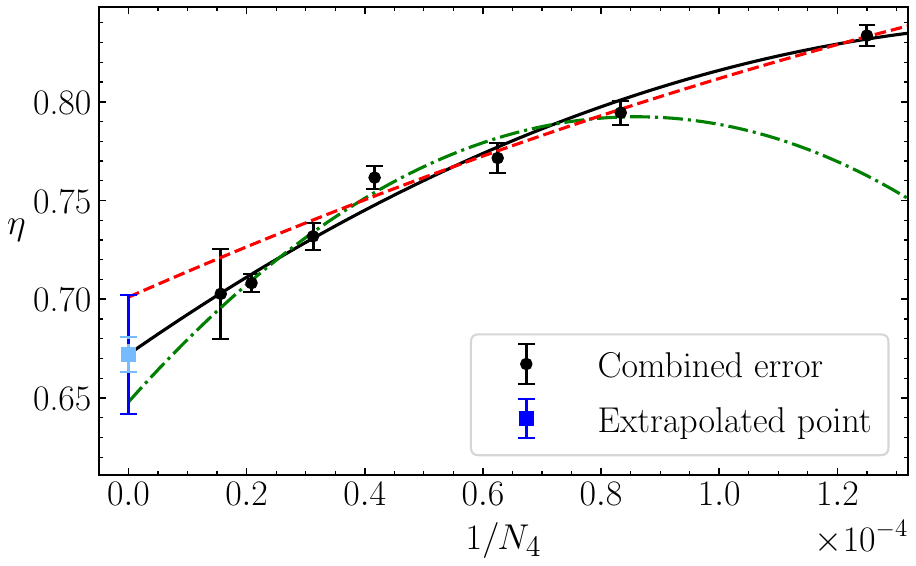}
\end{minipage}
\begin{minipage}{0.475\linewidth}
    \centering
    \includegraphics[width=\linewidth]{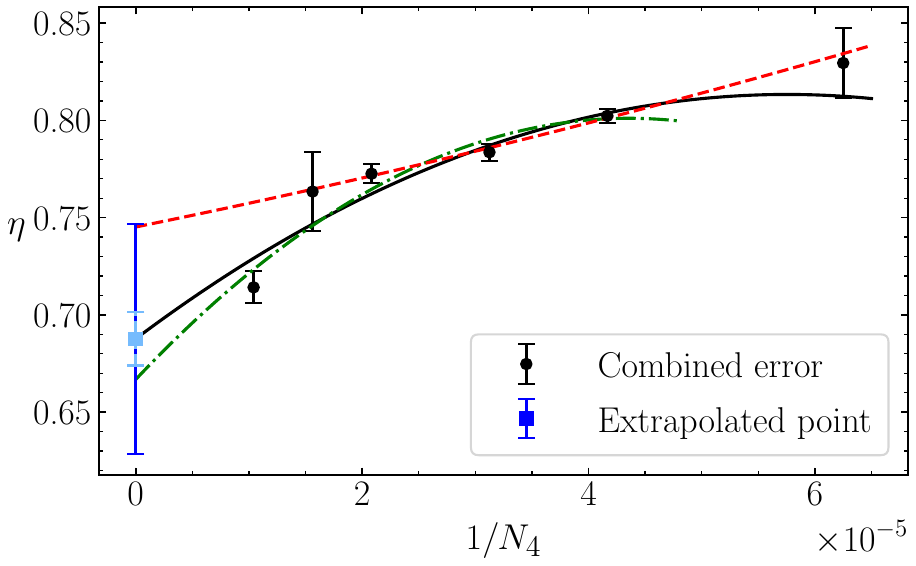}
\end{minipage}
\caption{
Plots of $\eta$ versus $1/N_4$ for four different lattice spacings, which from top left to bottom right are $\kappa_2=2.45$, $\kappa_2=3$, $\kappa_2=3.4$, and $\kappa_2=3.8$. The errors on the individual data points include the statistical error and a systematic error associated with varying the fit window, added in quadrature.  The central fit curve for the extrapolation to infinite volume is shown in black, with the statistical error on the extrapolated point of the central fit shown in light blue.  Alternative extrapolations dropping the largest or smallest volume ensembles from the fit are shown in dashed (red) and dot-dashed (green), respectively.  The infinite-volume extrapolated point (dark blue) includes a systematic error coming from these alternate fit variations.
} 
\label{fig:etaex}
\end{figure*}

The determination of $a_{\rm lat}/\ell$ from Eq.~(\ref{eq:aell2}) also requires as input the correction factor $\eta$, which accounts for the fraction of the four-volume under the classical de Sitter curve.  This factor is needed to correct for the asymmetric tail seen in Fig.~\ref{fig:dsfit}, and it gets closer to one as the continuum limit is approached.  The determination of $\eta$ proceeds from the same fits to the shelling function that yield $a_{\rm lat}/\ell$ using Method 1, where these fit results can be used to construct $\eta$ from Eq.~(\ref{eq:eta}); see \autoref{tab:etaaell} for our numerical results.  The values of $\eta$ are then extrapolated to their infinite volume limits at each lattice spacing.  Figure~\ref{fig:etaex} shows these extrapolations for all four lattice spacings.  We adopt the same procedure and method for estimating the systematic errors of these extrapolations as we did for the infinite volume extrapolation of $a_{\rm lat}/\ell$, where we take the spread in values from dropping the largest and smallest volumes from the extrapolation fit as an estimate of the systematic error.  As in the $a_{\rm lat}/\ell$ extrapolations, we use a three-parameter fit function that is quadratic in $1/N_4$.  At our two finest lattice spacings, where $\kappa_2=3.4$ and $\kappa_2=3.8$, we drop the largest two volumes as part of our systematic error estimate when extrapolating to infinite volume, just as we do for the $a_{\rm lat}/\ell$ analysis.  The results of the $\eta$ extrapolation fits are given in Tab.~\ref{tab:etaex}.

\begin{table}[ht]
\caption{Fit results for the infinite-volume extrapolation of $\eta$ at four lattice spacings, based on the values reported in \autoref{tab:etaaell}.}
\label{tab:etaex}
\centering
\begin{tabular}{m{3em} m{5em} m{5em} m{5em} }
\hline \hline
$\kappa_2$ & $\eta$ & $\chi^2/\mathrm{d.o.f.}$& $p$-value \\ \hline
2.45& 0.517(44) & 0.853 & 0.465\\
3.0& 0.627(33) & 1.498 & 0.214\\
3.4& 0.672(30) & 2.063 & 0.085\\
3.8& 0.688(59) & 3.382 & 0.018\\
\hline\hline
\end{tabular}
\end{table}

\subsubsection{Summary and combination of $a_{\rm lat}/\ell$ from different methods}
\label{sec:aell_comb}

The results for $a_{\rm lat}/\ell$ are in excellent agreement between the two different methods, as can be seen in Fig.~\ref{fig:aellex}, where the result for Method 2 is shown alongside the result from the Method 1 infinite-volume extrapolation.  The agreement is good at all lattice spacings, and for our most precise determinations, the agreement is within errors at the $1$-$2\%$ level.  This is a good cross-check of the determination, since each method suffers from different systematic errors and makes different assumptions.  

In order to make best use of these results for the downstream analyses that depend on $a_{\rm lat}/\ell$, we combine the results from the two methods.  The error bar for each of the two methods is dominated by the systematic error associated with an infinite-volume extrapolation; in Method 1 it is the infinite-volume extrapolation of $a_{\rm lat}/\ell$, and in Method 2 it is that of $\eta$.  Since the estimate of this extrapolation error uses the same approach, and these quantities come from the same fits to the de Sitter shelling function, we assume that the errors are $100\%$ correlated when combining the results for $a_{\rm lat}/\ell$ from the two methods.  This leads to a weighted average for the central value, with no reduction in error over the method with the smaller error \cite{Schmelling:1994pz}.  These results are summarized in Table~\ref{tab:aellinf}.

\begin{table}
\caption{$a_{\rm lat}/\ell$ for four lattice spacings from the two methods and their weighted average.}
\label{tab:aellinf}
\centering
\begin{tabular}{m{3em} m{5em} m{5em} m{5em}}
\hline \hline
$\kappa_2$ & Method 1 & Method 2 & Combined \\ \hline 
2.45&8.90(59)&9.03(19)& 9.02(19)\\
3.0&11.62(26)&11.67(16)& 11.66(16)\\
3.4&16.35(25)&16.34(19)& 16.34(19)\\
3.8&23.39(122)&23.72(52)& 23.67(52)\\
\hline\hline
\end{tabular}
\end{table}

\subsection{Determining Newton's constant and setting the lattice spacing}
\label{sec:G}

\begin{table}
\caption{Infinite-volume extrapolated values of Newton's constant at different lattice spacings, where the first error is statistical and the second is an estimate of the systematic error associated with the infinite-volume extrapolation.  Also quoted are the resulting relative direct and dual lattice spacings with errors.  The fit qualities for the infinite-volume extrapolations of Newton's constant are also included.}
\label{tab:Gtabs}
\centering
\begin{tabular}{c c c c c c } 
\hline \hline
$\kappa_2$& $G_a$ & $a_{\mathrm{rel}}$ &  $\ell_{\mathrm{rel}}$ & $\chi^2/\mathrm{d.o.f}$ & $p$-value \\
\hline
2.45 & 0.477(27)(57)  & 1.125(92)  &  1.45(12) &  0.61  & 0.61  \\
3.0  & 0.604(38)(35)  & 1.00  & 1.00  & 0.47 & 0.71   \\
3.4  & 0.805(81)(50)  &  0.866(67)  &  0.618(49)  &  0.39 & 0.76  \\
3.8  & 0.98(23)(06)  &  0.78(10)  &  0.386(51)   & 1.55  &  0.20  \\
\hline
\end{tabular}
\end{table}

In order to relate our lattice calculations to experiment it is necessary to determine the absolute lattice spacing in our simulations, thus allowing us to express our lattice lengths $a_{\rm lat}$ or $\ell$ in units of the Planck length.  This requires a calculation of Newton's constant $G$ in lattice units, which therefore requires that the simulations make contact with semiclassical physics.  In previous work using EDT simulations, $G$ was computed from a study of the binding energy of two scalar particles \cite{Dai:2021fqb} and from a study of the semiclassical fluctuations of the Euclidean partition function about the de Sitter instanton solution \cite{Bassler:2021pzt}.   In order to meet the precision requirements of the current analysis, we revisit the semiclassical de Sitter calculation of Ref.~\cite{Bassler:2021pzt}, following the theoretical method reviewed here in Sec.~\ref{sec:GNewtonanalysis}.  

Our new determination of Newton's constant requires that we generate a large number of new ensembles, in addition to those listed in Table~\ref{tab:ensembles}.  The de Sitter instanton analysis requires a finite-volume study for each nominal tuned volume at every lattice spacing.  Each finite-volume study requires ensembles at several different volumes, and the shift in the phase boundary with volume means that in order to minimize systematic errors due to wrong-phase contamination, the volumes in the finite-volume study should be larger than the nominal tuned volumes.  These new runs would not have been feasible without the rejection-free algorithm introduced in Ref.~\cite{Dai:2023tud}.

The analysis proceeds as follows.  For a given nominal volume, a given $\kappa_2$, and a tuned $\beta$ value, we generate ensembles at several volumes, all larger than the nominal volume, and measure the tuned $\kappa_4$ value of those runs with the same $\kappa_2$ and $\beta$ values.  The $\kappa_4$ values determined from these runs are given for all ensembles across four lattice spacings in \autoref{tab:kappa4MF}, \autoref{tab:kappa4F}, \autoref{tab:kappa4RF}, and \autoref{tab:kappa4SF}.  These results include the statistical error in $\kappa_4$, where the data is blocked before averaging to account for autocorrelations.  The first line in these tables at each distinct value of $\beta$ corresponds to one of the nominal volumes listed in Table~\ref{tab:ensembles}.  Note, however, that there is a slight mismatch in some of the tuned $\beta$ values between these ensembles and the ones listed in Table~\ref{tab:ensembles}.  This is because our estimates of the tuned $\beta$ values can shift as statistics is accumulated in the $\beta$ tuning analysis, and the long run times needed to generate the ensembles needed for the $\kappa_4$ finite-volume study make it difficult to redo these runs quickly when the tuned $\beta$ is adjusted.  In principle, one could calculate how the results of the $\kappa_4$ finite-volume study depend on small shifts in $\beta$ and correct for the slight mistunings.  We leave this as an excercise for future work, for now ignoring the error associated with these mistunings.

\begin{figure*}[htb]
\begin{minipage}{0.475\linewidth}
    \centering
    \includegraphics[width=\linewidth]{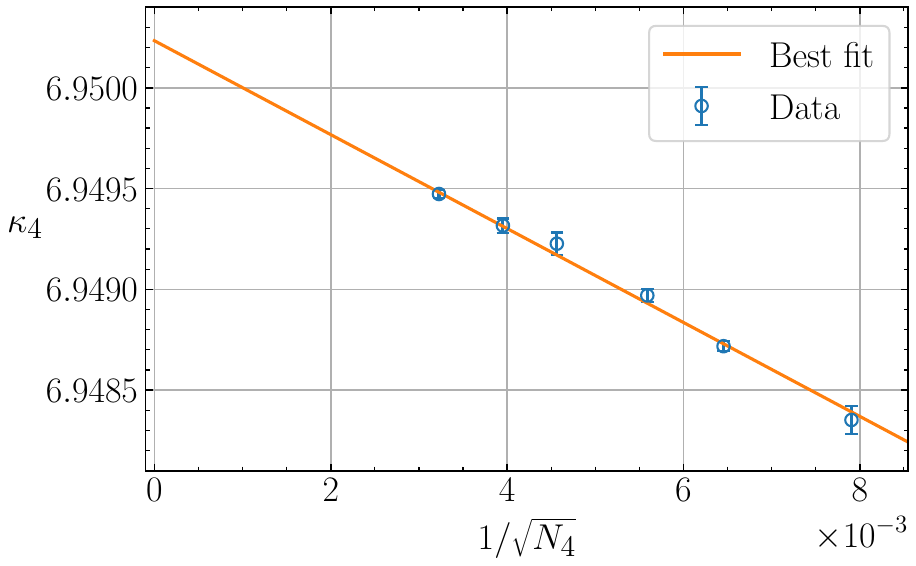}
\end{minipage}
\begin{minipage}{0.475\linewidth}
    \centering
    \includegraphics[width=\linewidth]{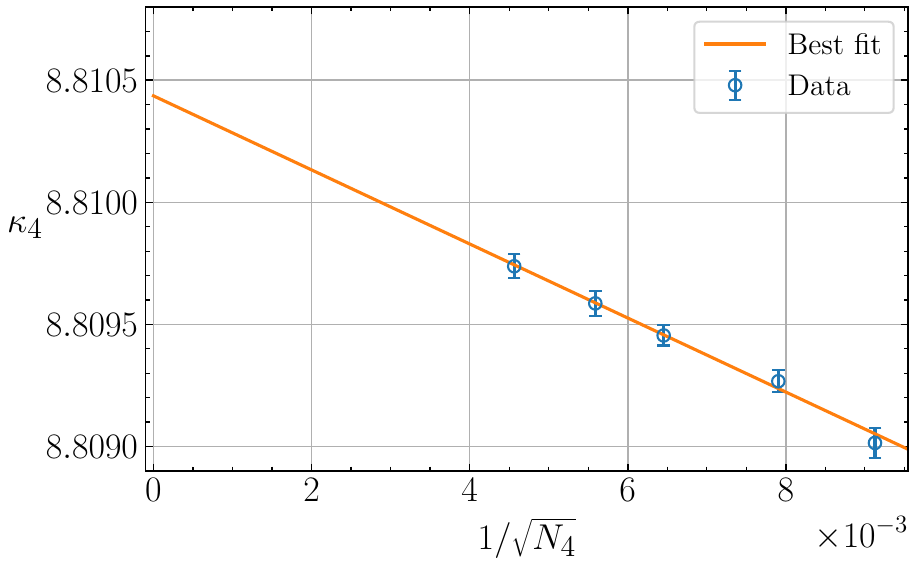}
\end{minipage}
\caption{ Tuned $\kappa_4$ plotted versus $1/\sqrt{N_4}$ for two representative $\kappa_2$ and tuned $\beta$ values, fitted to \eqref{eq:slopefit}. We show the fits for $\kappa_2=2.45$, $N_4=16000$ (left), and for $\kappa_2=3.4$, $N_4=12000$ (right). See also \autoref{tab:kappa4MF} and \autoref{tab:kappa4RF} for detailed fit results. These fits have $p$-value=0.53 (left), and $p$-value=0.84 (right). } 
\label{fig:kappa4exp}
\end{figure*}

 Our analysis requires that the combination $s_G=\sqrt{N_4}|\kappa_2-\kappa_4^{c}|$ be extracted from a finite-volume study with $\kappa_2$ and $\beta$ held fixed (see Eq.~\eqref{eq:slope}). The finite-volume study involves the linear fit
\begin{equation}
\label{eq:slopefit}
    \kappa_4(N_4)=A_{\kappa_4}+s_{G}\frac{1}{\sqrt{N_4}}\,,
\end{equation}
with $\kappa_4$ taken as a function of $1/\sqrt{N_4}$, and $A_{\kappa_4}$ and $s_G$ the free fit parameters.  The parameter $s_G$ thus obtained can be used to construct $G/a_{\rm lat}^2$ from Eq.~(\ref{eq:NCa}).  That the linear relation Eq.~(\ref{eq:slopefit}) is a good description of the $\kappa_4$ data and can be used to construct Newton's constant was shown in Ref.~\cite{Bassler:2021pzt}.  The fit results presented in \autoref{tab:slopesMF}, \autoref{tab:slopesF}, \autoref{tab:slopesRF}, and \autoref{tab:slopesSF} show that this trend continues to hold in the current analysis, as can be seen by the $p$-values of the fits.  For some of our scaling studies, mainly those at coarser lattice spacing or smaller nominal volume, we find that in order to get good fits we need to drop the smallest volume ensemble from the scaling fit.  This may be due to corrections to the linear relation Eq.~(\ref{eq:slopefit}) at smaller volumes and/or coarser lattice spacings, or to mistunings of the $\beta$ value that cause the simulations to be contaminated by their proximity to the phase transition.  Such effects are very small in absolute terms, but even small shifts in $\kappa_4$ could lead to deviations from the scaling of Eq.~(\ref{eq:slopefit}), given the precision needed for the analysis.  Figure~\ref{fig:kappa4exp} shows some plots of our fits to data for $\kappa_4$ versus $1/\sqrt{N_4}$ that are representative of our finite-volume scaling studies.

\begin{figure*}[htb]
\begin{minipage}{0.475\linewidth}
    \centering
    \includegraphics[width=\linewidth]{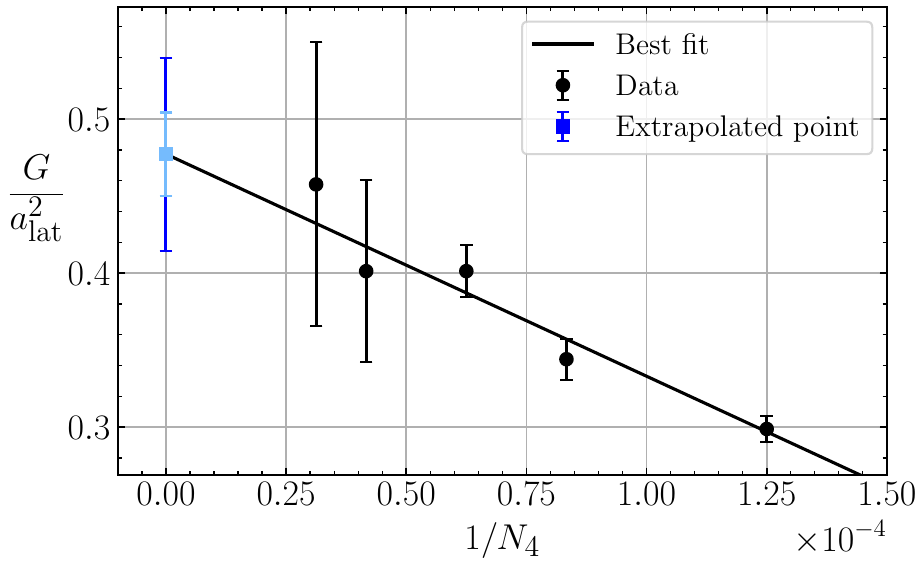}
\end{minipage}
\begin{minipage}{0.475\linewidth}
    \centering
    \includegraphics[width=\linewidth]{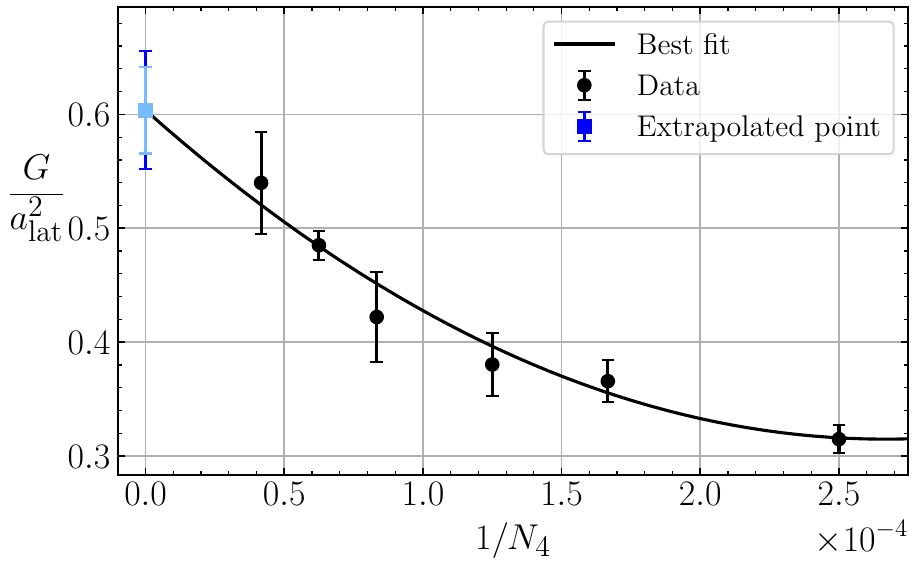}
\end{minipage}
\begin{minipage}{0.475\linewidth}
    \centering
    \includegraphics[width=\linewidth]{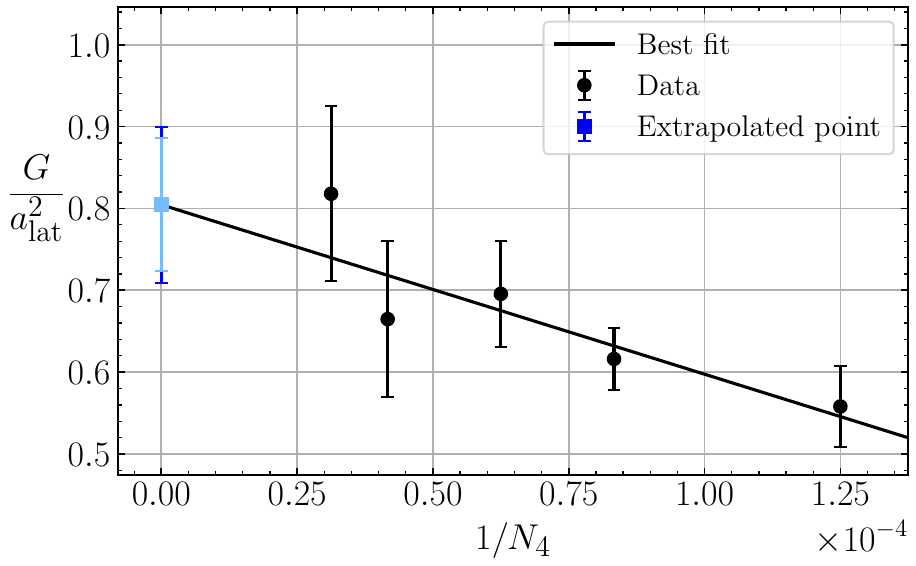}
\end{minipage}
\begin{minipage}{0.475\linewidth}
    \centering
    \includegraphics[width=\linewidth]{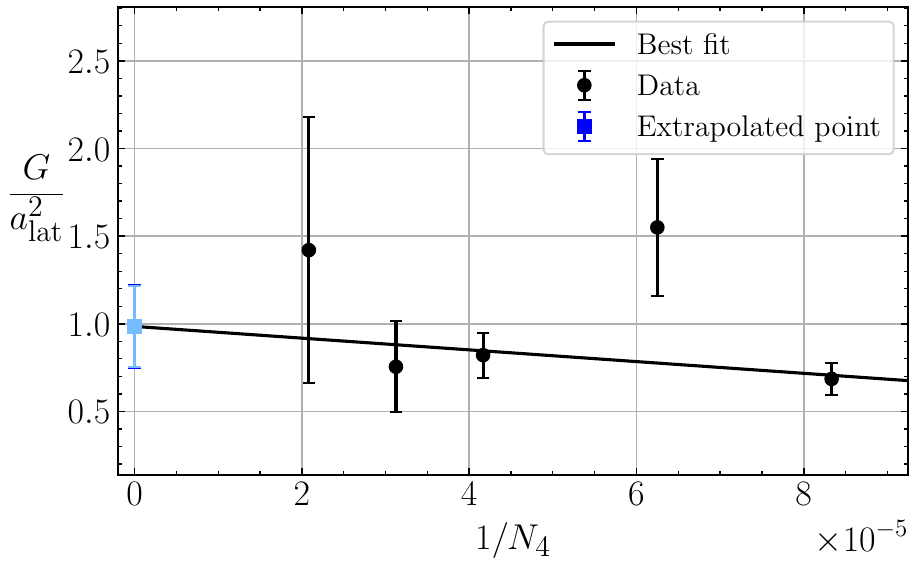}
\end{minipage}
\caption{
$G/a_{\rm lat}^2$ versus $1/N_4$ at all four lattice spacings, for data summarized in \autoref{tab:slopesMF}, \autoref{tab:slopesF}, \autoref{tab:slopesRF}, \autoref{tab:slopesSF}, and using the fit function Eq.~(\ref{eq:fitGa}) (with the quadratic term neglected unless required for a good fit). The resulting extrapolations and estimates of the fit quality are given in \autoref{tab:Gtabs}. We display our central fit here, though alternate fits are used to estimate an infinite-volume extrapolation error. The infinite-volume result is shown, with the statistical error in light blue and the total error in dark blue, where the total error includes the statistical error and an estimate of the systematic error due to the infinite-volume extrapolation added to it in quadrature.
} 
\label{fig:Gex}
\end{figure*}

\autoref{tab:slopesMF}, \autoref{tab:slopesF}, \autoref{tab:slopesRF}, and \autoref{tab:slopesSF} show the results for $|s_G|$ from all of our scaling studies.  The first error is statistical, and the second is an estimate of the systematic error due to varying the $\kappa_4$ data points included in the fits.  Adding these errors in quadrature and constructing $G/a_{\rm lat}^2$ from Eq.~(\ref{eq:NCa}) leads to the values of $G/a_{\rm lat}^2$ shown in Fig.~\ref{fig:Gex}, plotted as a function of $1/N_4$.  We have determined $G/a_{\rm lat}^2$ at multiple volumes at each lattice spacing, allowing us to use the infinite-volume limit of $G/a_{\rm lat}^2$, evaluated separately for each $\kappa_2$ value, to determine a relative lattice spacing.  In previous work \cite{Bassler:2021pzt} there was only one lattice spacing available with multiple volumes, so this was not possible.  In that case, the relative lattice spacing was determined using the diffusion process of a random walk.  In revisiting the return probability to obtain the relative lattice spacing on our new ensembles, we find that there is significant ambiguity in rescaling the return probability curve, given how much the shape changes with lattice spacing.  Using our $G/a_{\rm lat}^2$ for this purpose removes that ambiguity, while also being systematically improvable with more computing resources.  Compared to \cite{Bassler:2021pzt}, the relative lattice spacings change by up to a factor of $\sim1.7$.  We take the $\kappa_2=3.0$ ensembles to define our fiducial lattice spacing, so that $a_{\rm rel}$ at this lattice spacing is by definition one.  The other lattice spacings are measured relative to this, being inversely proportional to the square root  of the ratio of their $G/a_{\rm lat}^2$ values.  The value of $G/a_{\rm lat}^2$ at $\kappa_2=3.0$ then sets our absolute lattice spacing, allowing us to convert our measurements to units of the Planck length.

The infinite volume extrapolation is carried out using the fit function
\begin{equation}
\label{eq:fitGa}
\frac{G}{a_{\mathrm{lat}}^2}=G_a+\frac{H_G}{N_4}+\frac{I_G}{N_4^2}\,,
\end{equation}
where we only include the term quadratic in $1/N_4$ on the $\kappa_2=3.0$ lattice spacing, since the precision of the data demands it in order to get a good fit.  Figure~\ref{fig:Gex} shows the central fits for all of the infinite-volume extrapolations of $G/a_{\rm lat}^2$, along with their extrapolated values.  The dark blue riser line represents the statistical error on the infinite-volume result, while the light blue includes an extrapolation systematic error added to the statistical error in quadrature.  This systematic error is estimated by considering alternate fits for the extrapolation.  On the $\kappa_2=3.0$ ensemble, the central fit includes the term quadratic in $1/N_4$, while a linear fit neglecting this term leads to a larger reduced $\chi^2$ and a still acceptable $p$-value of 0.25.  The difference between the quadratic central fit and the linear fit is taken as an estimate of the infinite-volume extrapolation error.  For the $\kappa_2=2.45$ and $3.4$ lattice-spacing data, the central fit neglects the $1/N_4^2$ term, but the alternate fit includes it, and again the extrapolation systematic error is taken as the full difference between these two fit choices.  For the $\kappa_2=3.8$ lattice spacing data, the errors on $G/a_{\rm lat}^2$ are much larger, owing to the expense of simulating at this, our finest, lattice spacing.  A quadratic fit in $1/N_4$ gives a large difference from the central, linear in $1/N_4$ fit, and would lead to a $\sim 100\%$ error.  However, this is likely an overestimate of the error, since the data at other lattice spacings constrains the size of possible curvature in $1/N_4$.  We take the same estimate for the extrapolation systematic that we obtained for the extrapolation of our second finest lattice data (at $\kappa_2=3.4$) as an estimate of the systematic error at our finest lattice spacing.  This can be tested with more data at our finest lattice spacing, and these runs are in progress.

The results for $G/a_{\rm lat}^2$ in the infinite-volume limit, denoted by $G_a$ in the fit function Eq.~(\ref{eq:fitGa}), are given in Tab.~\ref{tab:Gtabs} for all four lattice spacings.  Taking $\kappa_2=3.0$ as our fiducial lattice spacing, the values of $a_{\rm rel}$ can be obtained from these results; these are also shown in Tab.~\ref{tab:Gtabs}.  We see that as $\kappa_2$ increases, $a_{\rm rel}$ decreases, so that larger $\kappa_2$ corresponds to finer lattice spacing, as expected. 
Knowing the dimensionless ratio $G/a_{\rm lat}^2$ allows us to convert $a_{\rm lat}$ into units of the Planck length, where $G/a_{\rm lat}^2>1$ would indicate that $a_{\rm lat}$ is sub-Planckian.  Thus, we see that our finest lattice spacing measured in $a_{\rm lat}$ units is very nearly the Planck length, though there appears to be no limitation other than practical for pushing it smaller.  Using our $a_{\rm lat}/\ell$ values from Sec.~\ref{sec:aellboth} to express our lattice spacings in renormalized dual lattice units, and again taking $\kappa_2=3.0$ as the fiducial spacing so that $\ell_{\rm rel}=1$ there, we find the values of $\ell_{\rm rel}$ given in Tab.~\ref{tab:Gtabs}.  Note that our results for $G/a_{\rm lat}^2$, combined with our results for $a_{\rm lat}/\ell$, imply that $\ell$ is significantly smaller than the Planck length.  The length $\ell$ is around 23 times smaller than the Planck length on our finest lattice spacing.  The relative length $\ell_{\rm rel}$ also decreases as $\kappa_2$ is taken larger, further confirming our picture that a continuum limit can be taken by following the first order line in the phase diagram of \autoref{fig:phaseDiagram} out to large $\kappa_2$ values.

\section{\label{sec:RVM}Running vacuum energy}

\subsection{Theory and motivation}
\label{sec:theorymotive}
\begin{figure}[ht]
\centering
\includegraphics[width=\linewidth]{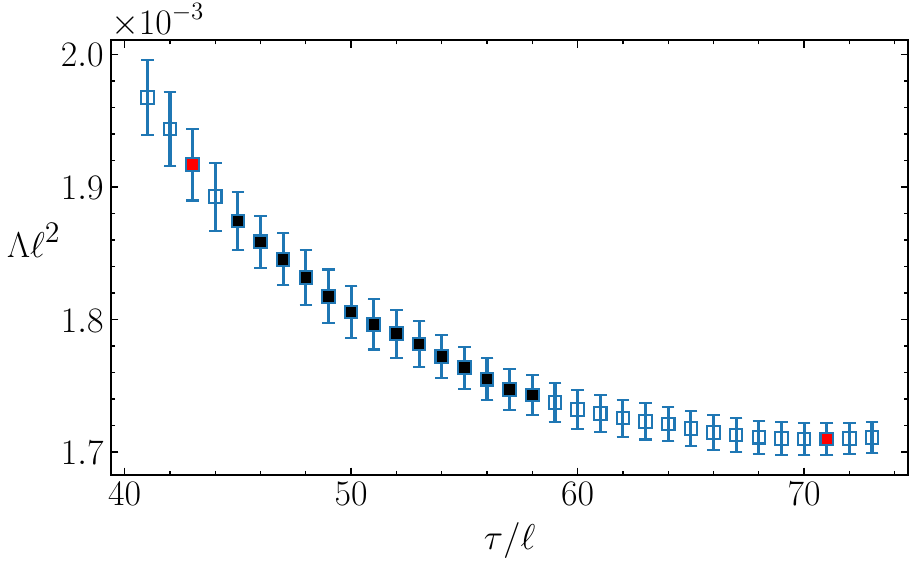}
\caption{ 
$\Lambda$ versus $\tau/\ell$ at our finest lattice spacing, with $\kappa_2 = 3.8$, $\beta=0.92$ and $N_4 = 16000$.  The points in black are the same ones used in the fit appearing in the lower right frame of Fig.~\ref{fig:lH2fit}. The inflection point discussed in \ref{sec:aell1} and the peak of the shelling function are located at $\tau/\ell=43$ and $\tau/\ell=71$, respectively. These points are highlighted in red.
} 
\label{fig:Lambda-tau}
\end{figure}

Our lattice geometries are well-described by the semi-classical Euclidean de Sitter solution, which is solid evidence that EDT could define a non-perturbative formulation of quantum gravity.  Interestingly, a more careful study shows that there are corrections to this picture.  As pointed out in Section~\ref{sec:aell2}, we find that the cosmological constant is not strictly constant as a function of Euclidean time, particularly on our smaller volume ensembles.  A deviation from the expected classical scaling of the geometries with volume could provide a window to quantum gravitational effects, assuming that they survive a continuum limit.  Although such deviations might be treated as finite-size effects to be extrapolated away in a flat space lattice QCD calculation, in quantum gravity the dynamics of the universe at small sizes is an interesting question in its own right.  Figure~\ref{fig:Lambda-tau} shows the cosmological constant inferred from Eq.~\ref{eq:latdeSitter} as a function of Euclidean time $\tau/\ell$ on one of our smaller physical-volume ensembles.  Eq.~\ref{eq:latdeSitter} requires as input the infinite-volume limit of $a_{\rm lat}/\ell$, which is needed to set $\ell_V/\ell$.  Although all of our ensembles are consistent with the classical de Sitter solution, at least within a chosen fit window, this implies a particular finite-volume value of $a_{\rm lat}/\ell$.  Because the lattice volumes at which we are simulating are rather small in Planck units, and the classical de Sitter solution that is used to fix $a_{\rm lat}/\ell$ only applies at large volumes, we must extrapolate our values of $a_{\rm lat}/\ell$ to the infinite-volume limit before we can use them as conversion factors.  When we use the appropriate infinite-volume extrapolated values of $a_{\rm lat}/\ell$ in Eq.~~\ref{eq:latdeSitter}, the behavior of $\Lambda$ versus $\tau$ on a finite volume shows significant deviations from the pure de Sitter solution, where a constant $\Lambda$ is expected.  We find that these results are broadly consistent across lattice spacings and that the Euclidean time dependence of the effective cosmological constant does not appear to be a lattice artifact.  In the remainder of this work we show that these deviations can be described by nontrivial vacuum dynamics and that they pass a number of important consistency checks in this interpretation.  

We find that our lattice geometries are well-described by assuming that the effective long-distance cosmological constant has a power law running with renormalization scale
\bea \label{eq:RVM} \Lambda(\mu) = \Lambda_0 + 3 \nu |\mu|^2,
\eea
where $\Lambda_0$ is a constant fixed by the physical volume of the lattice, which acts as an infrared cutoff, $\mu$ is the renormalization scale, and $\nu$ is a dimensionless coupling that absorbs the logarithmic running of the term quadratic in $\mu$, so that
\begin{equation}
    \nu(\mu^2_f) = \frac{\nu(\mu_i^2)}{1+b \log(\frac{\mu_i^2}{\mu_f^2})},
    \label{eq:nulog1}
\end{equation}
for some constant $b$.  We find Eq.~\ref{eq:RVM} is a good description of the lattice data if we identify the scale $\mu$ with the Hubble rate $H=\dot{A}/A$, or at the largest scales probed on a given ensemble, by the four-volume.  At these largest scales the presence of the classical maximum of the shelling function ensures that $H=0$ at that point.  Even so, we find that $\nu$ does not go to zero there, as Eq.~(\ref{eq:nulog1}) would imply if the scale $\mu$ were to be associated with the Hubble rate $H$ and set to zero.  In this case the (inverse) four volume serves as an alternative cut-off in Eq.~(\ref{eq:nulog1}).  The absolute value in Eq.~(\ref{eq:RVM}) reflects the fact that the renormalization scale should be identified with the magnitude of the Hubble rate, and that the $\mu^2$ term does not pick up a minus sign due to the Euclidean continuation of the time derivative appearing in $H$.  Note that in our simulations $H$ is not a constant as a function of Euclidean time because the spatial curvature term is not negligible, as it would be for pure de Sitter space in Lorentzian signature at large times.  Figure~\ref{fig:lH2fit} (bottom-right panel) shows the same data for $\Lambda(\tau)$ as in Fig.~\ref{fig:Lambda-tau}, now plotted as a function of $H^2$, which is reconstructed from the first derivative of the scale factor $A(\tau)$.  This set of data is well-described by a fit linear in $H^2$, hinting that our results could be consistent with a picture in which the vacuum energy runs according to Eq.~\ref{eq:RVM}, with a renormalization scale associated with the Hubble rate.  A more comprehensive study incorporating multiple lattice spacings and volumes shows that our results are well-described by this picture, allowing a first principles determination of the parameters $\nu$ and $b$.

A power law running of the cosmological constant was conjectured by Polyakov \cite{Polyakov:2000fk}, where conformal fluctuations of the metric provide the underlying mechanism.  Work by Sol\`a et al. \cite{Shapiro:2009dh, Sola:2013gha, Gomez-Valent:2015pia, Sola:2016jky, Moreno-Pulido:2023ryo} considered a series of models with a running of the vacuum energy given by Eq.~\ref{eq:RVM}; the various models differ in the identification of the scale $\mu$ and in the choice of the vacuum equation of state.  One of the main goals of that work was to constrain the possible models and their parameter space from fits to cosmological data \cite{Sola:2016jky, SolaPeracaula:2021gxi}.  In this work we turn this around and apply similar fits to the data we have generated from our lattice simulations.  We find strong constraints on this picture, as the lattice data is not compatible with a large swath of model choices and parameter values. The picture that emerges from the lattice calculations is in fact one of the simplest of the model variants, and it is compatible with observations, as discussed in Section~\ref{sec:cosmology}.

For our lattice study we consider the most general version of the running vacuum models of Ref.~\cite{Gomez-Valent:2015pia}, which has the following parameterization,
\bea\label{eq:RVM_gen} \Lambda(H) = \Lambda_0 + 3\nu H^2 + 3 \tilde{\nu} \dot{H}.
\eea
This model includes all of the terms in a derivative expansion to ${\cal O}(H^2)$ and reduces to Eq.~(\ref{eq:RVM}) if we identify $\mu$ with the Hubble scale and set $\tilde{\nu}=0$.  We allow for $\tilde{\nu}$ to be non-zero in our analysis, though the lattice data constrains it to be significantly smaller than $\nu$ and consistent with zero.

In order to determine the parameters $\nu$ and $\tilde{\nu}$ from the lattice, we first work through the consequences of the model defined in Eq.~(\ref{eq:RVM_gen}) for cosmological evolution in Euclidean space.
To do this, we take the vacuum energy to have the form of a perfect fluid, 
\bea T_{\mu\nu}= -p_\Lambda g_{\mu\nu} + (p_\Lambda + \rho_\Lambda)u_\mu u_\nu,
\eea
with $u=(1, 0, 0, 0)$ and $p_{\Lambda}$ the vacuum pressure.  With this, the Friedmann-Lema\^{\i}tre-Robertson-Walker (FLRW) equations in Euclidean space become
\begin{align} \label{eq:FRW}
    -\frac{\dot{a}^2}{a^2} + \frac{1}{a^2} &= \frac{8\pi G \rho_\Lambda}{3},\\
\label{eq:FRW2}    -\frac{\ddot{a}}{a} &= -\frac{4\pi G}{3}(\rho_\Lambda + 3 p_\Lambda),
\end{align}
where
\bea  \rho_{\Lambda} = \rho_{\Lambda_0} + \frac{3}{8\pi G}\left(\nu H^2 + \tilde{\nu}\dot{H}\right),
\label{eq:rvmrho}
\eea
with $\rho_{\Lambda_0}=\Lambda_0/(8\pi G)$ the constant contribution to the energy density of the vacuum energy model. 

In order to satisfy diffeomorphism invariance, the evolution must obey covariant energy conservation.  The various models of Sol\`a et al.~\cite{Gomez-Valent:2015pia, SolaPeracaula:2021gxi} differ in the way that this condition is satisfied, postulating different ways that the vacuum energy might transfer energy to the matter sector.  In our pure gravity simulations, we do not see evidence for an emergent matter sector in the cosmological dynamics, so that the vacuum sector independently satisfies covariant energy conservation, $T^{\mu\nu}_{\ \ ;\,\mu}=0$, leading to
\bea\dot{\rho}_\Lambda = -3H(\rho_\Lambda + p_\Lambda).
\label{eq:continuity}
\eea
Taking the model for vacuum energy defined by Eq.~\ref{eq:rvmrho}, covariant energy conservation then implies that the vacuum pressure is
\bea \label{eq:pLam}  p_\Lambda = \frac{1}{8\pi G}\left(-\Lambda_0 - (\nu-\tilde{\nu})H^2 - 2\nu \frac{\ddot{a}}{a} -\tilde{\nu}\frac{\dddot{a}}{\dot{a}} \right).
\eea
This therefore implies the existence of non-trivial dynamics for the vacuum equation of state $w\equiv p_\Lambda/\rho_\Lambda$.

Inserting the vacuum energy density of Eq.~(\ref{eq:rvmrho}) and the vacuum pressure of Eq.~(\ref{eq:pLam}) into the FLRW equations, we find
\begin{align}
    3\left(-\frac{\dot{a}^2}{a^2} + \frac{1}{a^2} \right) &= \Lambda_0 + 3(\nu - \tilde{\nu})H^2 + 3\tilde{\nu}\frac{\ddot{a}}{a}, \label{eq:Einstein00}\\
    -\frac{\ddot{a}}{a} &= \frac{\Lambda_0}{3} + \left(\nu - \frac{1}{2}\tilde{\nu}\right)\frac{\ddot{a}}{a} + \frac{1}{2}\tilde{\nu}\frac{\dddot{a}}{\dot{a}}. \label{eq:Einsteinii}
\end{align}
We can construct all of the terms in this pair of equations that involve $a$ and its time derivatives from the lattice shelling function.  This allows us to solve for the parameters $\nu$ and $\tilde{\nu}$ appearing in these equations.
For the purposes of comparing to lattice data, it is helpful to introduce a set of approximations, starting with
\bea  \frac{\ddot{a}}{a} \equiv S = S_0 + {\cal O}(H^2).\label{eq:defS0}
\eea
%
We also introduce
\bea \frac{\dddot{a}}{\dot{a}} \equiv T = T_0 + {\cal O}(H^2),\label{eq:defT0}
\eea 
where classically, $T=T_0=S_0=-\Lambda_0/3$.  We only apply these approximations to the terms in Eqs.~(\ref{eq:Einstein00}) and (\ref{eq:Einsteinii}) that contain a prefactor $\tilde{\nu}$.  This is a good approximation under the assumption that $\tilde{\nu}$ is much smaller than $\nu$, an assumption that is justified by the analysis of Sec.~\ref{sec:nutilde}.

We can then rewrite Eq. (\ref{eq:Einstein00}) as 
\bea \label{eq:FRW1}  3\left(-\frac{\dot{a}^2}{a^2} + \frac{1}{a^2} \right) = I_{H^2} + 3 \nu' H^2,
\eea
where
\begin{equation}
   \nu'\equiv\nu - \tilde{\nu},
\end{equation} 
and $I_{H^2}\equiv\Lambda_0 + 3\tilde{\nu}S$.  Our determination of the lattice scale factor allows us to construct the left side of this equation, thus providing a determination of the effective cosmological constant.  We can then extract the combination $\nu' = \nu - \tilde{\nu}$ from a linear fit to a reconstruction of the left side of Eq.~(\ref{eq:FRW1}) versus $H^2$.

We also combine $I_{H^2}$ determined from fits to Eq.~(\ref{eq:FRW1}) and $\ddot{a}/a$ constructed from the lattice scale factor, and making use of Eq.~(\ref{eq:Einsteinii}), construct the useful combination
\bea  I_{H^2} + 3 \frac{\ddot{a}}{a} = -3\left(\nu -\frac{3}{2}\tilde{\nu} \right)\frac{\ddot{a}}{a} - \frac{3}{2}\tilde{\nu}\frac{\dddot{a}}{\dot{a}}. 
\ \ \  
\label{eq:nu''1}
\eea
Eq.~(\ref{eq:nu''1}) provides an alternative determination of $\nu$ within this model for running vacuum energy under the assumption that we can neglect the terms proportional to $\tilde{\nu}$.  

\subsection{Determining model parameters from the lattice}
\label{sec:modelParams}

The calculation of the parameters of the running vacuum model $\nu$ and $\tilde{\nu}$ in Eq.~(\ref{eq:RVM_gen}) using our lattice results proceeds as follows.  The scale factor measured in $\ell$ units is obtained directly from the lattice scale factor from
\bea \label{eq:lattice_a}  \frac{a(\tau)}{\ell} = \frac{\ell_V}{\ell}\frac{A(\tau)}{\ell_V},
\eea
where we use the notation $A(\tau)/\ell_V\equiv (N_4^{\rm shell}(\tau))^{\frac{1}{3}}$ introduced in Sec.~\ref{sec:aell2} to emphasize the dependence on the lattice spacing conversion factors.  The factor $\ell_V/\ell$ is obtained from the results for $a_{\rm lat}/\ell$ from our combined analysis in Section~\ref{sec:desitter} using Eq.~(\ref{eq:ellVoverell}). 
The parameters of the dark energy model can then be constructed from the scale factor $A(\tau)$ and its derivatives with respect to $\tau$, including the Hubble rate $H = \dot{A}/A$, and the quantities $S = \ddot{A}/A$ and $T = \dddot{A}/\dot{A}$, where all three of these are measured in dual lattice units.

The numerical derivatives are all approximated using the five-point stencil
\begin{align*}
f^{\prime}(x) & \approx \frac{-f(x+2 d)+8 f(x+d)-8 f(x-d)+f(x-2 d)}{12 d}, \\
f^{\prime \prime}(x) & \approx \frac{-f(x+2 d)+16 f(x+d)-30 f(x)}{12 d^2} \\
& \; \;\;\;\;\;+ \frac{16 f(x-d)-f(x-2 d)}{12 d^2}, \\
f^{(3)}(x) & \approx \frac{f(x+2 d)-2 f(x+d)+2 f(x-d)-f(x-2 d)}{2 d^3},
\label{eq:fivepointstencil}
\end{align*}
and the derivatives are all taken under a jackknife to correctly propagate the errors through to the next stage of the analysis.  As an example, the first derivative of the scale factor with respect to Euclidean time is
\bea  \label{eq:ex1stDeriv}  \dot{A}(\tau/\ell) & \approx \frac{-A(\tau/\ell+2)+8 A(\tau/\ell+1)-8 A(\tau/\ell-1)+A(\tau/\ell-2)}{12}, \nonumber \\
\eea
where the explicit factors of $\ell$ show that $\tau$ is measured in terms of renormalized dual-lattice ($\ell$) units.

In order to determine the parameters $\nu$ and $\tilde{\nu}$, we first consider the parameter $\nu'\equiv \nu -\tilde{\nu}$ introduced in the previous subsection, as this combination is more convenient to extract from the simulations.  A value for $\nu'$ can be extracted from 
\bea \label{eq:latFLRW_2}  3\left(-\frac{\dot{A}^2}{A^2} + \frac{\ell^2}{\ell_V^2}\frac{\ell_V^2}{A^2} \right) = \Lambda(H)\ell^2 = (I_{H^2} + 3\nu' H^2)\ell^2. \nonumber \\
\eea
The sign difference between the two terms on the left side of Eq.~(\ref{eq:latFLRW_2}) leads to a substantial cancellation between these terms.  In order to construct this combination with sufficient precision to extract $\nu'$, we require a precise determination of $\ell_V/\ell$ (which itself depends on having a precise $a_{\rm lat}/\ell$).  This partial cancellation sets the precision level needed for the $a_{\rm lat}/\ell$ determination of Sect.~\ref{sec:desitter}, which must be at the percent level in order to resolve a $\nu'$ statistically and systematically different from zero.  The value of $\nu'$ is obtained from the slope of a linear fit to the left side of Eq.~(\ref{eq:latFLRW_2}) as a function of $H^2$, and $I_{H^2}\equiv\Lambda_0 + 3\tilde{\nu}S$ is obtained from the intercept.  As we show in the following subsections, $\tilde{\nu}$ is consistent with zero.  Taking $\tilde{\nu}=0$, we have $I_{H^2}=\Lambda_0$, and this becomes the input to our Method 2 determination of $a_{\rm lat}/\ell$, as detailed in Sect.~\ref{sec:aell2}.  

An alternative determination of $\nu'$ allows for a useful 
crosscheck of the model.  Forming the combination
\bea  I_{H^2} + 3 \frac{\ddot{A}}{A} \approx -3 \nu' \frac{\ddot{A}}{A},
\label{eq:nu''2}
\eea
which follows from Eq.~(\ref{eq:nu''1}).  Note that in the limit that $\tilde{\nu}\to 0$, this approximate equality becomes exact. As $\tilde{\nu}$ turns out to be much less than $\nu$ in our analysis, this equation is sufficient for our purposes.  In our calculation we also assume that $\ddot{A}/A$ is a constant, independent of $H^2$.  This follows identically from Eq.~(\ref{eq:Einsteinii}) in the limit that $\tilde{\nu} \to 0$ if we also ignore the logarithmic running of $\nu$.  This turns out to be a good approximation in the fit window of $H^2$ that we choose.

We are able to constrain $\tilde{\nu}$ from lattice data by considering $I_{H^2}$ directly.  The intercept of the $\Lambda(H)$ versus $H^2$ fit provides a determination of $I_{H^2}=\Lambda_0 + 3\tilde{\nu} S_0$, with $S_0$ the result of a constant fit to $\ddot{A}/A$.  Although we have no way to determine $\Lambda_0$ independently at fixed volume, if we make assumptions about the volume dependence of the two terms making up $I_{H^2}$, we can put constraints on the size of $\tilde{\nu}$.  We take the volume dependence of $\Lambda_0$ to be that of the semiclassical theory, proportional to $1/\sqrt{N_4}$.  As discussed in Sec.~\ref{sec:desitter}, this is a good description of our lattice data for $I_{H^2}$ at all four lattice spacings.  The volume dependence of $\tilde{\nu}$ is taken to be the logarithmic dependence of Eq.~(\ref{eq:nulog1}), with the scale associated with the fourth root of the four-volume of the lattice.  This is equivalent to using the curvature radius for the scale.  This logarithmic running turns out to be a good description of the volume dependence of $\nu'$, implying that it should also describe the running of $\tilde{\nu}$, since $\nu'$ is a linear combination of $\nu$ and $\tilde{\nu}$.  The result of fitting $I_{H^2}$ versus $1/\sqrt{N_4}$ puts strong constraints on any deviations from linearity, leading to bounds on the size of $\tilde{\nu}$.  Note that in Sec.~\ref{sec:aell2}, we identify $\Lambda_0$ with $I_{H^2}$, implicitly setting $\tilde{\nu}$ to zero in our Method 2 determination of $a_{\rm lat}/\ell$.  We also examine the constraints on $\tilde{\nu}$ when we do not make this assumption, using only the Method 1 determination of $a_{\rm lat}/\ell$, which does not depend on any assumptions about the size of $\tilde{\nu}$.  

In Sections~\ref{sec:nuPrime_determination}, \ref{sec:nutilde}, and \ref{sec:alt_nuPrime}, most quantities ($\Lambda$, $H$, $\tau$, etc.) are evaluated in $\ell$ units.  The exception to this is the four-volume, which is naturally expressed in units of $a_{\rm lat}^4$.  Some fit functions in the following sections include a prefactor for the four-volume, which should be assumed to carry the correct units to convert between $a_{\rm lat}$ and $\ell$ where necessary.  In these sections we suppress explicit factors of $\ell$ for clarity of the presentation.

\subsection{Calculation of $\nu'$}
\label{sec:nuPrime_determination}

\subsubsection{Fits to $\Lambda(H)$ versus $H^2$}

\begin{figure*}
\begin{minipage}{0.475\linewidth}
    \centering
    \includegraphics[width=\linewidth]{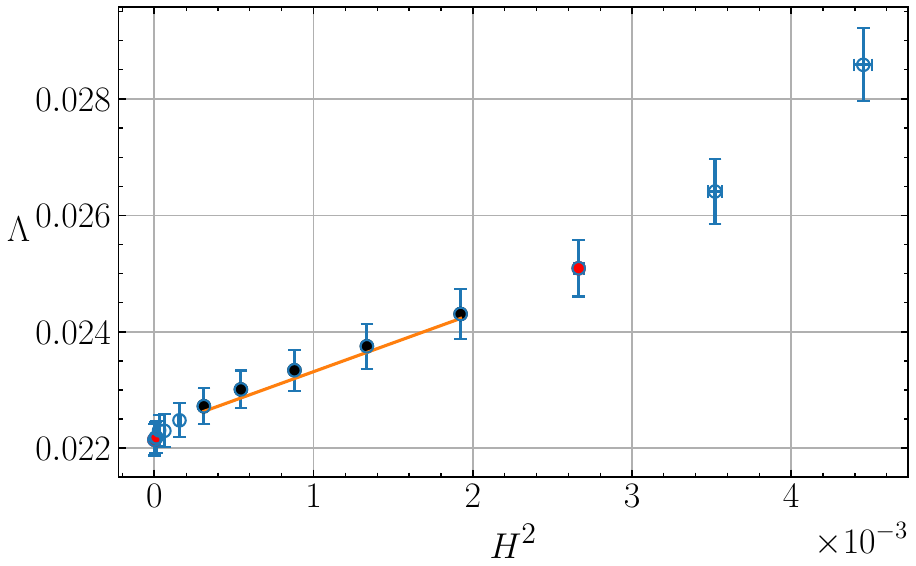}
\end{minipage}
\begin{minipage}{0.475\linewidth}
    \centering
    \includegraphics[width=\linewidth]{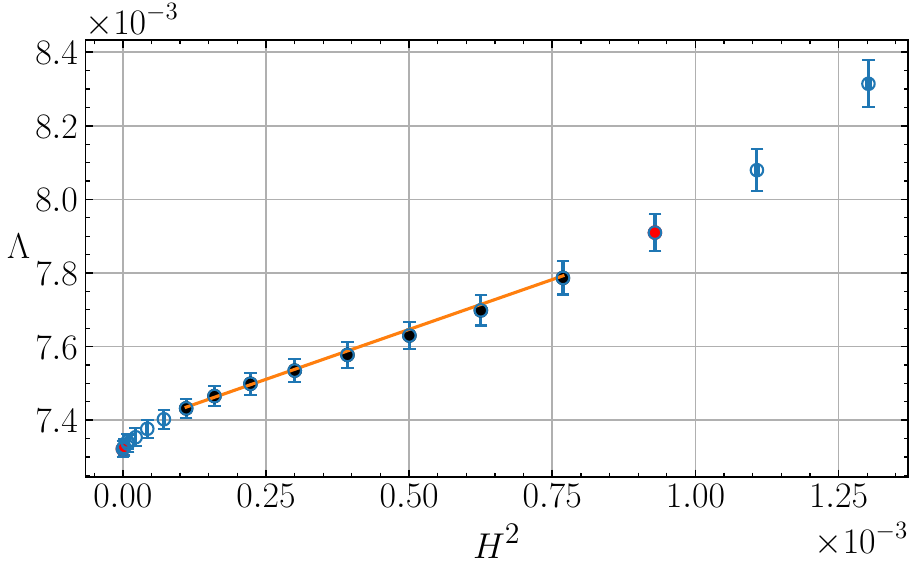}
\end{minipage}
\begin{minipage}{0.475\linewidth}
    \centering
    \includegraphics[width=\linewidth]{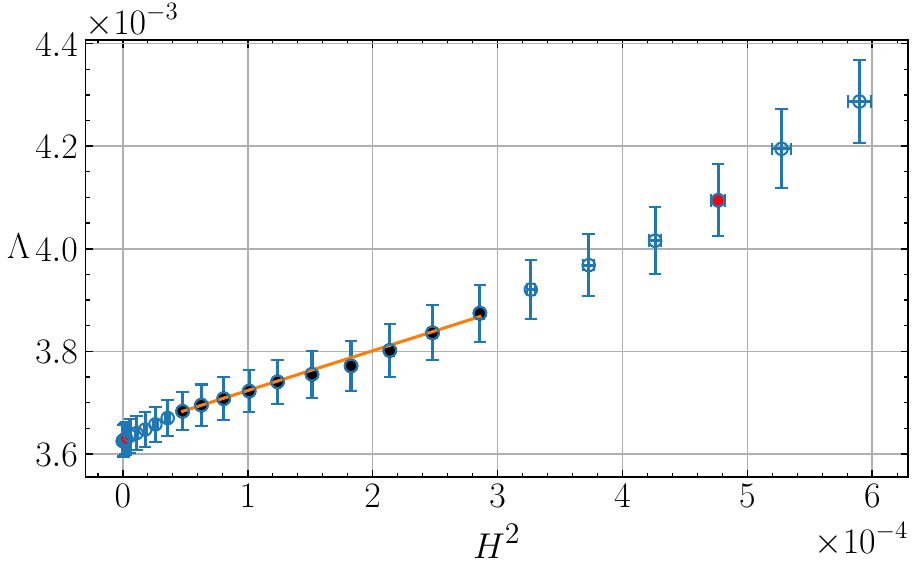}
\end{minipage}
\begin{minipage}{0.475\linewidth}
    \centering
    \includegraphics[width=\linewidth]{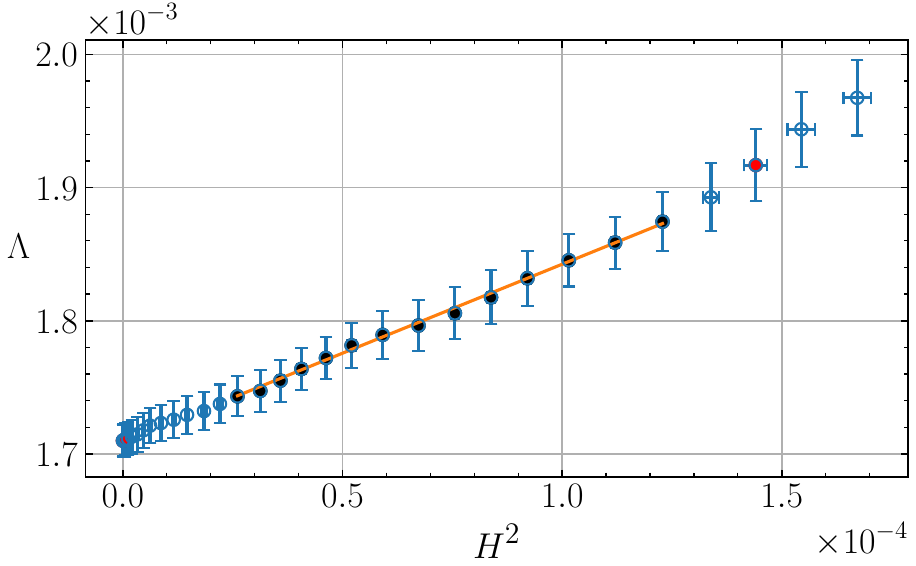}
\end{minipage}
\caption{ $\Lambda$ versus $H^2$ plotted for four lattice spacings, along with a linear fit to the form Eq.~(\ref{eq:lambdaH2linearfit}).  The points included in the fits are highlighted in black. 
Top left panel: $N_4 = 6$k,  $\kappa_2 = 2.45$.
Top right panel: $N_4 = 16$k,  $\kappa_2 = 3.0$.
Bottom left panel: $N_4 = 16$k, $\kappa_2 = 3.4$. 
Bottom right panel: $N_4 = 16$k, $\kappa_2 = 3.8$, $\beta=0.92$. The inflection point discussed in \ref{sec:aell1} and the point corrresponding to the peak of the shelling function are highlighted in red. 
} 
\label{fig:lH2fit}
\end{figure*}

As discussed in the previous subsection, the value of $\nu'$ is obtained by using the lattice shelling function to construct the left side of Eq.~(\ref{eq:latFLRW_2}).  This leads to an effective cosmological constant $\Lambda(H)$, which must be linear in $H^2$ if it is to be consistent with Eq.~(\ref{eq:RVM_gen}).  Fig. \ref{fig:lH2fit} shows the effective $\Lambda$ plotted versus $H^2$ for representative ensembles across four lattice spacings, where we find solid evidence that $\Lambda(H)$ is linear in $H^2$.  The linear fit region that we choose is marked with black points in Fig.~\ref{fig:lH2fit}.      

We parameterize the linear fit as
\bea \Lambda(H) = I_{H^2} + 3\nu'H^2
\label{eq:lambdaH2linearfit}
\eea 
where the slope gives a determination of $\nu' = \nu - \tilde{\nu}$, and the intercept gives the parameter $I_{H^2}$.  Eq.~(\ref{eq:lambdaH2linearfit}) should be a good approximation in the limit that $\tilde{\nu}$ is small compared to $\nu$.  The fit region is chosen to be to the left of the classical peak in order to avoid the asymmetric tail, which is most likely unphysical and contaminated by large lattice artifacts, while also not extending to too small a value of $\tau$, in order to avoid contamination from short-distance discretization effects.  

For the shelling function used to compute the effective cosmological constant $\Lambda(H)$, we use 60 sources per configuration on every ensemble.  There is a (correlated) statistical error on the horizontal axis, since $H^2$ is also a quantity derived from the lattice shelling function.  The fit to $\Lambda(H^2)$ versus $H^2$ uses orthogonal distance regression and includes the correlations between errors on the vertical and horizontal axes when minimizing $\chi^2$.

The range of the linear fit to $\Lambda(H)$ is varied in order to estimate a fitting systematic error in our determinations of $\nu'$ and $I_{H^2}$.  For the fits at $\kappa_2 = 2.45$ and $\kappa_2 = 3.0$, we vary either end of the fit range by plus and minus one or two.  For $\kappa_2 = 3.4$, we vary either end of the fit range by plus and minus 2 and 4.  For $\kappa_2 = 3.8$, we vary by plus and minus 3 and 6 if $N_4 \leq 32$k, and by plus and minus 4 and 8 if $N_4 > 32$k. Thus, for each ensemble, we have a total of 25 different fit ranges centered on our preferred window in the linear regime.  We estimate the fitting systematic error from the distribution of these fits. We take our central value and statistical error for $\nu'$ from a fit that is close to the middle of the distribution of fit values, and we take the systematic error from the standard deviation of the distribution of fit results.  For fits to be included in the systematic error estimate, the $\Lambda$ versus $H^2$ fit has to satisfy $p \geq 0.01$.
We also include an error from $a_{\rm lat}/\ell$ that is propagated through the analysis for $\nu'$ and $I_{H^2}$.  We vary $a_{\rm lat}/\ell$ by its upper and lower 1 $\sigma$ error and repeat the analysis used to extract $\nu'$ and $I_{H^2}$ with these values.  Although the error inferred from varying $a_{\rm lat}/\ell$ by its upper and lower $1\sigma$ range is not quite symmetric about its central value, the upper and lower errors are very similar. For simplicity, we symmetrize the $a_{\rm lat}/\ell$ error, taking the error from the larger of the two variations of ${a_{\rm lat}/\ell}$.

For volumes where the $\beta$ tuning error is non-negligible and we have a pair of tuned $\beta$ values, we do the fits on both ensembles separately to get the intercepts and slopes.  Then, instead of taking the weighted average of the two ensembles as we did for $\frac{a_{\rm lat}}{\ell}$, we use one ensemble as the central value of the quantity, and we take the difference between that and the result on the second ensemble as a systematic error and add the systematic error in quadrature to the original error.  This is different from our approach to combining the different tuned ensembles in the analysis of Sec.~\ref{sec:aell1}, since here it is important to preserve the correlations between observables measured on the same ensemble, especially for the $\tilde{\nu}$ analysis of the following subsection.

Table \ref{tab:lambdaH2} shows the fit parameters for the $\Lambda(H)$ analysis on all the ensembles considered in this work, where the first error includes the statistical and fit systematic error added in quadrature, and the second error is the $a_{\rm lat}/\ell$ error.
The good quality of the linear fits to $\Lambda(H)$ versus $H^2$ in Fig.~\ref{fig:lH2fit} shows that our lattice data is compatible with Eq.~(\ref{eq:RVM}), and it allows us to determine the running coupling $\nu'$, evaluated at a distance scale comparable to the de Sitter radius on any given ensemble across many different physical volumes.  

\subsubsection{Fits to $\nu'$ as a function of four-volume}

We expect the dimensionless coupling $\nu'$ to run logarithmically as a function of the renormalization scale as in Eq.~(\ref{eq:nulog1}).  In the de Sitter solution with Lorentz signature, the various measures of the size of the universe at large times are proportional, with the scalar curvature $R \propto \Lambda \propto H^2$.  In our Euclidean simulations, the Hubble rate vanishes at the classical peak (since $\dot{a}$ is zero at the maximum of the shelling function).  However, Fig.~\ref{fig:lH2fit} shows that the slope, and thus the parameter $\nu'$, does not appear to vanish in the limit that $H^2\to 0$.  This suggests that the appropriate cut-off scale at the longest distances on our lattices is not $H^2$ but the spatial curvature $1/a^2 \propto \Lambda \propto 1/\sqrt{V_4}$.  We thus associate the renormalization scale of the logarithmic running of $\nu'$ with the (fourth-root of the) physical lattice volume
\bea\label{eq:V4} V_4 =  \eta C_4 N_4  a^4_{\rm rel}.
\eea

Identifying the renormalization scale $\mu^2$ of Eq.~(\ref{eq:nulog1}) with $1/\sqrt{V_4}$, we parameterize the logarithmic running of Eq.~(\ref{eq:nulog1}) by
\bea\label{eq:nuPfit} \nu' = \frac{A'}{\log (B' \sqrt{V_4})},
\eea
where $A'$ and $B'$ are free fit parameters.  We find that a large subset of our data set for $\nu'$ is compatible with Eq.~(\ref{eq:nuPfit}).  This is true across many volumes and multiple lattice spacings, and the parameters determined from our fits are fairly stable under different fit variations that consider smaller subsets of the data or introduce lattice spacing dependence into Eq.~(\ref{eq:nuPfit}).  In the remainder of this subsection we discuss those fit variations and our best estimates of the resulting fit parameters.  We also discuss the motivation for our cut on the data for $\nu'$ and why we believe that the points that have been left out of the fits are likely contaminated by significant lattice artifacts.  Finally, we discuss how the variation of $\nu'$ with the volume of the lattice is also visible in the linear fits of $\Lambda(H)$ versus $H^2$ by shifting the fit window on our larger volume lattices to larger values of $H^2$ (smaller distance scales).  This shows that the running of $\nu'$ with volume is not merely an artifact of the way the tuning is performed to produce ensembles at different lattice volumes and the same nominal lattice spacing, since a similar running of $\nu'$ with scale is seen on individual lattice ensembles.

\begin{figure}
\centering
\includegraphics[width=\linewidth]{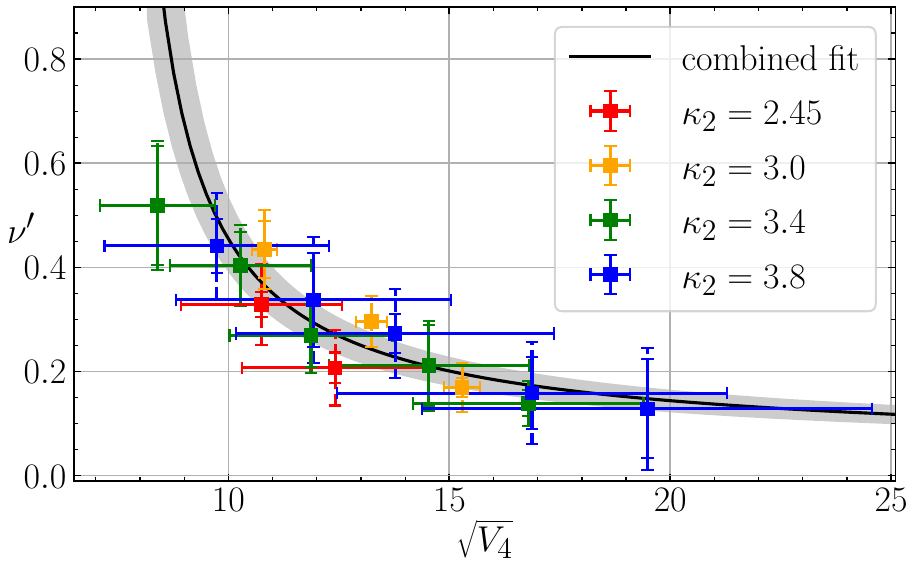}
\caption{ 
A combined fit of $\nu'$ versus $\sqrt{V_4}$ to data at all four lattice spacings for all ensembles that pass our cut, as discussed in the body of this subsection.  The vertical errors in $\nu'$ are the statistical and fitting errors added in quadrature (solid riser line), and the sum of the statistical/fitting errors and the $a_{\rm lat}/\ell$ systematic error added in quadrature (dashed riser line).  The fit of this data to the form Eq.~\eqref{eq:nuPfit}, which assumes a logarithmic running, is shown with its $1\sigma$ statistical error band. \newline
} 
\label{fig:nu'fit}
\end{figure}

Figure~\ref{fig:nu'fit} shows $\nu'$ plotted as a function of $\sqrt{V_4}$ for the subset of the $\nu'$ data that we consider for our main analysis.  The errors on $\nu'$ are those quoted in Table~\ref{tab:lambdaH2}, where the solid riser line represents the statistical and fitting error, and the dashed riser line includes the $a_{\rm lat}/\ell$ error added in quadrature.  The horizontal errors stem from the dependence of $\sqrt{V_4}$ on $a_{\rm rel}$ and $\eta$.  The central values and uncertainties on these quantities are taken from Section~\ref{sec:desitter}, and we assume that $a_{\rm rel}$ and $\eta$ are uncorrelated when combined to produce $V_4$, since they come from an almost completely non-overlapping set of ensembles and different analyses.  The statistics and fitting errors on each ensemble are treated as uncorrelated across ensembles, but the $a_{\rm lat}/\ell$ errors are treated as $100\%$ correlated across ensembles at the same lattice spacing and uncorrelated for ensembles at different lattice spacings.  The same is true of the $\eta$ and $a_{\rm rel}$ errors.

In order to correctly account for correlated systematic errors in our fits to $\nu'$ as a function of $\sqrt{V_4}$, we find it convenient to generate a synthetic data set that contains the information extracted from the $\Lambda$ versus $H^2$ fits.  The statistical errors on the individual data points for $\nu'$ are uncorrelated, since each point comes from a different ensemble.  The fit error is also assumed to be uncorrelated across ensembles, so just like the statistical error, it only contributes to the diagonal elements of the covariance matrix.  

Our synthetic data sample is obtained from 10,000 random Gaussian draws for every ensemble, with a mean given by the central value of $\nu'$ and the error given by the statistical and fit systematic error added in quadrature.  We account for the error coming from $a_{\rm lat}/\ell$ to the arrays as follows:  We assume that when $a_{\rm lat} / \ell$ for a given lattice spacing deviates from its central value by some percentage of one $\sigma_{a_{\rm lat}/\ell}$, all $\nu'$ values at the same lattice spacing also shift by the same percentage of its $a_{\rm lat}/\ell$ systematic error, $\sigma_{\nu',a_{\rm lat}/\ell}$. Therefore, for each lattice spacing, we generate an array of $10,000$ samples from a Gaussian random number generator, $A_{\rm ran} \thicksim \mathcal{N}(0,1)$. This allows us to construct the array $\left(\frac{a_{\rm lat}}{\ell}\right)_{\rm avg} + A_{\rm ran} \sigma_{\frac{a_{\rm lat}}{\ell}}$, which gives a distribution for $\frac{a_{\rm lat}}{\ell}$ with the correct central value and standard deviation at each lattice spacing.  To obtain the final correlated array for $\nu'$ we use the fact that for two independent normal distributions, $\mathcal{N}_{1}(\mu_{1}, \sigma_{1})$, $\mathcal{N}_{2}(\mu_{2}, \sigma_{2})$, their sum $\mathcal{N}_{1} + \mathcal{N}_{2} = \mathcal{N}(\mu_{1} + \mu_{2}, \sqrt{\sigma_{1}^{2} + \sigma_{2}^{2}})$.  We then get $\nu'$ for each ensemble by subtracting $A_{\rm ran}\sigma_{\nu',\frac{a_{\rm lat}}{\ell}}$ from the original $\nu'$ sample, thereby including the $a_{\rm lat}/\ell$ systematic error in quadrature with the statistical and fit systematic error in $\nu'$, and maintaining the correlation with $a_{\rm lat}/\ell$.  The minus sign is due to the fact that the larger $a_{\rm lat}/\ell$ is, the smaller $\nu'$ is. Because the random array for $a_{\rm lat}/\ell$ is shared by all ensembles at the same lattice spacing, the $a_{\rm lat}/\ell$ error is fully correlated across different volumes at a single lattice spacing.  The fully correlated $a_{\rm lat}/\ell$ error covariance matrix is added to the diagonal matrix containing the statistical and fitting errors of $\nu'$ in quadrature.  This gives the full covariance matrix for $\nu'$, which is reused when performing the fits over a jackknife array of $10,000$ random samples.

Figure~\ref{fig:nu'fit} shows the fit of $\nu'$ data to Eq.~(\ref{eq:nuPfit}), including a 1 $\sigma$ statistical error band on the fit curve.  We obtain $A'=0.146(28)$ and $B'=0.138(10)$, with a $\chi^2/{\rm d.o.f.}$ of 1.100 and $p$-value of 0.355.  There is no need to include discretization effects in order to describe the data using Eq.~(\ref{fig:nu'fit}), at least for this subset of the data.  Because the agreement of the running of $\nu'$ across multiple lattice spacings is very good, it is unlikely that this effect is simply a lattice artifact.  Also, the statistical error of the normalization parameter $A'$ is $\sim 5\sigma$ from zero, so the non-trivial running of $\Lambda$ we observe in our EDT simulations is very unlikely to be due to a statistical fluctuation.

\begin{figure*}
\begin{minipage}{0.475\linewidth}
    \centering
    \includegraphics[width=\linewidth]{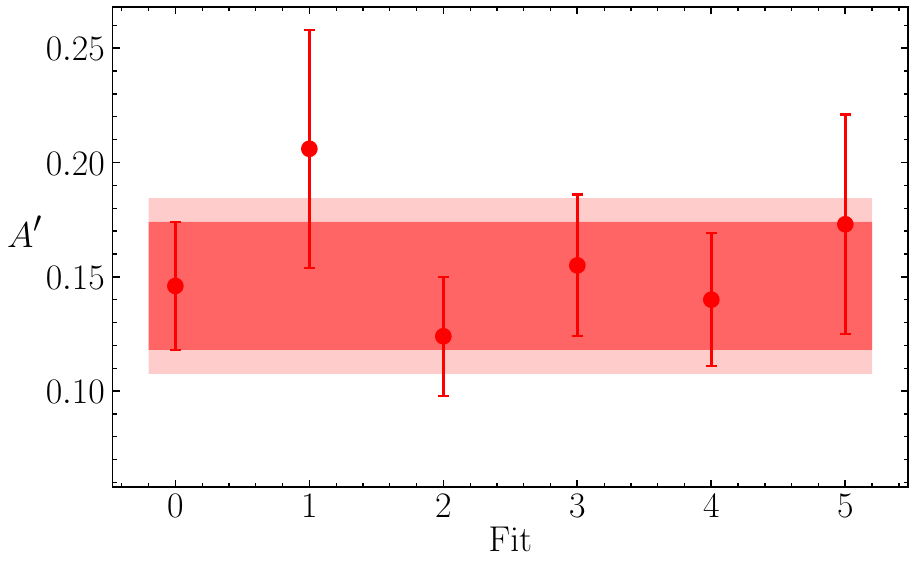}
\end{minipage}
\begin{minipage}{0.475\linewidth}
    \centering
    \includegraphics[width=\linewidth]{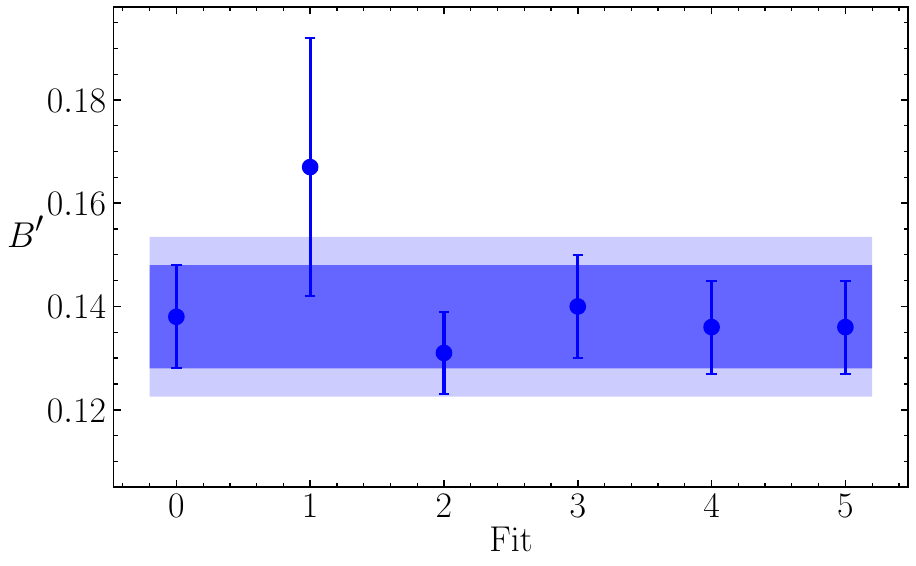}
\end{minipage}
\caption{ 
A stability plot for visualizing the fits of Table~\ref{tab:nupresults}. The fits are labeled (left to right) in the same order as Table \ref{tab:nupresults} (up to down).  The central value and $1\sigma$ error band is shown in the plot as the darker shaded region. The lighter bands represent the total error, which includes a systematic error accounting for the size of the variations of the alternative fits. \newline
} 
\label{fig:logfitstability}
\end{figure*}

To test the stability of this fit, we perform a number of fit variations.  These variations include dropping the largest volume at each lattice spacing, the smallest two volumes across all ensembles, and the finest or coarsest lattice spacing in the analysis.  We also consider a fit where we add $\ell_{\rm rel}^2$ dependence to the normalization of the coupling,
\bea
\label{eq:logfitell} \nu' = \frac{A'+k'\ell_{\rm rel}^2}{\log (B' \sqrt{V_4})}.
\eea
These fit results are shown in Table \ref{tab:nupresults}, and the variation of the parameters $A'$ and $B'$ across different fits is shown in Fig. \ref{fig:logfitstability}.  This illustrates the stability of our fits to dropping subsets of the data.  The analysis is not very sensitive to dropping either the finest or the coarsest of our lattice spacings, as the lattice spacing dependence is quite small.  The fits are more sensitive to dropping the smallest or largest volumes, since these are needed to constrain the shape of the fit.  The coefficient $k'$ of the $\ell_{\rm rel}^2$ dependence in the fit form of Eq.~(\ref{eq:logfitell}) that we include to estimate discretization effects is found to be $-0.032(42)$, so consistent with zero.  
We take the standard error of the distribution of these fit variations in Table \ref{tab:nupresults} as an estimate of the systematic error in our result for $\nu'$ due to fit-range variations and discretization effects. 
We thus quote the values $A'=0.146(28)(27)$ and $B'=0.138(10)(2)$ as the inputs for our discussion on the implications for cosmology in Secs.~\ref{sec:Implications_RVM} and \ref{sec:cosmology}.  The first error quoted is statistical, and the second is an estimate of the fitting and discretization errors.

\begin{table*}
\caption{Results for various fits to the $\nu'$ versus $\sqrt{V_4}$ data set. The first row is the result of the central fit to the full data set that passes our cut on volume and lattice spacing. This fit is shown with $1\sigma$ statistical error band in Fig. \ref{fig:nu'fit}. The other rows show the results for different fit variations.  The second row shows the fit where the largest volume at each lattice spacing is dropped from the fit. The third row shows a fit that drops the smallest volume at $\kappa_2=3.4$ and at $\kappa_2=3.8$. The fourth and fifth rows show the fits that drop the coarsest ($\kappa_2=2.45$) and finest ($\kappa_2=3.8$) lattice spacings, respectively. The sixth row shows the fit to Eq. \ref{eq:logfitell}, using all of the same data as the central fit.}
\label{tab:nupresults}
\centering
\begin{tabular}{m{7em} m{5em} m{5em} m{5em} m{5em} m{5em}}
\hline \hline
Fit scheme & $A'$ & $B'$ & $k'$ & $\chi^2/\mathrm{d.o.f.}$ & $p$-value\\ \hline 
Fig. \ref{fig:nu'fit}&0.146(28)&0.138(10)& NA & 1.100 & 0.355\\
w/o largest&0.206(52)&0.167(25)& NA & 0.178 & 0.996\\
w/o smallest 2& 0.124(26) & 0.131(8) & NA & 0.934  & 0.507\\
w/o $\kappa_2 = 2.45$& 0.155(31) & 0.140(10) & NA & 1.197  & 0.283\\
w/o $\kappa_2 = 3.8$& 0.140(29) & 0.136(9) & NA & 1.589 & 0.123 \\
$\ell_{\rm rel}^2$ term &0.173(48)&0.136(9)& -0.032(42) & 1.140 &0.323\\\hline
\hline
\end{tabular}
\end{table*}

\begin{figure}
\centering
\includegraphics[width=\linewidth]{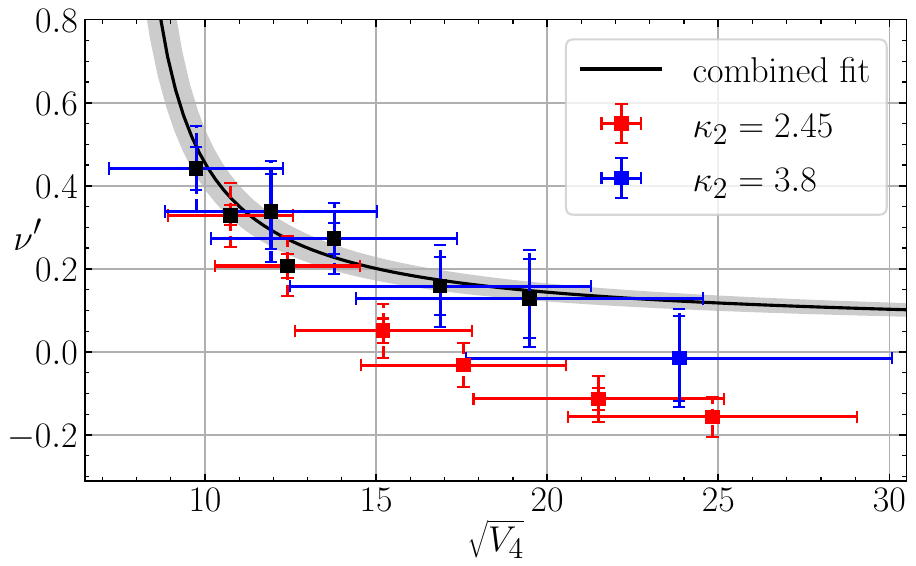}
\caption{
The full data set of all volumes at our coarsest lattice spacing, $\kappa_2=2.45$, and at our finest lattice spacing, $\kappa_2=3.8$. The black curve is the combined fit from Fig.~\ref{fig:nu'fit}. The data points highlighted in black are the ones that are included in the combined fit. \newline
} 
\label{fig:midsuperfine}
\end{figure}

\subsubsection{Tests of the model}

As pointed out above, we exclude some of our data from the fits to determine $\nu'$, due to the fact that they are likely contaminated by lattice artifacts.  To illustrate this, we show the full set of data for our coarsest and finest lattice spacings in Fig.~\ref{fig:midsuperfine}, along with the best fit curve from the analysis described above and its $1\sigma$ error band.  As can be seen in Fig.~\ref{fig:nu'fit} and Table~\ref{tab:nupresults}, the fit function Eq.~(\ref{eq:nuPfit}) is an excellent description of the subset of the data that makes our cut.  All of the data points in this subset are in good agreement not only with the model fit function but with each other across four lattice spacings.  However, as we go to larger volumes, particularly at coarser lattice spacings, this agreement is not as good, as can be seen in Fig.~\ref{fig:midsuperfine}.  Despite the large errors, a trend can be seen. Both lattice spacings begin to deviate from the model fit function and from each other at larger volumes, but the deviation from the model fit function is bigger and happens at smaller physical volumes for the coarsest lattice spacing, suggesting non-universal behavior.  Even the finest lattice spacing begins to deviate from the model fit at a sufficiently large volume, though the errors are still quite large on the largest volume point at the finest lattice spacing.  Modeling the discretization effects in the large volume regime where there are deviations from the model fit function (and between the lattice data at different lattice spacings) is difficult, so we restrict our main analysis to the region where the overlap between different lattice spacings is better.  

This hint of non-universal behavior at long distance scales is reminiscent of the deviation of our lattice data for the shelling function from the classical de Sitter solution, as seen in Fig.~\ref{fig:deSitteraAymTail}.  There as well, the long-distance physics displays non-universal behavior, which approaches theoretical expectations as the continuum limit is approached, with the notable difference that the effect in the de Sitter comparison is much more significant, given the smaller errors there.  This type of long-distance lattice artifact appears when the lattice regulator breaks an underlying symmetry of the theory, requires a fine-tuning of bare lattice parameters, and only vanishes in the continuum limit.  Regardless of its origin, it is plausible that other long-distance quantities, like $\nu'$, may similarly be contaminated by lattice artifacts.  Not only is there an indication that the behavior of $\nu'$ at long distances and coarse lattices is non-universal, but we argue that it is also not physical, as it turns out that a negative value of $\nu'$ for the model picked out by our lattice calculations would lead to a vacuum equation of state $w < -1$, a result in violation of the null energy condition.  Such a result would be severely constrained from theoretical expectations and from observation \cite{Carroll:2003st, Dubovsky:2005xd}.  However, the trend in our data suggests that this behavior corrects itself as the continuum limit is approached.  Further reduction in the errors of $\nu'$ and in the relative lattice spacings would help to clarify this picture.

\begin{figure}
\centering
\includegraphics[width=\linewidth]{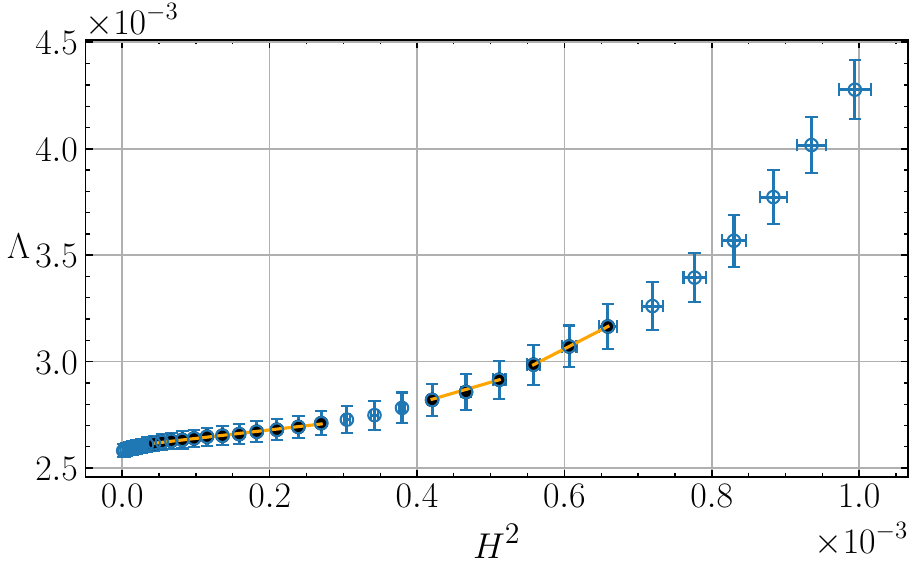}
\caption{$\Lambda$ versus $H^2$ for one of our larger volumes at our second finest lattice spacing, with $\kappa_2 = 3.4$, $N_4 = 32$k.  The range in $H^2$ extends further than in Figure~\ref{fig:lH2fit} in order to show the nonlinear behavior at larger $H^2$ values.  There are three fit windows, highlighted in black, where a linear fit is performed.  The left-most one is the central value fit used to extract $\nu'$ for this ensemble.  The two to the right are alternative fits used to study the scale dependence of $\nu'$ as $H^2$ is increased.
} 
\label{fig:32krf-nonlinear}
\end{figure}
\begin{figure}
\centering
\includegraphics[width=\linewidth]{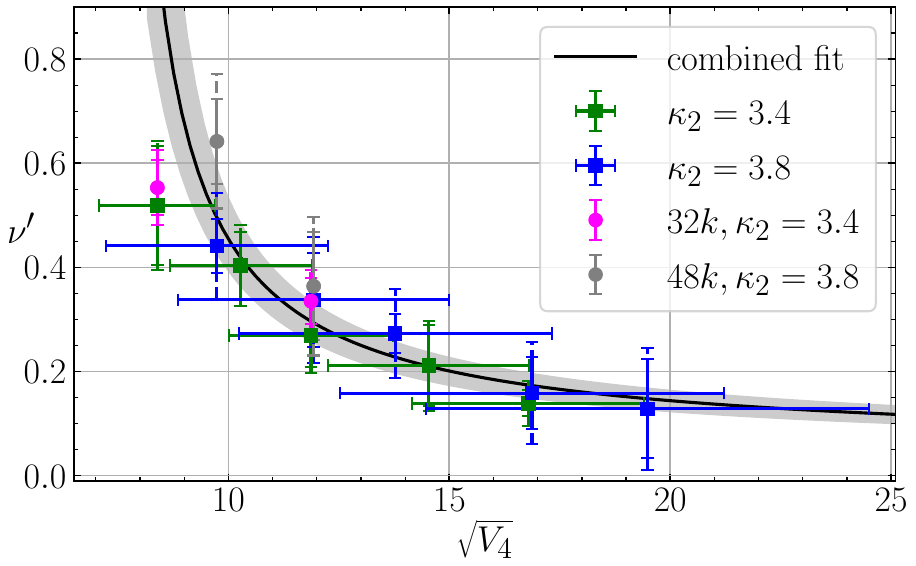}
\caption{ Plot of $\nu'$ versus $\sqrt{V_4}$ showing only data at the two finest lattice spacings ($\kappa_2=3.4$ and $3.8$), as well as the central fit to all four lattice spacings with $1\sigma$ error band.  Also shown are values of $\nu'$ measured at larger $H^2$ values beyond the linear regime on some of our larger volume ensembles.  These alternate values of $\nu'$ are measured at larger $H^2$ than our central fit values. They are plotted at smaller values of $\sqrt{V_4}$, which are taken to be in the same proportion as $H_{\rm cutoff}^2/H^2$, where $H_{\rm cutoff}^2$ marks the upper bound on the linear regime for $\nu'$.  This gives a rough estimate of the scale at which $\nu'$ in the non-linear regime should be evaluated in comparison to other determinations of $\nu'$ in the linear regime on smaller volumes.  The $\nu'$ points in pink are measured on the $\kappa_2=3.4$, $N_4=32$k ensemble in the non-linear regime.  The $\nu'$ points in gray are measured on the $\kappa_2=3.8$, $N_4=48$k ensemble in the non-linear regime.     
} 
\label{fig:nup-Hwindow}
\end{figure}
We further test the running of $\nu'$ with scale by studying its extraction from fits to different windows in $H^2$ from the $\Lambda(H)$ versus $H^2$ data.  We restrict the study to our two finest lattice spacings, since these are expected to have the smallest systematic errors from lattice artifacts.  We also restrict this study to the largest or second largest volumes at these lattice spacings.  This choice follows straightforwardly from the fact that we can probe scales shorter than the maximum distance scale on our largest volumes, but we cannot probe distances longer than this on our smaller volumes.  We assume that in the linear regime at large distances that the value of $\nu'$ is effectively determined at a scale of order the minimum spatial curvature ($\sim 1/a^2$) on a given volume ensemble.  If the renormalization scale of the logarithmic running in the model is to be identified with the Hubble scale $H$, just as the quadratic running of $\Lambda(H)$ is, then we must see a variation of $\nu'$ with $H^2$ on individual ensembles that is similar to the variation of $\nu'$ across ensembles at different physical volumes.

To illustrate this, we consider the largest volume at our second finest lattice spacing, with $N_4=32$k and $\kappa_2=3.4$.
As can be seen in Fig.~\ref{fig:32krf-nonlinear}, the linear regime extends out to values of $H^2$ around $3\times 10^{-4}$, which we denote by $H_{\rm cutoff}^2$.  Beyond that, the slope increases with increasing $H^2$.  If we consider larger values of $H^2$, for example $H^2 = \sqrt{2} H_{\rm cutoff}^2$ and $H^2 = 2 H_{\rm cutoff}^2$, these correspond to scales that are 1/2 and 1/4, respectively, the four-volume of the lattice.  Thus, we might compare the value of $\nu'$ extracted from linear fits at these larger $H^2$ values on the 32k, $\kappa_2=3.4$ ensemble to the values of $\nu'$ determined from the standard linear fits probing the largest distance scales on smaller volumes, in particular the 16k and 8k volumes at the same lattice spacing.  To extract the values of $\nu'$ from $\Lambda(H)$ at higher $H^2$ values where the trend is no longer linear, we have to restrict the fit range to just a few points, as shown in Fig.~\ref{fig:32krf-nonlinear}.  To estimate a fit systematic error for these additional $\nu'$ points, we consider variations of the fit window where the starting point of the window is fixed and the ending point is allowed to vary by $\pm 1$ or $\pm 2$. We keep only the fits with $p \geq 0.01$ when using the range of these alternate fits to estimate a systematic error.  The values of $\nu'$ extracted from the linear fits at these larger $H^2$ values on the 32k, $\kappa_2=3.4$ ensemble are shown in pink in Fig.~\ref{fig:nup-Hwindow} at the horizontal points corresponding to four-volumes that are 1/2 and 1/4 that of the 32k ensemble.  The data for $\nu'$ from lattice spacings at $\kappa_2=3.4$ and $3.8$ are also shown in Fig.~\ref{fig:nup-Hwindow}, along with the combined fit across volumes and lattice spacings discussed above, for comparison.  The running of $\nu'$ with $H^2$ on an individual ensemble is thus in good overall consistency with the running of the coupling with scale set by the physical volume.  We find a similar trend when comparing the determination of $\nu'$ on the 48k volume of our finest lattice spacing ($\kappa_2=3.8$) for values of $H^2= \sqrt{2} H_{\rm cutoff}^2$ and $H^2= \sqrt{3} H_{\rm cutoff}^2$, which should be compared to the smaller volumes at 24k and 16k at the same lattice spacing, respectively.  These two points appear in grey in Fig.~\ref{fig:nup-Hwindow}, and they are also in good agreement with the overall trend from the combined fit across volumes and lattice spacings.

\subsection{\label{sec:nutilde} Calculation of $\tilde{\nu}$}

\subsubsection{The method}

As discussed in Sec.~\ref{sec:modelParams}, we can constrain $\tilde{\nu}$ from $I_{H^2}= \Lambda_0 + 3\tilde{\nu}S$.  It is not possible to separately determine $\Lambda_0$ at a fixed volume, so the determination of $\tilde{\nu}$ must exploit the difference in finite-volume scaling between the two terms making up $I_{H^2}$.  We associate $\Lambda_0$ with the classical cosmological constant, which is strictly constant as a function of Euclidean time, and in our simulations it is fixed by the four-volume of our geometries.  It scales as in Eq.~(\ref{eq:Lambdaell2eta}), so that its volume dependence is $\propto \frac{1}{\sqrt{V_4}}$, with $V_4$ given by Eq.~(\ref{eq:V4}).  Since $\tilde{\nu}$ is a dimensionless coupling, its running can be parameterized analogously to Eq.~(\ref{eq:nuPfit}) for $\nu'$, leading to
\bea \tilde{\nu} = \frac{\tilde{A}}{\log(\tilde{B} \sqrt{V_4})},
\eea
with free parameters $\tilde{A}$ and $\tilde{B}$.  Within the family of running vacuum models considered in Refs.~\cite{Gomez-Valent:2015pia, SolaPeracaula:2021gxi, Moreno-Pulido:2023ryo}, it is assumed that $\nu$ and $\tilde{\nu}$ are dimensionless couplings that run with the same functional form.  Given that $\nu'=\nu - \tilde{\nu}$, if this linear relation is to be maintained across scales, the couplings must share the same scale dependence.  We thus assume that $\tilde{B}=B'$ and use the value of $B'$ from our $\nu'$ fits as an input to our determination of $\tilde{\nu}$.  The normalization parameter, $\tilde{A}$, we leave free.  Thus, we expect the intercept $I_{H^2}$ to be described by a function of $\sqrt{V_4}$ of the form
\bea I_{H^2} = \frac{\tilde{C}}{\sqrt{V_4}} + \frac{3\tilde{A}S}{\log(B' \sqrt{V_4})}.
\label{eq:nutildefit1}
\eea
where $\tilde{A}$ and $\tilde{C}$ are free parameters.

Before presenting this finite-size analysis, we note a complicating factor due to our method for extracting $a_{\rm lat}/\ell$ and how we work around it.  The determination of $I_{H^2}$ requires as input a value for $a_{\rm lat}/\ell$.  In our Method 2 determination of $a_{\rm lat}/\ell$, its value is computed using Eq.~(\ref{eq:aell2}) by adjusting the input $a_{\rm lat}/\ell$ value that enters $I_{H^2}$ to match the output value, so that $a_{\rm lat}/\ell$ is determined in a self-consistent way.  One of the implicit assumptions entering Eq.~(\ref{eq:aell2}) is that $I_{H^2}=\Lambda_0$, with $\tilde{\nu}$ set to zero.  

It is possible to use the determination of $I_{H^2}$ that comes from the combined (Methods 1 and 2) result for $a_{\rm lat}/\ell$ in the finite-size scaling analysis of Eq.~(\ref{eq:nutildefit1}), and we do find that $\tilde{\nu}$ is compatible with zero in this analysis.  This provides a consistency check that the assumption that $\tilde{\nu}=0$ in our Method 2 $a_{\rm lat}/\ell$ determination is valid.  Still, we would like to test the quality of the bound on $\tilde{\nu}$ when we do not make the assumption that it is zero at an intermediate step of the analysis.  We do this by using only the Method 1 determination of $a_{\rm lat}/\ell$ to obtain $I_{H^2}$, since this makes no implicit assumptions about the value of $\tilde{\nu}$.  This does not significantly worsen the bounds on $\tilde{\nu}$, which remains  consistent with zero, and is constrained to be at most only a few percent of the magnitude of $\nu$.

\subsubsection{Determining $S$}

\begin{figure*}
\begin{minipage}{0.475\linewidth}
    \centering
    \includegraphics[width=\linewidth]{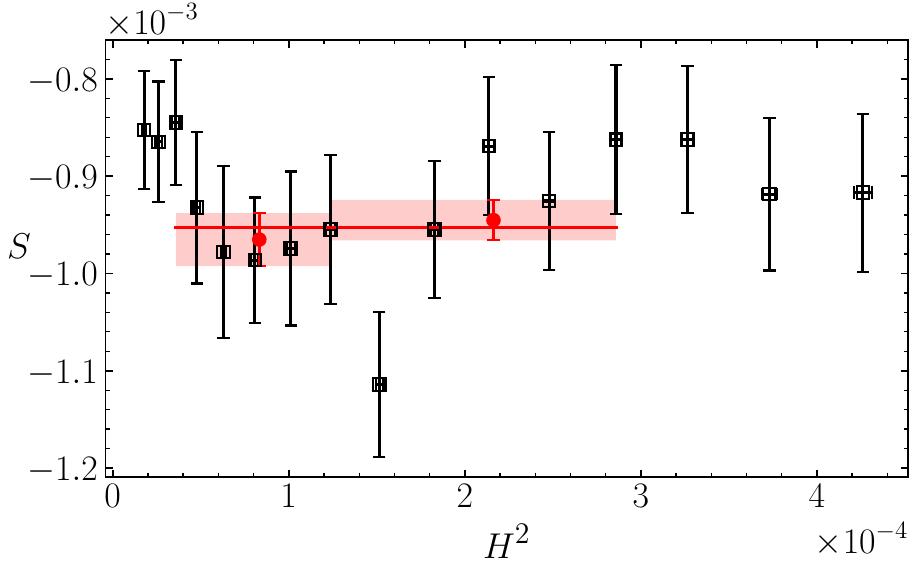}
\end{minipage}
\begin{minipage}{0.475\linewidth}
    \centering
    \includegraphics[width=\linewidth]{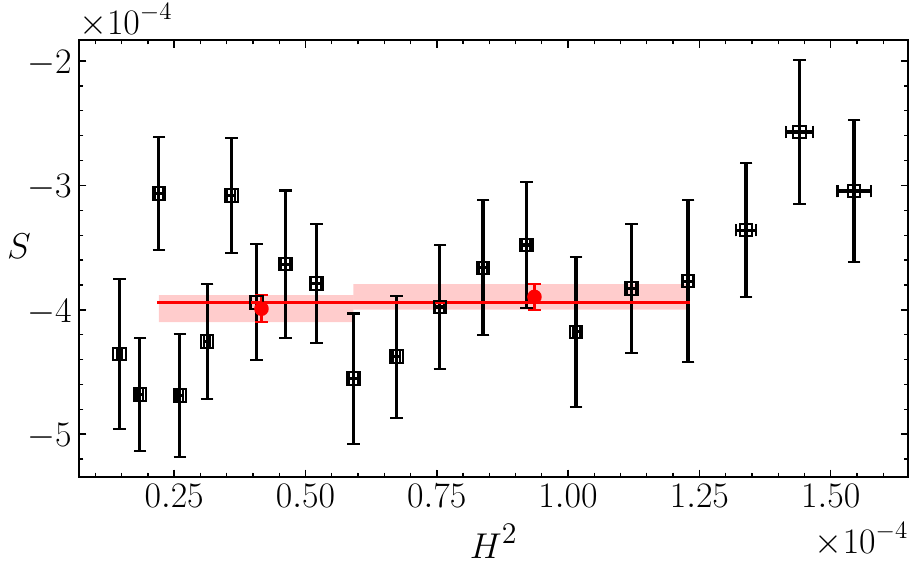}
\end{minipage}
\caption{
Left panel: $S$ versus $H^2$ for $\kappa_2 = 3.4$, $N_4 = 16$k. 
Right panel: $S$ versus $H^2$ for $\kappa_2 = 3.8$, $N_4 = 16$k. 
Red circles represent the weighted averages of the data points in the first half and second half of the fit range.  The fit ranges over which these averages are performed are denoted by the light red bands. The final result for $S_0$, given by the red horizontal line, comes from a constant fit to the smoothed points (red circles).
} 
\label{fig:sH2}
\end{figure*}

In order to carry out the finite-size scaling analysis of $I_{H^2}$ suggested by Eq.~(\ref{eq:nutildefit1}), it remains to specify how we determine $S\equiv \ddot{A}/A$, and how we parameterize its volume dependence.  One way to determine $S$ is to extract it directly from the second derivative of the lattice scale factor.  Figure~\ref{fig:sH2} shows $S$ plotted as a function of $H^2$ in the same fit region that is used for the $\nu'$ analysis, which is the region of the shelling function that is best described by the classical de Sitter solution.  Figure~\ref{fig:sH2} shows $S$ versus $H^2$ at our two finest lattice spacings, the $\kappa_2=3.4$ and 3.8 ensembles, for $N_4=16$k.  

Within the running vacuum model with $\tilde{\nu}$ small, the corrections to a constant $S$ as a function of $H^2$ are also small.  There are complicating factors in testing this expectation, which include the substantial statistical noise involved when evaluating a second derivative numerically, and the possible presence of lattice artifacts at coarse lattice spacings.  On our two finest ensembles, the value of $S$ is compatible with a constant as a function of $H^2$ within the standard fit window chosen for other observables.  On our two coarsest lattice spacings, the value of $S$ versus $H^2$ has some nontrivial structure.  In order to incorporate the coarser data, we smooth over these effects by averaging the first half and second half of the data points in the $H^2$ range.  These smoothed data points are shown as red circles in Fig.~\ref{fig:sH2}.  We fit these two points to a constant on each ensemble.  The results for $S_0$ from these constant fits are shown in Table~\ref{tab:s0values}, along with the $p$-values of the fits, showing that the smoothed data is indeed consistent with a constant over this $H^2$ range.  On our two finest lattice spacings it is not necessary to do the smoothing, and we get consistent results between the smoothed fits and the fits to a constant across the $H^2$ window. The numbers quoted in \ref{tab:s0values} are all from the fits to the smoothed data.

\begin{figure}
\centering
\includegraphics[width=\linewidth]{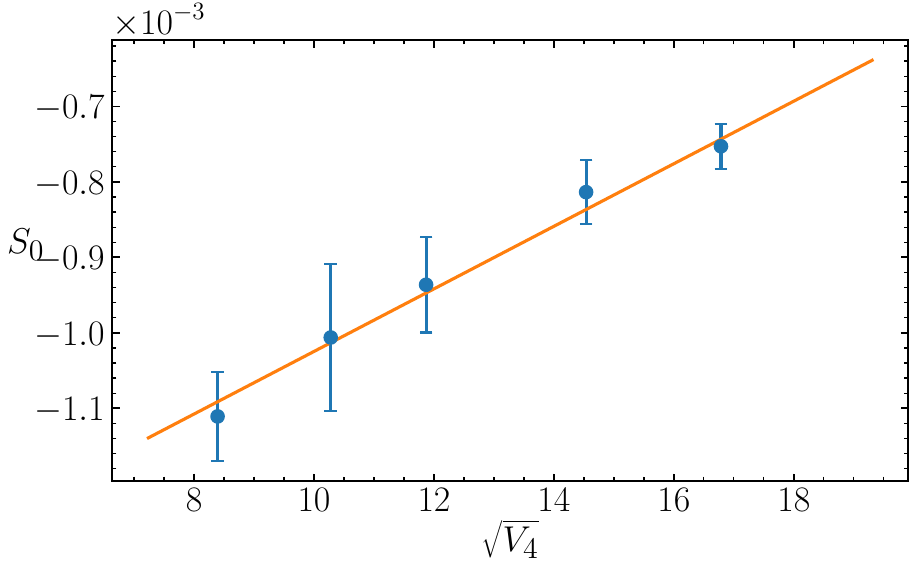}
\caption{ 
$S_0$ versus $\sqrt{V_4}$ at lattice spacing $\kappa_2 = 3.4$, with linear fit $S_0 = a_0 + b_0 \sqrt{V_4} $. 
} 
\label{fig:linears}
\end{figure}

Once $S_0$ is determined from these fits, we must parameterize its volume dependence.  We chose the fit function
\bea  S_0 = a_0+b_0\sqrt{V_4}, 
\eea
with $a_0$ and $b_0$ free parameters.  This simple interpolating function does an excellent job of describing the $S_0$ data across volumes in the region where we have data.  
Table~\ref{tab:s0fits} shows the results of our $S_0$ versus volume fits at all four of our lattice spacings, and Fig.~\ref{fig:linears} illustrates one of these fits.  We only include data on ensembles that were used in the $\nu'$ analysis.  This means that we only have two points at the coarsest lattice spacing, so we are unable to estimate a $\chi^2/$d.o.f. for this two-parameter fit.

\begin{table}
\caption{Results of the linear fits to $S_0$ versus $\sqrt{V_4}$, such as the one shown in Fig. \ref{fig:linears}.  At $\kappa_2=2.45$ there are only two data points in the acceptable region, so there is not enough data to estimate a $\chi^2/{\rm d.o.f.}$  }
\label{tab:s0fits}
\centering
\begin{tabular}{m{2em} m{5em} m{5em} m{4em} m{4em}}
\hline \hline$\kappa_2$ & $a_0\times 10^3$  & $b_0\times10^3$ & $\chi^2/\mathrm{d.o.f.}$ & $p$-value\\ \hline 
$2.45$&-7.13(169) & 0.144(148) & NA & NA \\
$3.0$&-2.94(24) & 0.055(17) & 0.356 & 0.583 \\
$3.4$&-1.44(10) & 0.042(7) & 0.182 & 0.909 \\
$3.8$&-0.50(3) & 0.013(2) & 0.217 & 0.884 \\
\hline\hline
\end{tabular}
\end{table}

An alternative determination of $S_0$ is possible, if we allow ourselves to use theory input as a guide.  We estimate $S_0$ using the FLRW equations, particularly Eq.~(\ref{eq:Einsteinii}), which has the dynamics of the running vacuum model inserted; this requires as input our values for $I_{H^2}$ and $\nu'$.  In the spirit of a consistency check, we assume that $\tilde{\nu}$ is zero when constructing $S_0$.  Eq.~(\ref{eq:Einsteinii}) then reduces to
\begin{equation}
    S \equiv \frac{\ddot{A}}{A} = -\frac{\Lambda_0}{3} -\nu S
\end{equation} 
and we can solve for $S$, replacing $\Lambda_0$ with the intercept $I_{H^2}$ from the $\Lambda$ versus $H^2$ fits. We find
\begin{equation}
    S_0 = -\frac{1}{3} I_{H^2}\frac{1}{1+\nu}.
\label{eq:S_model}
\end{equation} 
Using extracted values for $I_{H^{2}}$ and $\nu'$ we can then obtain $S_{0}$.

\subsubsection{Fits to $I_{H^2}$ versus four-volume and constraints on $\tilde{\nu}$}

We now return to the finite-volume scaling of $I_{H^2}$ using the minimal set of assumptions about $\tilde{\nu}$.  We use only the Method 1 value of $a_{\rm lat}/\ell$ as input, and the data driven approach to interpolate $S_0$ as a function of lattice volume.  Our fit function becomes
\bea I_{H^2} = \frac{\tilde{C}}{\sqrt{V_4}} + \frac{3\tilde{A}(a_0+b_0\sqrt{V_4})}{\log(B' \sqrt{V_4})},
\label{eq:nutildefit2}
\eea
where $B'$ is determined from our $\nu'$ fits, as detailed in the previous subsection, and $a_0$ and $b_0$ are determined from our fits to the volume dependence of $S_0$.

We take two approaches to the finite-volume scaling.  In the first, we fit each lattice spacing separately to Eq.~(\ref{eq:nutildefit2}).  In this case, we do not include the horizontal error, since this comes from converting the number of four-simplices to a physical volume, a conversion factor that is common across data points at the same lattice spacing.  We include only the data points on ensembles at volumes that were found to be consistent with the model form of the logarithmic running in the $\nu'$ analysis.  The fit results are shown in Table~\ref{tab:nutresults}.  We see that the fits have acceptable confidence levels, although the coarsest lattice spacing ($\kappa_2=2.45$) has only two data points, so a two parameter fit does not allow a determination of the $\chi^2/$d.o.f.  The values of $\tilde{A}$ (which normalizes $\tilde{\nu}$) for the finest two lattice spacings are compatible with zero well within the statistical errors, while the coarsest two lattice spacings are compatible with zero at the 1-1.5$\sigma$ level. This indicates that $\tilde{\nu}$ is consistent with zero on our lattice geometries.

\begin{table}
\caption{Fit results for $I_{H^2}$ versus $\sqrt{V_4}$ using the fit function, Eq.~(\ref{eq:nutildefit1}), for each lattice spacing separately.  At $\kappa_2=2.45$ there are only two data points in the acceptable region, so there is not enough data to estimate a $\chi^2/{\rm d.o.f.}$}
\label{tab:nutresults}
\centering
\begin{tabular}{m{3em} m{5em} m{6em} m{5em} m{5em}}
\hline \hline$\kappa_2$ & $\tilde{C}$  & $\tilde{A}$ & $\chi^2/\mathrm{d.o.f.}$ & $p$-value\\ \hline 
$2.45$&0.2325(389) & -0.041(38) & NA & NA \\
$3.0$&0.1107(72) & -0.018(13) & 2.502 & 0.155 \\
$3.4$&0.0433(18) & 0.003(4) & 0.488 & 0.691 \\
$3.8$&0.0171(26) & 0.005(20) & 0.203 & 0.894 \\
\hline\hline
\end{tabular}
\end{table}

In the second approach to the finite-volume scaling, we include the horizontal errors associated with converting the lattice volumes to common physical units, and we do a combined fit across all lattice spacings, similar to our approach to fitting $\nu'$.  For an ensemble with $N$ configurations, we have three one dimensional arrays of $N$ entries for $\nu'$, $I_{H^2}$, and $S_0$. The $i$th element of all three arrays corresponds to the fit results from the same configurations, so correlations among them are preserved. A $3$-by-$3$ covariance matrix $C$ that encodes the correlation among the three variables can be calculated from these $3$ arrays.
Since we add a fitting systematic error to the statistical error, the square of the fitting error is added to the diagonal elements of this $3$-by-$3$ covariance matrix $C$. We do not add the $a_{\rm lat}/\ell$ systematic error directly to the covariance matrix $C$ because the $a_{\rm lat}/\ell$ systematic errors are correlated across volumes at a given lattice spacing, and must be accounted for more carefully.  

Given the covariance matrix, we can generate correlated arrays of the three variables $\nu'$, $I_{H^2}$, and $S_0$ that reproduce the correct correlations between them. To do this, we first generate an independent Gaussian random vector $X = (x_1,x_2,x_3)^T$, where $x_i \thicksim \mathcal{N}(0,1)$, and $i = 1,2,3$. Then we multiply the lower triangular matrix $L$ coming from the Cholesky decomposition of the covariance matrix $C$ ($C = LL^T$) to $X$ to get $Y = LX$. The random vector $Y$ has mean value zero for all three variables, and its fluctuations reproduce the covariance matrix $C$. Adding the central value of the three variables to the vector $Y$, we get the vector of variables $Z=(\nu',I_{H^2},S_0)^T$ that reproduces the correct mean value and the original covariance matrix $C$. We account for the correlated errors of $a_{\rm lat}/\ell$ in a similar way to what is done for the $\nu'$ analysis in the previous subsection.  We get the final array $V = Z - A(\sigma_{\nu',\frac{a_{\rm lat}}{\ell}},\sigma_{I_{H^2},\frac{a_{\rm lat}}{\ell}},0)^T$, where $A$ is the array that gives the deviation of $\frac{a_{\rm lat}}{\ell}$ from its central value for the corresponding lattice spacing.

\begin{figure}
\centering
\includegraphics[width=\linewidth]{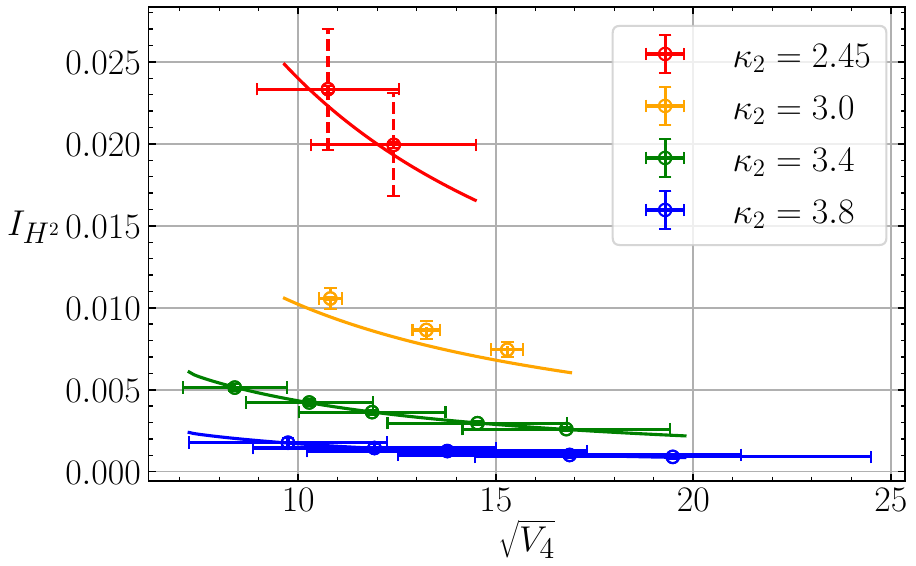}
\caption{Combined fit of $I_{H^2}$ versus $\sqrt{V_4}$ across all four lattice spacings to Eq.~(\ref{eq:nutildefit1}).  This fit does not assume $\tilde{\nu}$ is zero at an intermediate step of the analysis.  Fit results are shown in the left column of Table \ref{tab:method1nutcombined}.
} 
\label{fig:nutildefit}
\end{figure}

For the $I_{H^{2}}$ fit, the value of $\tilde{C}$ is different across lattice spacings, but the value of $\tilde{A}$ is fixed across lattice spacings, analogously to what is done for the parameter $A'$ in the $\nu'$ fits of the previous subsection.  The quality of the combined fit is good, as can be seen in Tab.~\ref{tab:method1nutcombined} under the column, ``Data Driven $S_0$.''  Again we find $\tilde{A}$ to be compatible with zero, this time with an error that is only around $3\%$ the size of our central value for the normalization of $\nu'$, thus constraining $\tilde{\nu}$ to be much less than $\nu$.  This combined fit is shown in Fig.~\ref{fig:nutildefit}.  This fit avoids an implicit or explicit setting of $\tilde{\nu}$ to zero anywhere in the analysis, and as such, it represents the most reliable constraint on $\tilde{\nu}$ coming from our data.

\begin{table}
\caption{Fit results for the combined fit of $I_{H^2}$ versus $\sqrt{V_4}$ to the fit function Eq.~(\ref{eq:nutildefit1}) for data across all four lattice spacings, using the Method 1 approach for obtaining $\frac{a_{\rm lat}}{\ell}$.  The ``Data Driven $S_0$'' approach, in combination with the Method 1 approach for obtaining $\frac{a_{\rm lat}}{\ell}$, avoids the assumption that $\tilde{\nu}$ (and thus $\tilde{A}$) is zero.  The ``Model Driven $S_0$'' approach does introduce the assumption that $\tilde{\nu}=0$. The zero result for $\tilde{A}$ using this set of assumptions should be viewed as a consistency check, since it ensures that $\tilde{\nu}$ remains consistent with zero.}
\label{tab:method1nutcombined}
\centering
\begin{tabular}{m{6em} m{8em} m{8em}}
\hline \hline Parameter & Data Driven $S_0$ & Model Driven $S_0$ \\ \hline 
$\tilde{C}_{2.45}$&0.240(55)&0.239(55)\\
$\tilde{C}_{3.0}$ &0.102(7)&0.102(7)\\
$\tilde{C}_{3.4}$ &0.043(6)&0.043(6)\\
$\tilde{C}_{3.8}$ &0.017(5)&0.017(5)\\
$\tilde{A}$&-0.00036(393)&-0.00068(266)\\
$\chi^2/\mathrm{d.o.f.}$ &0.785&0.780\\
$p$-value& 0.644&0.649\\
\hline\hline
\end{tabular}
\end{table}

We also consider a few additional fits that make the assumption $\tilde{\nu}=0$ somewhere in the analysis chain.  Even so, the finite-volume study of $I_{H^2}$ shows that $\tilde{A}$ (and thus $\tilde{\nu}$) remains compatible with zero in all of these fits.  This represents a useful consistency check on the $\tilde{\nu}$ constraint.  Table~\ref{tab:method1nutcombined} shows, under the column labeled, ``Model Driven $S_0$,'' a fit using $S_0$ determined from Eq.~(\ref{eq:S_model}).  We substitute the values of $I_{H^2}$ and $\nu'$ obtained in the previous subsection into Eq.~(\ref{eq:S_model}) to construct $S_0$ as a function of $V_4$.  This becomes possible once the assumption that $\tilde{\nu}=0$ is made.  We see from the results of the fit that $\tilde{A}$ is once again consistent with zero, with a small error. Table~\ref{tab:nutcombined} shows similar fits to that of Table~\ref{tab:method1nutcombined}, but where the value of $a_{\rm lat}/\ell$ was taken from the combination of Methods 1 and 2.  As stated above, the Method 2 determination of $a_{\rm lat}/\ell$ assumes that $\tilde{\nu}=0$.  As can be seen by comparing the results for $\tilde{A}$ in Tables~\ref{tab:method1nutcombined} and \ref{tab:nutcombined}, the constraint on $\tilde{A}$ from using the combined $a_{\rm lat}/\ell$ is barely different from that of using only the Method 1 determination of $a_{\rm lat}/\ell$.  The fit results strongly suggest that $\tilde{A}$ is consistent with zero.

\begin{table}
\caption{Fit results for the combined fit of $I_{H^2}$ versus $\sqrt{V_4}$ to the fit function Eq.~(\ref{eq:nutildefit1}) for data across all four lattice spacings, using the combination of Methods 1 and 2 to obtain $\frac{a_{\rm lat}}{\ell}$.  The constraints on $\tilde{\nu}$ are slightly tighter here, but the assumption that $\tilde{\nu}=0$ is made in the Method 2 determination of $\frac{a_{\rm lat}}{\ell}$.  However, these fits still provide a useful consistency check, since the value of $\tilde{A}$ (which normalizes $\tilde{\nu}$) determined from the fits to Eq.~(\ref{eq:nutildefit1}) is compatible with zero. }
\label{tab:nutcombined}
\centering
\begin{tabular}{m{6em} m{8em} m{8em}}
\hline \hline Parameter & Data Driven~$S_0$ & Model Driven~$S_0$ \\ \hline 
$\tilde{C}_{2.45}$&0.239(42)&0.238(42)\\
$\tilde{C}_{3.0}$ &0.102(5)&0.102(5)\\
$\tilde{C}_{3.4}$ &0.043(6)&0.043(6)\\
$\tilde{C}_{3.8}$ &0.017(4)&0.017(4)\\
$\tilde{A}$& -0.00037(358)&-0.00061(241)\\
$\chi^2/\mathrm{d.o.f.}$&0.804&0.800\\
$p$-value& 0.625&0.630\\
\hline\hline
\end{tabular}
\end{table}

\subsection{Alternative calculation of $\nu'$}
\label{sec:alt_nuPrime}

\begin{figure}
\centering
\includegraphics[width=\linewidth]{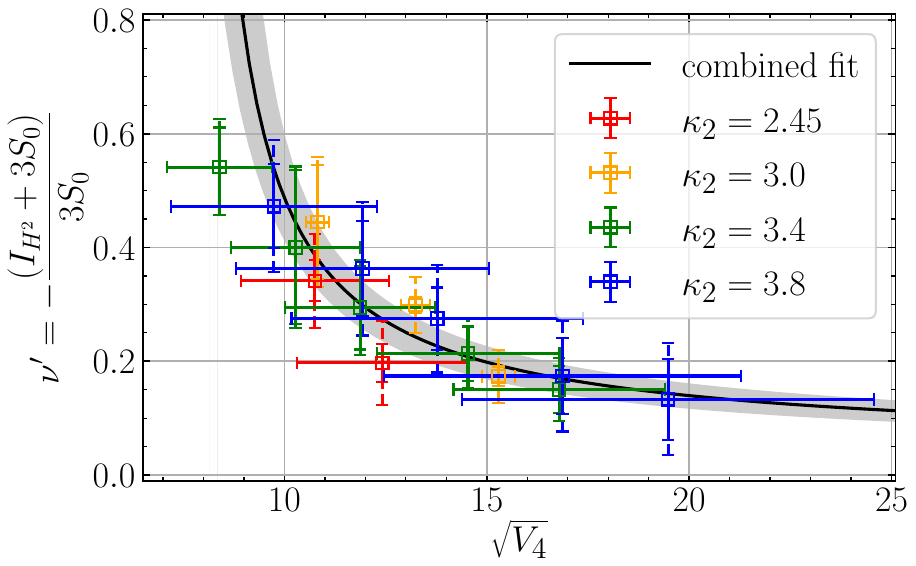}
\caption{ A combined fit of our alternate determination of $\nu'$ versus $\sqrt{V_4}$ to data at all four lattice spacings for all ensembles that pass our cut.  These are the same ensembles included in the original $\nu'$ analysis (see Fig.~\ref{fig:nu'fit}).  As in Fig.~\ref{fig:nu'fit}, the vertical errors in $\nu'$ are the statistical and fitting errors added in quadrature (solid riser line), and the sum of the statistical/fitting errors and the $a_{\rm lat}/\ell$ systematic error added in quadrature (dashed riser line).  The fit to this data assumes the same logarithmic running and the result is shown with its $1\sigma$ statistical error band.
The fit to Eq.~(\ref{eq:nuPfit}) yields $A' = 0.136(28), B' = 0.132(10)$, with $\chi^2/\mathrm{d.o.f.} = 0.937$, and $p$-value$ = 0.514$.
} 
\label{fig:nuppfit}
\end{figure}

As discussed in subsection~\ref{sec:modelParams}, the FLRW equations suggest an alternative determination of $\nu'$ from the one presented in subsection~\ref{sec:nuPrime_determination}.  The second FLRW equation [Eq.~(\ref{eq:FRW2})]
can be solved for $\nu'$ to obtain 
\bea  \nu' = \frac{I_{H^2} + 3 S }{-3 S}.
\label{eq:nu''3}
\eea
This equation is exact within the running vacuum model defined by Eqs.~(\ref{eq:rvmrho}) and (\ref{eq:pLam}) if $\tilde{\nu}=0$.  It then becomes an excellent approximation to assume that $S=S_0$, as long as we can ignore the logarithmic running of $\nu'$ with $H^2$ in the chosen fit window, which is seen to be the case in our extraction of $\nu'$ from the slope of $\Lambda(H)$ versus $H^2$.  Using $S=S_0$ in Eq.~(\ref{eq:nu''3}), we construct this alternative determination of $\nu'$ and show the results in Fig.~\ref{fig:nuppfit}.

As a comparison of Figs.~\ref{fig:nu'fit} and \ref{fig:nuppfit} shows, the agreement between these two determinations of $\nu'$ as a function of $\sqrt{V_4}$ is excellent, with every point in agreement within errors.  Although the individual points are highly correlated, having been determined on the same ensembles, they do have different systematic errors.  The solid riser line in Fig.~\ref{fig:nuppfit} shows the statistical plus fitting error in the second method for computing $\nu'$, while the dashed riser line shows the statistical plus fitting error combined in quadrature with the $a_{\rm lat}/\ell$ error.  The $a_{\rm lat}/\ell$ error enters in the determination of the intercept $I_{H^2}$ used to construct $\nu'$ from Eq.~(\ref{eq:nu''3}).  

A combined fit to the data of Fig.~\ref{fig:nuppfit} to the form of Eq.~(\ref{eq:nuPfit}) leads to a good fit, with a $\chi^2/{\rm d.o.f.}$ of $0.937$ corresponding to a $p$-value of $0.514$.  The values of the fit parameters, $A'=0.136(28)$ and $B'= 0.132(10)$, are in excellent agreement with those from the first method of determining $\nu'$, as can be seen by comparing to the results of Table~\ref{tab:nupresults}.  This agreement between the two methods for determining $\nu'$ is a useful 
test of the model defined by Eqs.~(\ref{eq:rvmrho}) and (\ref{eq:pLam}).  

If we relax the assumption that $\tilde{\nu}=0$, we might attempt to use this second method for determining $\nu'$ to constrain $\tilde{\nu}$ as well.  As can be seen from Eq.~(\ref{eq:Einsteinii}) and the definitions of 
$I_{H^2}$, $S$, and $T$, we find
\bea I_{H^2} + 3 \frac{\ddot{a}}{a} = -3\left(\nu -\frac{3}{2}\tilde{\nu} \right)S - \frac{3}{2}\tilde{\nu}T.
\eea
Unfortunately, the third derivative $\dddot{A}$ is only poorly determined by the data, given the large statistical errors associated with numerical derivatives of high order.  There is also expected to be some cancellation between the $\tilde{\nu}$ term proportional to $S$ with that proportional to $T$, since $S$ and $T$ are equal up to corrections of ${\cal O}(H^2)$.  Thus, a meaningful constraint on $\tilde{\nu}$ is not possible with this approach.

\subsection{Implications for the running vacuum model}
\label{sec:Implications_RVM}

Our analysis shows that the EDT lattice formulation of gravity singles out the running vacuum model represented by
\bea\label{eq:RVM_summary} \Lambda(H) = \Lambda_0 + 3\nu H^2,
\eea
where we have neglected $\tilde{\nu}$ in Eq.~(\ref{eq:RVM_gen}), given our strong bounds on this parameter.  Within this model, the dimensionless coupling $\nu$ runs logarithmically,
\bea\label{eq:nu_running_summary} \nu(H^2_f) = \frac{\nu(H_i^2)}{1+b \log(\frac{H_i^2}{H_f^2})}.
\eea
This dynamics leads to a vacuum energy density
\bea  \rho_{\Lambda} = \rho_{\Lambda_0} + \frac{3\nu H^2}{8\pi G}.
\label{eq:rvmrho_summary}
\eea
The vacuum pressure follows from the additional assumption that energy is covariantly conserved.  The vacuum pressure 
\bea \label{eq:pLam_summary}  p_\Lambda = -\rho_{\Lambda_0} - \frac{\nu}{8\pi G} \left(H^2 + 2\frac{\ddot{a}}{a}\right),
\eea
leads to an equation of state $w$ that ensures this is the case.  
The absence of an emergent matter sector in our simulations suggests that the vacuum sector independently maintains covariant energy conservation, and that the vacuum pressure adjusts itself so that this is the case.  This has important implications when applying this model to cosmology, as discussed in the next section.

If we take the lattice evidence in favor of this vacuum model seriously, we can use our numerical results to obtain a value for $\nu$ in the present universe.  To do this, we first express the renormalization scale in terms of the cosmological constant.  This is the most convenient choice, since we have determined our lattice $\nu$ as a function of $\sqrt{V_4}$, which is proportional to $1/\Lambda_0$, and our universe in the current dark energy dominated epoch is nearly de Sitter, so that we have the approximate proportionality $H^2\propto \Lambda_{0, {\rm phys}}$, with $\Lambda_{0,{\rm phys}}$ the cosmological constant in our universe today.  Thus, we can write
\bea \label{eq:final_nu}  \nu(\Lambda_0) = \frac{\nu(M_{\rm pl}^2)}{1 + b \ln{\left(\frac{M_{\rm pl}^2}{\Lambda_0}\right)}}=
              \frac{A'}{\ln{\left(\frac{B'}{M_{\rm pl}^2} \frac{\sqrt{V_4}}{a_{\rm lat}^2}M_{\rm pl}^2\right)}}.
\eea
Using the relation 
\bea\label{eq:final_Lam0}  \Lambda_0 a_{\rm lat}^2 = \sqrt{\frac{24 \pi^2}{\eta C_4 N_4}},
\eea
and using Eq.~(\ref{eq:V4}) for $V_4$, as well as the observation that in our fiducial lattice units, $1/M_{\rm pl}^2 = G/a_{\rm fid}^2$, it follows that
\bea\label{eq:final_b}  b = \frac{1}{\ln{\left(\sqrt{24}\pi B' \frac{G}{a_{\rm fid}^2}  \right)}},
\eea
\bea\label{eq:final_nuMpl}  \nu(M_{\rm pl}^2)= b A'.
\eea
Since $B'$ is measured in lattice units where the fiducial lattice spacing is the one at $\kappa_2=3.0$, we take $G/a_{\rm fid}^2$ at that lattice spacing as well.  Substituting our computed values of $A'$, $B'$, and $G/a_{\rm fid}^2$ into Eqs.~(\ref{eq:final_nu}), (\ref{eq:final_b}), and (\ref{eq:final_nuMpl}), and taking $\Lambda_{0,{\rm phys}} = 10^{-123} M_{\rm pl}^2$ from observations, we find that 
\begin{equation}
    \nu(\Lambda_{0,{\rm phys}})=(5.1\pm 1.3) \times 10^{-4}\,.
\end{equation}

\section{Making contact with Cosmology}
\label{sec:cosmology}

We discuss some of the implications of our results for observational cosmology.  We draw on the work of Sol\`a and collaborators \cite{Gomez-Valent:2015pia, Sola:2016jky, Moreno-Pulido:2023ryo}, which involved studying a family of vacuum models with a number of free parameters, including $\nu$ and $\tilde{\nu}$, that are constrained by fits to cosmological data.  Here we restrict ourselves to reviewing the consequences of the model that is singled out by our lattice calculation, namely that defined by Eqs.~(\ref{eq:nu_running_summary}), (\ref{eq:rvmrho_summary}), and (\ref{eq:pLam_summary}), with the numerical value of $\nu$ fixed by our calculation and $\tilde{\nu}$ set to zero.  We assume that the forms of the vacuum energy density and the vacuum pressure determined in our calculation are valid throughout cosmic history and are not altered by the presence of matter in the universe, nor by the virtual quantum effects of matter back-reacting on quantum geometry.  The latter assumption could be tested by unquenching our calculation.  We further assume that the Euclidean continuation (and thus its inverse) maintains the same form of the running of $\Lambda(\mu)$, where the renormalization scale $\mu$ is associated with the absolute value of the Hubble scale (see Eq.~(\ref{eq:RVM}) and the surrounding discussion).  In this exercise, it is not necessary to include any additional free parameters beyond those of the $\Lambda$CDM model, since $\nu(\Lambda_{0,{\rm phys}})\approx 5\times 10^{-4}$ is determined from our {\it ab initio} calculation and is thus a prediction of the dynamical triangulations formulation.  

To begin our analysis we consider the FLRW equation in Lorentz signature,
\bea\label{eq:FLRW_lorentz}  \frac{\dot{a}^2}{a^2} + \frac{1}{a^2} = \frac{8\pi G \rho}{3},
\eea
where we assume that the total energy density $\rho$ includes the energy density of our running vacuum model $\rho_\Lambda$, the energy density $\rho_m$ of nonrelativistic matter, and the energy density $\rho_r$ of radiation.
Solving Eq.~(\ref{eq:FLRW_lorentz}) for $H=\dot{a}/a$ assuming that the spatial curvature is negligible and substituting this into Eq.~(\ref{eq:rvmrho_summary}) for the vacuum energy density, if we also take $\nu$ as a small expansion parameter, we find
\bea  \rho_\Lambda = (1+\nu)\rho_{\Lambda_0} + \nu(\rho_m + \rho_r) + {\cal O}(\nu^2).
\eea
In the current epoch, where dark energy is $\approx70\%$ of the energy density, the $\rho_{\Lambda_0}$ term dominates the vacuum energy density $\rho_\Lambda$.  The $\nu$ in the prefactor of the $\rho_{\Lambda_0}$ term is a small correction that is not well constrained with the current precision on observations.  As we run the expansion of the universe backwards in time, first $\rho_m$ and then eventually $\rho_r$ dominates the energy density of the universe.  In that case $\rho_{\Lambda_0}/(\rho_m+\rho_r) \ll \nu$, so that $\rho_\Lambda \approx \nu (\rho_m + \rho_r)$.  The smallness of the coefficient $\nu$ ensures that the vacuum energy density stays a small fraction of the energy density of the universe, since it is suppressed by a factor of $\nu$ compared to $\rho_m$ and $\rho_r$.  However, the total vacuum energy density in the early universe is still much larger than the $\rho_{\Lambda_0}$ term, so that the vacuum energy density is predicted to be much larger in the early universe in this model.
The current best constraints on Big Bang Nucleosynthesis (BBN) fix the Hubble scale in that era at the percent level \cite{Yeh:2022heq}.  Because the parameter $\nu$ is ${\cal O}(10^{-3})$, the corrections from this model are compatible with BBN constraints within the existing errors.  Note that the logarithmic running of $\nu$ leads to an increase in $\nu$ of less than $20\%$ at the time of BBN as compared to today, so our order of magnitude estimate of $\nu$ in the late universe still applies in this earlier era.

An important feature of this model is that the vacuum pressure given by Eq.~(\ref{eq:pLam_summary}) ensures that the vacuum energy is covariantly conserved independently of the matter sector.  If this were not the case and the decrease in vacuum energy was accompanied by a transfer of energy into matter or radiation, this would lead to strong violations of experimental constraints.  For example, if the vacuum decayed to radiation, this would lead to a large deviation of the cosmic microwave background from the $\Lambda$CDM expectation and would not be consistent with observations.  If the vacuum decayed to cold dark matter during the radiation era, then far too much dark matter would be produced to be compatible with the current energy budget of the universe.  Thus, the separate covariant energy conservation of the vacuum sector ensures that the vacuum dynamics does not significantly modify the successes of the standard $\Lambda$CDM picture.  

Although this model does not lead to large deviations from $\Lambda$CDM, it does predict ${\cal O}(10^{-3})$ deviations throughout cosmology, such that if this running vacuum effect were to be realized in nature, it should be testable to a high degree of confidence when cosmological observations reach a sufficient precision in the hopefully not-too-distant future.  As an example of such a deviation, we consider how the vacuum equation of state $w\equiv p_\Lambda/\rho_\Lambda$ in the running vacuum model differs from the constant $\Lambda$CDM value of $-1$.  Substituting Eqs.~(\ref{eq:rvmrho_summary}) and (\ref{eq:pLam_summary}) into the definition of $w$ and expanding in $\nu$, we find
\bea  w \approx -1 + \frac{\nu}{4\pi G \rho_{\Lambda_0}}\left(H^2 - \frac{\ddot{a}}{a} \right) = -1 + \nu \frac{\rho_m}{\rho_{\Lambda_0}},
\eea
where we have assumed that the radiation density can be ignored in the present era.  This expression can be rewritten as \cite{Moreno-Pulido:2022phq, Moreno-Pulido:2022upl}
\bea  w \approx -1 + \nu \frac{\Omega_{m,0}}{\Omega_{\Lambda_0}}\left(1+z \right)^3,
\eea
where $\Omega_i$ is the fraction of the total energy density of the universe, $z$ is the redshift due to the cosmological expansion, and we have employed a small-parameter expansion such that this formula is valid for $z \lesssim 5$.  The subscript 0 denotes the fraction of the energy density of matter today.  With $\nu$ of ${\cal O}(10^{-3})$, it is not possible to discriminate between this model and that of $\Lambda$CDM using the current data.  However, the positive sign that we have determined for $\nu$ means that $w$ does not fall below $-1$, and therefore does not violate the null-energy condition.  This would lead to quantum instabilities that might be hard to reconcile with our long-lived universe \cite{Carroll:2003st}.  Other cosmological observables, like the matter power spectrum and the present-day Hubble rate, also receive small corrections in this model.  For the full mathematical treatment of the corrections to the matter power spectrum within a family of models including the one singled out by the present work, see Ref.~\cite{Gomez-Valent:2015pia}.

\section{Conclusion and Outlook}
\label{sec:conclude}

This work presents a detailed analysis of a new set of EDT ensembles at larger volumes and finer lattice spacings than was previously possible to simulate.  These ensembles have been generated using a new rejection free algorithm \cite{Dai:2023tud}, which led to a speed-up of around a factor of one-hundred on our finest ensembles when compared with our older parallel rejection algorithm.  We use this algorithm to generate several volumes across four lattice spacings.  In order to meet the precision goals of this work, we introduce a new set of methods for obtaining the ratio of the direct to (renormalized) dual lattice spacing, $a_{\rm lat}/\ell$.  Although this is a derived lattice quantity without a direct physical meaning, it is needed for constructing physical observables.  The two methods introduced here to construct $a_{\rm lat}/\ell$ each assume that the emergent geometries are well-described by the classical de Sitter solution in the large volume limit.  A comparison of the results from the two methods for computing $a_{\rm lat}/\ell$ show that they are in excellent agreement with each other at the percent level, thus providing a solid test that this lattice specific quantity is correctly determined.   

We revisit our calculation of the absolute lattice spacing using the method introduced in Ref.~\cite{Bassler:2021pzt} and a large new data set generated for this purpose.  The renormalized value of Newton's constant can be inferred from the finite-volume scaling of the tuned bare parameter $\kappa_4$ of the lattice action, as reviewed in Sec.~\ref{sec:GNewtonanalysis}.  The scaling of $\kappa_4$ with volume is related to Newton's constant through a derivation assuming the Euclidean partition function is dominated by the de Sitter instanton.  Using this approach to determine Newton's constant, we are able to establish the Planck length in units of our lattice spacing.  By computing Newton's constant independently at each of our four lattice spacings, we are also able to measure the relative lattice spacing of our ensembles against a chosen fiducial spacing, defined to be the one at $\kappa_2=3.0$.  The relative lattice spacings are in agreement with the expected behavior, with the lattices getting finer as $\kappa_2$ is increased.  This is true for both the direct and renormalized dual lattice spacings.  This result suggests that there is no barrier in principle to taking the lattice spacing arbitrarily small within the EDT formulation.

New ensembles and methods for determining the lattice spacing enable us to test the de Sitter nature of our geometries at a level of precision not previously possible, and a careful study shows a deviation from this picture.  Non-trivial vacuum dynamics is seen, with a power law running of the cosmological constant.  Good agreement across four lattice spacings suggests that the running is not a lattice artifact.  Consistency checks also provide strong evidence that the running seen in the simulations is not an artifact. 
The running vacuum model that is singled out by our lattice calculation is summarized by Eqs.~\eqref{eq:nu_running_summary}, \eqref{eq:rvmrho_summary}, and \eqref{eq:pLam_summary}.  This is a one-parameter model, since $\Lambda_0$ must be taken from experiment, while $\nu$ and its running are fixed from first principles by the analysis presented here.  An additional parameter, $\tilde{\nu}$, appearing in a more general version of the running vacuum model, is consistent with zero and is constrained to be much smaller than $\nu$ by our analysis.

If we take this model seriously and make the bold assumption that it holds throughout cosmic history, we see that it leads to potentially observable effects.  As Sec.~\ref{sec:cosmology} discusses, the vacuum energy is nearly constant in the present dark energy dominated era, but it is larger as one follows the evolution of the universe backwards in time.  The smallness of $\nu$ ensures that the vacuum energy density during matter or radiation domination is always $\sim1000$ times smaller than the dominant energy density component of either era, so that the model remains consistent with bounds from the early universe.  Thus, the deviations from $\Lambda$CDM are predicted to be small, at the ${\cal O}(10^{-3})$ level.  Even so, the model is highly falsifiable, since the deviations would appear across many cosmic observables if they could be measured to sufficient precision. 

The lattice calculation presented here could be improved in a number of ways.  Going to larger, finer lattices would give better control over discretization effects, especially the trending of $\nu$ towards negative values if the volume is taken too large at too coarse of a lattice spacing, as seen in \autoref{fig:midsuperfine}.  A better determination of the relative lattice spacings would also help, and additional runs to help pin these down are in progress.  Unquenching the calculation is also a high priority for future work.  Including the quantum fluctuations of light fields, for example those of the Standard Model, would lead to a back-reaction on the geometry, which could potentially alter $\nu$.  We expect that this would change $\nu$ by an ${\cal O}(1)$ constant, not by orders of magnitude, but it is important to do the simulations in order to test this assumption.  According to a semiclassical calculation \cite{Moreno-Pulido:2022phq, Moreno-Pulido:2023ryo}, heavy particles would contribute directly to $\nu$ in proportion to their mass squared.  Thus, the existence of a large number of GUT-scale particles could potentially modify $\nu$ from our result, which ought to be regarded as the pure gravity prediction.  Unquenching could also impact the predictions of the pure-gravity result in a more dramatic way if threshold effects are present, as discussed below.   

It is worth commenting on the possibility that the running vacuum effect calculated here might actually be present in nature, especially in light of the conventional wisdom from effective field theory that the vacuum energy does not run \cite{Foot:2007wn}.  We appeal to an argument by Polyakov that nonperturbative fluctuations of the scale factor might be important and that perturbation theory about a fixed background could be misleading in this instance \cite{Polyakov:2006bz}.  The nonperturbative calculation presented here does not require an expansion around a classical background, and it gives a concrete realization of Polyakov's sketch of running vacuum energy \cite{Polyakov:2000fk}, without suffering from the serious drawbacks of his model, which neglected the traceless gravitational degrees of freedom, and found a logarithmic running that is not strong enough to be compatible with observations.  As such, our calculation suggests that there is non-trivial infrared dynamics driving the screening, and not a conspiracy of cancellations in the ultra-violet physics.  

This picture also has implications for the value of $\Lambda_0$.  Note that the value of $\Lambda_0$ in our model is not explained in this work, as it is identified as the consequence of a finite-volume cut-off on the simulations, which can be freely adjusted.  However, if the screening is driven by infra-red physics, $\Lambda_0$ might naturally be explained as an offset due to a threshold-crossing effect in the running of the vacuum energy, where its value would be determined by the light particle spectrum, taking on different values in different eras.  If so, the unquenched version of this theory may lead to a model rich enough to explain not only our present vacuum dominated era, but to earlier ones that might help resolve some of the big emerging tensions in observational cosmology, especially the well-known Hubble tension \cite{Abdalla:2022yfr}.  The investigation of this possibility requires including unquenched fermions in the simulations, since neutrinos are the lightest non-zero mass particles in the known spectrum.  If this picture were to be realized, the predictions of Sec.~\ref{sec:cosmology} might well be subdominant to vacuum dynamics that includes unquenching effects.  Work in this direction is also in progress.

\begin{acknowledgments}
The authors thank Mark Alford, Claude Bernard, Simon Catterall, Jay Hubisz, Alex Maloney, and Scott Watson for valuable discussions and Mark Alford, Claude Bernard, Simon Catterall, and Scott Watson for comments on the manuscript.  MD and JL were supported by the U.S. Department of Energy (DOE), Office of Science, Office of High Energy
Physics under Award Number DE-SC0009998.  
JUY was supported by the U.S. Department of Energy grant DE-SC0019139, and by Fermi
Research Alliance, LLC under Contract No. DE-AC02-
07CH11359 with the U.S. Department of Energy, Office
of Science, Office of High Energy Physics, and by the Department of Energy through the Fermilab Theory QuantiSED program in the area of ``Intersections of QIS and Theoretical Particle Physics.''
MS acknowledges support by Perimeter Institute for Theoretical Physics. Research at Perimeter Institute is supported in part by the Government of Canada through the Department of Innovation, Science and Economic Development and by the Province of Ontario through the Ministry of Colleges and Universities. Computations were performed in part on the Syracuse University HTC Campus Grid and were supported by NSF award ACI-1341006. This research was enabled in part by support provided by ACENET, Calcul Québec, Compute Ontario, and the Digital Research Alliance of Canada. The authors acknowledge support by the state of Baden-Württemberg through bwHPC.
\end{acknowledgments}

\appendix
\clearpage
\section*{Tabulated data}
\label{app:tables}

\begin{center}
\begin{table}
\caption{Method 1 determinations of $a_{\rm lat}/\ell$ and $\eta$ on ensembles used in our analysis, see Eqs.~\eqref{eq:deSitterMod}-\eqref{eq:aoverell}. The first error quoted is the statistical error, and the second is a systematic error associated with varying the fit window.  For volumes with two tuned $\beta$ values, the third error is an estimate of the error due to mistuning $\beta$.
}
\label{tab:etaaell}
\begin{tabular}{c c c c} 
\hline \hline
$\kappa_2$ & $N_4$ & $a_{\rm lat}/\ell$ & $\eta$ \\
\hline
\multirow{6}{4em} {\centering 2.45} & $6000$ & 10.44(8)(9) & 0.639(5)(8)\\ 
& $8000$ & 9.96(6)(4) & 0.610(5)(8)\\ 
& $12000$ & 9.69(14)(8) & 0.609(15)(7)\\ 
& $16000$ & 9.90(17)(18) & 0.565(12)(12)\\ 
& $24000$ & 9.32(18)(14) & 0.580(15)(28)\\ 
& $32000$ & 9.12(6)(8) & 0.543(4)(11)\\ 
\hline
\multirow{6}{4em} {\centering 3.0} & $8000$ & 13.49(5)(6) & 0.770(3)(7)\\ 
& $12000$ & 13.02(7)(5) & 0.732(5)(6)\\ 
& $16000$ & 12.61(2)(1) & 0.717(1)(0)\\ 
& $24000$ & 12.35(3)(4) & 0.699(2)(4)\\ 
& $32000$ & 12.37(3)(26) & 0.681(1)(13)\\ 
& $48000$ & 11.84(1)(15) & 0.648(0)(9)\\ 
\hline
\multirow{7}{4em} {\centering 3.4} & $8000$ & 19.16(9)(3) & 0.834(5)(1)\\ 
& $12000$ & 18.69(7)(3) & 0.794(3)(4) \\ 
& $16000$ & 18.10(7)(1) & 0.772(3)(6)\\ 
& $24000$ & 17.69(7)(5) & 0.762(3)(4)\\ 
& $32000$ & 17.50(7)(9) & 0.732(3)(5)\\ 
& $48000$ & 16.84(4)(6)(45) & 0.708(2)(5)(4)\\ 
& $64000$ & 16.80(6)(8)(17) & 0.703(3)(5)(22)\\ 
\hline
\multirow{6}{4em} {\centering 3.8} & $16000$ &27.22(8)(2)(21) & 0.830(2)(1)(17)\\ 
& $24000$ & 26.46(8)(7) & 0.802(2)(2) \\ 
& $32000$ & 26.16(4)(6) & 0.784(1)(4)\\ 
& $48000$ & 25.32(5)(6) & 0.773(1)(4)\\ 
& $64000$ & 25.06(4)(8)(41) & 0.763(1)(4)(20)\\ 
& $96000$ & 24.02(2)(4)(93) & 0.714(0)(4)(8)\\ 
\hline\hline

\end{tabular}
\end{table}
\end{center}

\begin{table}
\caption{$\kappa_4$ values for de Sitter finite-size analysis at tuned $\beta$ values for $\kappa_2=2.45$. The first line for each distinct value of $\beta$ is the ensemble we consider tuned. When marked with an $*$, the ensemble has not been included in the central fit, but it has been used to estimate a systematic uncertainty for the slope.}
\label{tab:kappa4MF}
\centering
\begin{tabular}{c c c c c } 
\hline \hline
$\kappa_2$& $\beta$ & $N_4$ & $N_{\rm config}$ & $\kappa_4$ \\
\hline
\multirow{6}{4em} {\centering 2.45}& \multirow{6}{4em} {\centering -0.590} & 6000 & 9172 & 6.811743(64)*\\
&  & 8000 & 24819 & 6.812738(38) \\
&   & 12000 & 29801 & 6.813441(22) \\
&  & 16000 & 14174 & 6.813807(48) \\
&   & 24000 & 26331 & 6.814285(29) \\
&   & 32000 & 25656 & 6.814513(23) \\
\hline
\multirow{6}{4em} {\centering 2.45}& \multirow{6}{4em} {\centering -0.575} & 8000 & 4232 & 6.856277(69)\\
&   & 12000 & 38106 &  6.857112(16) \\
&   & 16000 & 32171 &  6.857477(28) \\
&   & 24000 & 33825 &  6.857937(31) \\
&   & 32000 & 23531 &  6.858235(39) \\
&   & 48000 & 28092 &  6.858364(46) \\

\hline
\multirow{6}{4em} {\centering 2.45}& \multirow{6}{4em} {\centering -0.555} & 12000 & 2285 & 6.91550(11)* \\
&   & 16000 & 32157 &  6.916028(25)   \\
&   & 24000 & 39487 &  6.916411(23)   \\
&   & 32000 & 28582 &  6.916691(32)   \\
&   & 48000 & 24329 &  6.916925(26)   \\
&   & 64000 & 18123 &  6.917029(54)   \\
\hline
\multirow{6}{4em} {\centering 2.45}& \multirow{6}{4em} {\centering -0.544} & 16000 & 2834 &  6.948352(68) \\
&   & 24000 & 32690 & 6.948718(24)     \\
&   & 32000 & 6396 &  6.948970(33)   \\
&   & 48000 & 29416 & 6.949227(55)  \\
&   & 64000 & 28516 & 6.949316(35)   \\
&   & 96000 & 29005 &  6.949473(18)   \\
\hline
\multirow{5}{4em} {\centering 2.45}& \multirow{5}{4em} {\centering -0.530} & 24000 & 3358 &  6.990061(71)  \\
&   & 32000 & 29743 &  6.990197(21)    \\
&   & 48000 & 33759 &  6.990530(33) \\
&   & 64000 & 28766 &  6.990599(54)   \\
&   & 96000 & 27946 &  6.990728(41)  \\
\hline
\multirow{4}{4em} {\centering 2.45}& \multirow{4}{4em} {\centering -0.520} & 32000 & 2485 & 7.01976(13)   \\
&   & 48000 & 36183 &  7.020030(47)  \\
&   & 64000 & 28996 &  7.020201(41)    \\
&   & 96000 & 33879 &  7.020309(18)  \\
\hline
\end{tabular}
\end{table}

\begin{table}
\caption{$\kappa_4$ values for de Sitter finite-volume analysis for tuned $\beta$ values at $\kappa_2=3.0$. The first line for each distinct value of $\beta$ is the ensemble we consider tuned. The ensemble marked with an $*$ is considered an outlier and is not included in the analysis.
}
\label{tab:kappa4F}
\centering
\begin{tabular}{c c c c c } 
\hline \hline
$\kappa_2$& $\beta$ & $N_4$ & $N_{\rm config}$ & $\kappa_4$ \\
\hline
\multirow{6}{4em} {\centering 3.0}& \multirow{6}{4em} {\centering -0.859} & 4000 & 9168 & 7.651190(55) \\
&  & 6000 & 16869 &  7.652089(24) \\
&  & 8000 & 10397 &  7.652556(77) \\
&   & 12000 & 8536 & 7.653214(98) \\
&  & 16000 & 9310 &  7.653603(71) \\
&   & 24000 & 8552 & 7.65389(10)  \\
\hline
\multirow{6}{4em} {\centering 3.0}& \multirow{6}{4em} {\centering -0.820} & 6000 & 1441  &  7.76272(16)\\
&  & 8000 & 18768 & 7.763078(37)    \\
&   & 12000 & 19771 &  7.763601(21) \\
&  & 16000 & 14157 &  7.763998(82)  \\
&   & 24000 & 17632 &  7.764330(39)  \\
&   & 32000 & 12744 &  7.764476(39) \\
\hline
\multirow{5}{4em} {\centering 3.0}& \multirow{5}{4em} {\centering -0.800} & 8000 & 1974  & 7.82031(11) \\
&   & 12000 & 18375 & 7.820737(53)  \\
&  & 16000 & 23811 & 7.821109(38) \\
&   & 24000 & 17967 &  7.821482(57) \\
&   & 32000 & 12580 &  7.821628(45) \\
\hline
\multirow{5}{4em} {\centering 3.0}& \multirow{5}{4em} {\centering -0.782} & 12000 &  8548 & 7.872552(33) \\
&  & 16000 & 23668 & 7.872799(36) \\
&   & 24000 & 23683 &  7.873145(33) \\
&   & 32000 & 18805 &  7.873359(24) \\
&   & 48000 & 15664 &  7.873485(49) \\
\hline
\multirow{5}{4em} {\centering 3.0}& \multirow{5}{4em} {\centering -0.771} & 16000 & 11215  &  7.904573(30) \\
&   & 24000 & 22577 & 7.90486(10)  \\
&   & 32000 & 19734 & 7.905131(29)*  \\
&   & 48000 & 24729 & 7.905222(19)  \\
&   & 64000 & 13080 & 7.905327(32)  \\
\hline
\multirow{6}{4em} {\centering 3.0}& \multirow{6}{4em} {\centering -0.756} & 24000 &  1108 & 7.94825(12) \\
&   & 32000 & 21101 & 7.948549(38)  \\
&   & 48000 & 20625 & 7.948704(23)  \\
&   & 64000 & 19182 & 7.948784(33)  \\
&   & 96000 & 16204 & 7.948937(38)  \\
&   & 128000 & 23718 & 7.949014(33)  \\
\hline
\end{tabular}
\end{table}

\begin{table}
\caption{$\kappa_4$ values for de Sitter finite-volume analysis for tuned $\beta$ values at $\kappa_2=3.4$. The first line for each distinct value of $\beta$ is the ensemble we consider tuned. 
}
\label{tab:kappa4RF}
\centering
\begin{tabular}{c c c c c } 
\hline \hline
$\kappa_2$& $\beta$ & $N_4$ & $N_{\rm config}$ & $\kappa_4$ \\
\hline
\multirow{6}{4em} {\centering 3.4}& \multirow{6}{4em} {\centering -0.910} & 8000 & 6255 &  8.695358(63) \\
&   & 12000 & 22333 & 8.695751(25)  \\
&  & 16000 & 11941 &  8.695982(40)  \\
&   & 24000 & 11296 & 8.696250(41)   \\
&   & 32000 & 8596 &  8.696278(57) \\
&   & 48000 & 7983 &  8.696497(40) \\
\hline
\multirow{5}{4em} {\centering 3.4}& \multirow{5}{4em} {\centering -0.870} & 12000 & 3100 &  8.809014(60)  \\
&  & 16000 & 14226 &  8.809267(44)   \\
&   & 24000 & 14977 & 8.809454(41)   \\
&   & 32000 & 10020 & 8.809586(51)  \\
&   & 48000 & 16766 & 8.809738(49)  \\
\hline
\multirow{5}{4em} {\centering 3.4}& \multirow{5}{4em} {\centering -0.853} & 16000 & 3069 &  8.857898(86)  \\
&   & 24000 & 15734 &  8.858044(25)  \\
&   & 32000 & 13430 &  8.858196(45) \\
&   & 48000 & 20829 &  8.858293(17) \\
&   & 64000 & 15318 &  8.858406(27) \\
\hline
\multirow{5}{4em} {\centering 3.4}& \multirow{5}{4em} {\centering -0.839} & 24000 & 1034 & 8.89835(12)   \\
&   & 32000 & 16171 & 8.898427(36)  \\
&   & 48000 & 18318 & 8.898515(52)  \\
&   & 64000 & 21547 & 8.898674(16)  \\
&   & 96000 & 22711 & 8.898771(16)  \\
&   & 128000& 13941 & 8.898744(57)  \\
\hline
\multirow{5}{4em} {\centering 3.4}& \multirow{5}{4em} {\centering -0.830} & 32000 & 1149 &  8.92439(14)  \\
&   & 48000 & 29601 & 8.924501(45)  \\
&   & 64000 & 18946 & 8.924595(45) \\
&   & 96000 & 15979 & 8.924640(21)  \\
&   & 128000& 13412 & 8.924728(37)  \\
\hline
\end{tabular}
\end{table}

\begin{table}
\caption{$\kappa_4$ values for de Sitter finite-volume analysis for tuned $\beta$ values at $\kappa_2=3.8$. The first line for each distinct value of $\beta$ is the ensemble we consider tuned. When marked with an $*$, the ensemble has not been included in the central fit, but it has been used to estimate a systematic uncertainty for the slope.}
\label{tab:kappa4SF}
\centering
\begin{tabular}{c c c c c } 
\hline \hline
$\kappa_2$& $\beta$ & $N_4$ & $N_{\rm config}$ & $\kappa_4$ \\
\hline
\multirow{7}{4em} {\centering 3.8}& \multirow{7}{4em} {\centering -0.931} & 12000 & 3458 & 9.830392(67)   \\
&  & 16000 & 12365 & 9.830563(56)   \\
&  & 20000 & 9554 &  9.830727(50)  \\
&   & 24000 & 6757 & 9.830815(64)    \\
&   & 28000 & 5474 & 9.830811(54)    \\
&   & 32000 & 3844 & 9.830903(97)   \\
&   & 40000 & 2621 & 9.830947(82)   \\
\hline
\multirow{10}{4em} {\centering 3.8}& \multirow{10}{4em} {\centering -0.920} & 16000 & 3087 & 9.861485(72)*   \\
&  & 20000 & 5584 &  9.861684(87)  \\
&   & 24000 & 10725 &  9.861881(26)   \\
&   & 28000 & 3794 &  9.861944(73)   \\
&   & 32000 & 10429 & 9.861924(32)   \\
&   & 40000 & 4765 &  9.861973(99)  \\
&   & 48000 & 6715 &  9.861972(86)  \\
&   & 56000 & 4296 &  9.862041(57)  \\
&   & 64000 & 3696 &  9.862038(72)  \\
&   & 80000 & 2387 &  9.862107(79)  \\
\hline
\multirow{8}{4em} {\centering 3.8}& \multirow{8}{4em} {\centering -0.894} & 24000 & 1736 & 9.935649(92)   \\
&   & 28000 & 10098 &  9.935648(39)   \\
&   & 32000 & 6723 &   9.935721(73) \\
&   & 40000 & 5614 &   9.935818(53) \\
&   & 48000 & 5053 &   9.935795(50) \\
&   & 56000 & 4981 &   9.935812(60) \\
&   & 64000 & 5138 &   9.935879(50) \\
&   & 80000 & 2827 &   9.936012(71) \\
\hline
\multirow{5}{4em} {\centering 3.8}& \multirow{5}{4em} {\centering -0.870} & 32000 & 4506 & 9.975799(67)   \\
&   & 40000 & 5736 &  9.975915(50)  \\
&   & 48000 & 14283 & 9.975926(65)   \\
&   & 56000 & 5914 &  9.975984(75)  \\
&   & 64000 & 5395 &  9.975963(40)  \\
&   & 80000 & 4146 &  9.976005(73)  \\
\hline
\multirow{4}{4em} {\centering 3.8}& \multirow{4}{4em} {\centering -0.880} & 48000 & 3781 &  10.004601(78)  \\
&   & 56000 & 7773 &  10.004685(42)   \\
&   & 64000 & 7083 &  10.004707(58)  \\
&   & 80000 & 4236 &  10.004723(67)  \\
\hline
\end{tabular}
\end{table}

\begin{table}
\caption{Slopes according to \eqref{eq:slopefit} and fit-details for the $\kappa_2=2.45$ ensembles. Ensembles marked with $*$ were not included in the fit for $G_a$ following \eqref{eq:fitGa}, but were used to estimate a systematic uncertainty in the infinite-volume extrapolation.}
\label{tab:slopesMF}
\centering
\begin{tabular}{c c c c c c } 
\hline \hline
$\kappa_2$& $\beta$ & $N_4$ &  $|s_G|$ & $\chi^2/\mathrm{d.o.f}$ & $p$-value \\
\hline
\multirow{6}{4em} {\centering 2.45}&  -0.590 & 6000 & 0.3139(69)(69)  & 1.07 & 0.36 *\\
& -0.575 & 8000 & 0.3128(53)(73)  & 0.31 & 0.73 \\
& -0.555 & 12000 & 0.2716(90)(54) & 0.77 & 0.46 \\
& -0.544 & 16000 & 0.2329(72)(66)  & 0.79 & 0.53\\
& -0.530  & 24000 & 0.233(26)(22)  & 2.38 & 0.067\\
& -0.520  & 32000 & 0.204(22)(35)  & 0.57 & 0.56\\
\hline
\end{tabular}
\end{table}

\begin{table}
\caption{Slopes according to \eqref{eq:slopefit} and fit-details for the $\kappa_2=3.0$ ensembles. 
}
\label{tab:slopesF}
\centering
\begin{tabular}{c c c c c c } 
\hline \hline
$\kappa_2$& $\beta$ & $N_4$ &  $|s_G|$ & $\chi^2/\mathrm{d.o.f}$ & $p$-value \\
\hline
\multirow{6}{4em} {\centering 3.0}&  -0.859 & 4000 & 0.2967(61)(100)   & 0.49 & 0.75  \\
& -0.820 & 6000 & 0.2555(71)(105)   &  0.78 & 0.54   \\
& -0.800 & 8000 & 0.246(13)(13)  & 0.70  & 0.55  \\
& -0.782 & 12000 & 0.221(11)(18) & 1.28  & 0.28  \\
& -0.771  & 16000 & 0.1926(24)(44)   & 0.063  & 0.94  \\
& -0.756  & 24000 & 0.1731(107)(94)   & 0.51  & 0.73 \\
\hline
\end{tabular}
\end{table}

\begin{table}
\caption{Slopes according to \eqref{eq:slopefit} and fit-details for the $\kappa_2=3.4$ ensembles.}
\label{tab:slopesRF}
\centering
\begin{tabular}{c c c c c c } 
\hline \hline
$\kappa_2$& $\beta$ & $N_4$ &  $|s_G|$ & $\chi^2/\mathrm{d.o.f}$ & $p$-value \\
\hline
\multirow{5}{4em} {\centering 3.4}&  -0.910 & 8000 &   0.1674(83)(123) & 1.02 & 0.39   \\
& -0.870 & 12000 &  0.1517(78)(53)   & 0.29   & 0.84  \\
& -0.853 & 16000 & 0.1343(92)(85) & 0.57  & 0.64   \\
& -0.839 & 24000 & 0.141(15)(14)  & 1.05   & 0.38   \\
& -0.830  & 32000 & 0.1143(96)(114) & 0.37 & 0.77  \\
\hline
\end{tabular}
\end{table}

\begin{table}
\caption{Slopes according to \eqref{eq:slopefit} and fit-details for the $\kappa_2=3.8$ ensembles.}
\label{tab:slopesSF}
\centering
\begin{tabular}{c c c c c c } 
\hline \hline
$\kappa_2$& $\beta$ & $N_4$ &  $|s_G|$ & $\chi^2/\mathrm{d.o.f}$ & $p$-value \\
\hline
\multirow{5}{4em} {\centering 3.8}&  -0.931 & 12000 &  0.136(11)(14)   &  0.29 &  0.91  \\
& -0.920 & 16000 & 0.0696(50)(33) & 0.11  &  1.00 \\
& -0.894 & 24000 &  0.1139(161)(77)  &  0.52  & 0.79  \\
& -0.880  & 32000 & 0.084(18)(30)  & 0.22 & 0.99  \\
& -0.870  & 48000 & 0.066(17)(31)  & 0.025 & 0.88  \\
\hline
\end{tabular}
\end{table}

\begin{center}
\begin{table}
\caption{Fit results for linear fits to $\Lambda$ versus $H^2$ data sets.  The first quoted error on the fit parameters is the statistical error combined in quadrature with the systematic error obtained from varying the fit window.  In the instances where we have a second tuned ensemble, in order to estimate a $\beta$ tuning systematic error (see Table \ref{tab:ensembles}), the first error also includes a systematic error from comparing the results of the two different tuned ensembles. The second quoted error is the parametric error due to the uncertainty in $\frac{a_{\rm lat}}{\ell}$. 
}
\label{tab:lambdaH2}
\begin{tabular}{c c c c c c} 
\hline \hline
$\kappa_2$ & $N_4$ & $I_{H^2}\times 10^3$ & $\nu'$ &$\chi^2/{\rm d.o.f.}$&$p$-value\\
\hline
\multirow{6}{2em} {\centering 2.45} & 6000 & 22.50(19)(124) & 0.33(2)(7) & 0.396 & 0.756 \\
& 8000 & 19.22(16)(105) & 0.21(2)(6) & 1.899 & 0.127 \\
& 12000 & 15.68(16)(86) & 0.05(2)(5) & 0.409 & 0.746 \\
& 16000 & 13.90(2)(76) & -0.03(0)(5) & 0.087 & 0.784 \\
& 24000 & 11.27(11)(62) & -0.11(2)(4) & 0.751 & 0.557 \\
& 32000 & 10.14(3)(56) & -0.16(0)(4) & 0.106 & 0.899 \\
\hline
\multirow{6}{2em} {\centering 3.0} & 8000 & 10.49(7)(38) & 0.43(5)(5) & 0.028 & 1.000 \\
& 12000 & 8.58(1)(31) & 0.30(1)(4) & 0.594 & 0.667 \\
& 16000 & 7.39(1)(27) & 0.17(1)(4) & 0.383 & 0.821 \\
& 24000 & 6.04(2)(21) & 0.07(2)(3) & 0.591 & 0.738 \\
& 32000 & 5.33(2)(19) & 0.01(2)(3) & 0.088 & 0.997 \\
& 48000 & 4.35(3)(15) & -0.14(1)(3) & 0.102 & 0.992 \\
\hline
\multirow{7}{2em} {\centering 3.4} & 8000 & 5.14(6)(15) & 0.52(11)(4) & 0.625 & 0.711 \\
& 12000 & 4.23(4)(13) & 0.40(6)(4) & 0.313 & 0.949 \\
& 16000 & 3.64(2)(11) & 0.27(6)(3) & 0.259 & 0.990 \\
& 24000 & 2.96(3)(9) & 0.21(7)(3) & 0.087 & 1.000 \\
& 32000 & 2.60(1)(8) & 0.14(2)(3) & 0.043 & 1.000 \\
& 48000 & 2.07(7)(6) & -0.01(9)(3) & 0.043 & 1.000 \\
& 64000 & 1.82(7)(5) & -0.04(8)(2) & 0.029 & 1.000 \\
\hline
\multirow{6}{2em} {\centering 3.8} & 16000 & 1.71(3)(10) & 0.44(5)(8) & 0.128 & 0.999 \\
& 24000 & 1.40(2)(8) & 0.34(8)(8) & 0.216 & 1.000 \\
& 32000 & 1.22(0)(7) & 0.27(3)(7) & 0.016 & 1.000 \\
& 48000 & 0.99(1)(5) & 0.16(6)(6) & 0.009 & 1.000 \\
& 64000 & 0.87(4)(5) & 0.13(9)(6) & 0.036 & 1.000 \\
& 96000 & 0.70(3)(4) & -0.01(10)(5) & 0.014 & 1.000 \\
\hline\hline

\end{tabular}
\end{table}
\end{center}

\begin{center}
\begin{table}
\caption{Results of fits to $S$ versus $H^2$ on all ensembles.  The error quoted for $S_0$ is the statistical error combined in quadrature with a systematic error that comes from varying the choice of fit range.  For the ensembles that have a second tuned $\beta$ value, in order to estimate a $\beta$ tuning systematic error (see Table \ref{tab:ensembles}), the quoted error on $S_0$ also includes an error based on the difference between the $S_0$ results on the two tunings.  The $\chi^2/{\rm d.o.f.}$ and $p$-value are quoted for the fits that produce the central values.
}
\label{tab:s0values}
\begin{tabular}{c c c c c } 
\hline \hline
$\kappa_2$ & $N_4$ & $S_0$ &$\chi^2/{\rm d.o.f.}$&$p$-value\\
\hline
\multirow{6}{2em} {\centering 2.45} & 6000 & -5.59(16) & 3.736 & 0.054 \\
& 8000 & -5.35(18) & 1.894 & 0.210 \\
& 12000 & -4.96(10) & 2.645 & 0.145 \\
& 16000 & -4.73(2) & 3.458 & 0.104 \\
& 24000 & -4.20(16) & 4.574 & 0.032 \\
& 32000 & -3.97(5) & 4.858 & 0.028 \\
\hline
\multirow{6}{2em} {\centering 3.0} & 8000 & -2.42(14) & 0.247 & 0.646 \\
& 12000 & -2.20(2) & 0.477 & 0.527 \\
& 16000 & -2.10(3) & 0.037 & 0.859 \\
& 24000 & -1.87(6) & 6.389 & 0.011 \\
& 32000 & -1.76(2) & 0.554 & 0.497 \\
& 48000 & -1.69(4) & 3.070 & 0.121 \\
\hline
\multirow{7}{2em} {\centering 3.4} & 8000 & -1.11(5) & 0.828 & 0.404 \\
& 12000 & -1.01(9) & 0.740 & 0.431 \\
& 16000 & -0.94(6) & 0.246 & 0.647 \\
& 24000 & -0.81(4) & 0.809 & 0.410 \\
& 32000 & -0.75(2) & 1.185 & 0.317 \\
& 48000 & -0.69(4) & 0.515 & 0.512 \\
& 64000 & -0.62(4) & 0.302 & 0.613 \\
\hline
\multirow{6}{2em} {\centering 3.8} & 16000 & -0.39(2) & 1.485 & 0.264 \\
& 24000 & -0.34(2) & 3.377 & 0.107 \\
& 32000 & -0.32(1) & 1.175 & 0.319 \\
& 48000 & -0.28(2) & 0.481 & 0.525 \\
& 64000 & -0.26(1) & 2.640 & 0.145 \\
& 96000 & -0.24(2) & 3.019 & 0.123 \\
\hline\hline

\end{tabular}
\end{table}
\end{center}

%

\end{document}